\documentclass[%
 reprint,
 amsmath,amssymb,
 aps,
]{revtex4-2}

\usepackage{bbm}
\usepackage[utf8]{inputenc}
\usepackage{amssymb}
\usepackage{graphicx}
\usepackage{dcolumn}
\usepackage{bm}

\usepackage[table]{xcolor} 
\usepackage{colortbl}

\usepackage{array}
\usepackage{adjustbox}
\usepackage{booktabs}
\usepackage{multirow}

\begin{document}

\preprint{APS/123-QED}

\title{Graph-based Descriptors for Condensed Matter}%

\author{An Wang}
\affiliation{Department of Chemistry, University of Warwick, Coventry CV4 7AL, United Kingdom}

\author{Gabriele C. Sosso}
\affiliation{Department of Chemistry, University of Warwick, Coventry CV4 7AL, United Kingdom}%
\email{G.Sosso@warwick.ac.uk}%

\date{\today}

\begin{abstract}
Computational scientists have long been developing a diverse portfolio of methodologies to characterize condensed matter systems. Most of the descriptors resulting from these efforts are ultimately based on the spatial configurations of particles, atoms, or molecules within these systems. Noteworthy examples include symmetry functions and the smooth overlap of atomic positions (SOAP) descriptors, which have significantly advanced the performance of predictive machine learning models for both condensed matter and small molecules. However, while graph-based descriptors are frequently employed in machine learning models to predict the functional properties of small molecules, their application in the context of condensed matter has been limited. In this paper, we put forward a number of graph-based descriptors (such as node centrality and clustering coefficients) traditionally utilized in network science, as alternative representations for condensed matter systems. We apply this graph-based formalism to investigate the dynamical properties and phase transitions of the prototypical Lennard-Jones system. We find that our graph-based formalism outperforms symmetry function descriptors in predicting the dynamical properties and phase transitions of this system. These results demonstrate the broad applicability of graph-based features in representing condensed matter systems, paving the way for exciting advancements in the field of condensed matter through the integration of network science concepts.
\end{abstract}

\maketitle


\section{Introduction}
\label{sec:intro}

Characterising the structural or dynamical properties of a condensed matter system requires the usage of one or more metrics capable to capture the state of the system - and, ideally, to identify any change from one state to the other. Finding such metrics is a fundamental challenge in the field, bacause it requires a deep understanding of the physical properties and behavior of the system, as well as innovative methods and tools to measure and analyze these characteristics~\cite{bari2021fundamental}. Pair correlations functions and diffusion coefficients are perhaps the most prominent examples of metrics that can be used to investigate the structural and dynamical properties of a condensed matter system, respectively. Many alternatives to these metrics exists, such as order parameters (Steinhardt~\cite{steinhardt1983bond} and ten Wolde~\cite{ten1995numerical}) or the intermediate scattering function (ISF)~\cite{van1954correlations,rahman1964correlations,kob1995testing1,kob1995testing2}. 

In the last few years we have witnessed a proliferation of ``descriptors", or ``features", or ``fingerprints" (all terms that we are going to consider interchangeable with ``metrics" for the sake of this discussion, albeit these ``metrics" must not be confused from the functions used to define distances in metric spaces) as a result of the success of machine learning(ML)-based methods in physical chemistry. Notable examples include symmetry functions (SF)~\cite{behler2007generalized,behler2011atom}, smooth overlap of atomic positions (SOAP)~\cite{bartok2013representing}, coulomb matrices~\cite{rupp2012fast,montavon2013machine} and tensorial representations~\cite{nye1985physical,schutt2017quantum} (particularly popular in the context of equivariant neural networks~\cite{cohen2016group,park2021accurate,schutt2017quantum,gilmer2017neural,anderson2019cormorant,thomas2018tensor}). Some hybrid approach have also recently been proposed, such as: (i.) the TimeSOAP method, which combines SOAP descriptors, time-dependent variation analysis, and ML to track and quantify high-dimensional fluctuations in complex molecular systems~\cite{caruso2023timesoap}; (ii.) a method that combines machine learning with quantum mechanics to predict the properties of materials and molecules through Gaussian process regression and SOAP~\cite{bartok2017machine}, and; (iii.) a comprehensive structural and chemical similarity metric that combines SOAP, regularized entropy matching, and machine learning~\cite{de2016comparing}.

The vast majority of these descriptors are ultimately built starting from the distances and or angles between the particles, atoms, or molecules - that is, by means of information derived from Euclidean geometry. On the contrary, graph-based descriptors - which do not necessarily relies on metric structures to offer a representation of the system, are rarely used in condensed matter research. A significant exception is the VGOP method~\cite{chapman2022efficient}, which is used for atomic structure characterization. The primary goal of introducing a graph-based feature is to eliminate metric structures from the geometric information, preserving solely the system's topological structure. However, the VGOP method, which depends on truncations of an ordered series of distances and the entropy of degree distribution, falls short of fully accomplishing this. We will revisit this method in Section~\ref{sec:graph} and discuss its limitations in detail in Appendix~\ref{app:alternative}.
Interestingly, in the context of small molecules, graph-based descriptors are much more common: a non-comprehensive list would include the topological polar surface area (TPSA)~\cite{ertl2000fast}, the Hosoya index (molecular complexity)~\cite{hosoya1971topological} of the Wiener index~\cite{wiener1947structural}, which is derived from well-established concepts of chemical graph theory~\cite{bonchev2018chemical,bonchev2018chemical,diudea2018energetics}.

Here, we leverage descriptors traditionally used to deal with complex networks to put forward a portfolio of novel, graph-based descriptors that are capable to accurately represent the structural and dynamical properties of condensed matter systems. To this end, we focus on the challenging task to build an interpretable machine learning model capable of characterising flow defects in condensed matter. This is a non-trivial endeavour, as this dynamical property of the system stems from only a small fraction of its particles - namely those that are majorly prone to a structural re-arrangement within a certain timescale~\cite{alsayed2005premelting,taylor1934mechanism}. 

For a many-body system composed of $\mathcal{N}$ particles $\bold{x}'=(x_1, \cdots, x_{\mathcal{N}})^T$ evolving over time $t$, $N$ particle environments $\bold{x}=(x_1, \cdots, x_N)^T$ can be sampled along its evolution over time ($t$). Finding a set of descriptors $\bold{X}^{(l)}=h_l(\bold{x}), l \in [1,n]$, that characterize the properties of each particle (a process often referred to as ``feature engineering") is key to studying the local properties of the system via machine learning approaches. For convenience, we shall group the various descriptor vectors $\bold{X}^{(l)}$ into a feature matrix $\bold{X} = (\bold{X}^{(1)}, \cdots, \bold{X}^{(n)})$, where each row $\bold{X}_{(z)}$ represents a complete sample, and each column $\bold{X}^{(l)}$ represents the value of a feature across all samples. In supervised  learning methods, a composite function $\bold{y}=f(h_1(\bold{x}), \cdots, h_n(\bold{x}))$ needs to be fitted, where these $n$ inner functions $h_{1},\cdots,h_{n}$ are derived from feature engineering, the outer function $\bold{y}=f(\bold{X})$ represents models such as neural networks~\cite{mcculloch1943logical,rosenblatt1958perceptron,rumelhart1986learning} or support vector machines (SVM)~\cite{boser1992training,cortes1995support}, and $\bold{y}$ is the dependent variable indicating the target property. 

Cubuk~\textit{et al.} previously employed~\cite{cubuk2015identifying} a combination of radial ($G$) and angular ($\Psi$) symmetry functions~\cite{behler2007generalized} as descriptors ($\bold{X}$) to build a SVM ($f$) model capable to classify a given particle within a condensed matter system as either rearrangement-prone or stable. Specifically, the $i$-th particle can either be rearrangement-prone (or ``soft", $y_i = 1$) or stable (``hard", $y_i = 0$) - and this can be measured for all $i \in [1,N]$, where $y_i$ is the $i$-th element of $\bold{y}$. At which point, one can measure the ``softness" of each particle by quantifying the distance to the decision plane. By using a linear kernel function, this distance can be interpreted in the original input space.~\cite{cubuk2015identifying}. Whether a particle $i$ is soft at time $t$ can thus be assessed according to the magnitude of the non-affine displacement $D_{\text{min},0}^2$ at time $t$~\cite{cubuk2015identifying}, or the degree of positional change $p_{\text{hop},i}(t)$ within the neighborhood of time $t$, i.e., the interval of $[t-\Delta t,t+\Delta t]$, where $\Delta t$ is a small positive time scale~\cite{schoenholz2016structural,smessaert2013distribution}. In particular, if $p_{\text{hop},i}(t)$ exceeds a certain threshold value $p_c$ (which can be set according to the system and/or timescale under investigation) the particle is considered to be soft. The softness is a measure of how the local structure of a given system influence its local dynamics (it has been shown that the softness is influenced by the first coordination number~\cite{schoenholz2016structural}), and provides an enticing observable to characterise complex dynamical processes within disordered systems~\cite{cubuk2015identifying}. For example, the concept of softness has been successfully used to explain the non-exponential relaxation dynamics observed in glassy liquids~\cite{schoenholz2016structural} and to predict crystallization pathways in soft colloidal glasses~\cite{ganapathi2021structure}.

Although the SF $G$ and $\Psi$ can be used to analyse and predict the dynamical behavior of particles in disordered solids, their applicability under different thermodynamic conditions, e.g., temperature or pressure, is rather limited (see sections~\ref{sec:trad} and ~\ref{subsec:results_ml} for some relevant theoretical considerations and results, respectively). Besides, in supervised learning, assuming that data is independent and identically distributed (i.i.d.) is key for model generalization, evaluation, and selection. However, in many-body systems - where the probability that a particle belongs to a certain category (soft or hard) is influenced by its neighbors, the i.i.d. assumption does not hold. An appropriate feature engineering can mitigate this issue by ensuring features are diverse and heterogeneous, covering a broad range of system properties. However, it is important to note that it cannot fully overcome the inherent non-i.i.d. nature characteristic of many-body interactions.

Here, rather than utilising approaches based on Euclidean geometry in $\mathbb{R}^3$, we model particles and their interactions within a many-body system as a graph $\mathcal{G}(\mathcal{V},\mathcal{E})$. This approach enables us to construct heterogeneous descriptors directly from the graph structure, transforming the $l$-th descriptor $\bold{X}^{(l)}, l \in [1,n]$ into a function $h_l: (\mathcal{V},\mathcal{E}) \rightarrow \mathbb{R}$ that operates on vertices and edges. In particular, we aim to develop a scale-independent approach designed to identify properties that remain consistent despite geometric variations like stretching or compression (which in turns makes them robust with respect to variations in terms of, e.g., the temperature or the pressure of the physical system). This approach is detailed in section~\ref{sec:graph}.

To demonstrate the applicability of our graph-based approach, we focused on the prototypical Lennard-Jones system. Specifically, we built a machine learning model analogous to that of Cubuk \textit{et al.}~\cite{cubuk2015identifying} with the purpose of labelling each particle as either soft or hard. This allowed us to directly compare the performance of our graph-based descriptors with the results previously obtained by Cubuk \textit{et al.} utilising SF instead. This comparison utilized datasets generated from molecular dynamics (MD) simulations, which are described in detail in section~\ref{sec:md}, and employed a set of control (computational) experiments detailed in section~\ref{sec:ml}. The results, summarised in section~\ref{subsec:results_ml}, reveal that our graph-based approach not only outperforms the SF-based approach utilised by Cubuk \textit{et al.} but also significantly surpasses it in terms of the transferability of the model. These outcomes demonstrate that our graph-based descriptors can effectively capture the structural invariance of the system across different thermodynamic conditions, which in turn leads to a more robust structural characterization.

Overall, our graph-based approach allows for an intuitive representation and analysis of complex interaction networks, which is especially beneficial for disordered materials. By leveraging graph theory, we can accurately pinpoint the key structural characteristics of the system, as discussed in section~\ref{subsec:results_graph_metrics}.

Finally, we have used our best-performing soft-hard classifier to construct the hard particle counter $\mathfrak{N}(t)$ and derive the average softness $\mathfrak{S}(t)$ as functions of time $t$, thus linking the local environments with a global property of the system. This is discussed in section~\ref{subsec:applications}, where we demonstrate that $\mathfrak{N}(t)$ and $\mathfrak{R}(t)$ can be used to accurately describe the crystallization process of supercooled liquids over $t$.

\section{Theory and Methods}

\subsection{Symmetry functions-based approach}
\label{sec:trad}

The approach previously utilised by Cubuk \textit{et al.}~\cite{cubuk2015identifying,schoenholz2016structural} leverages the coordinates of the $\mathcal{N}$ particles in either
$\mathbb{R}^2$ (when dealing with a two-dimensional system) or $\mathbb{R}^3$ (when dealing with a three-dimensional system) to obtain radial ($G$) and angular ($\Psi$) symmetry functions as follows:

\vspace{0.25cm}
\paragraph{Radial structural function ($G$ function)}

\begin{equation}
    G_I (i;r,\delta) = \sum_{j\in I} \text{e}^{-\frac{1}{2 \delta^2} (r-R_{ij})^2},
    \label{equ:g_function}
\end{equation}

\noindent where $R_{ij}$ represents the distance between particles $i$ and $j$ in $\mathbb{R}^3$ and $I$ denotes a specific chemical species whose density we aim to examine. The parameter $r$ determines the ``detection radius" around the particle $i$, while $\delta$ defines the ``resolution", so that the radial distribution at different distances can be smoothly obtained by adjusting the values of $r$ and $\delta$.

\vspace{0.25cm}
\paragraph{Angular structural function ($\psi$ function)}

\begin{equation}
    \Psi_{IJ} (i; \xi,\lambda,\zeta) = \sum_{j \in I} \sum_{k \in J} \text{e}^{-\frac{R_{ij}^2+R_{jk}^2+R_{ik}^2}{\xi^2}} (1+\lambda \text{cos} \theta_{ijk})^{\zeta},
    \label{equ:psi_function}
\end{equation}

\noindent where $\theta_{ijk}$ refers to the angle formed between two vectors $\overrightarrow{\bold{R}_{ij}}$ and $\overrightarrow{\bold{R}_{ik}}$, $\lambda = \pm1$ indicates whether small or large bond angles are under consideration, $\zeta$ is a parameter that defines the angular resolution and $I$ and $J$ represent distinct chemical species. Note that the interaction between particles $i$, $j$, $k$ is considered significant only when $R_{ij}^2+R_{jk}^2+R_{ik}^2<\xi^2$. Thus, the $\Psi$ function can explore different angular features of the local environment around the $i$-th particle by adjusting the values of $\xi$, $\lambda$ and $\zeta$.

In fact, by varying $I$, $J$, $r$, $\delta$, $\xi$, $\lambda$ and $\zeta$, we can get different $G$ and $\Psi$ functions for all $N$ particles, which can then be used to construct the set of response variables $\bold{X}$ for a given machine learning model $\bold{y}=f(\bold{X})$. These structural functions were originally designed to create high-dimensional potential energy surfaces, by capturing the local environment of particles via encoding distances and angles among particles in a way that is invariant to rotation, translation, or permutation~\cite{behler2007generalized}. Prior research has shown that $G$ functions play a very important role in capturing the structural properties of a given system~\cite{cubuk2015identifying,schoenholz2016structural,ganapathi2021structure}. Indeed, $G$ functions essentially represents a discretized (or even, parameterized) analogous of the radial distribution function $g(r)$~\cite{cubuk2015identifying}, as:

\begin{equation*}
    \lim_{\delta \rightarrow 0} \text{e}^{-\frac{(r-R_{ij})^2}{2 \delta^2}} = \delta_D(R_{ij}-r),
\end{equation*}

\noindent where $\delta_D$ represents the Dirac delta. The radial distribution function $g(r)$ quantifies the likelihood of finding another particle at distance $r$, as a ratio to the probability relative to a random distribution of particles at the same density. It is derived by summing $\delta_D (r-R_{ij})$ over index $j$ averaging across the index $i$, and normalizing the result. In two dimensions, normalization involves the annulus's circumference, i.e., $2\pi r$, around particle $i$ at distance $r$. In three dimensions, it uses the spherical shell's surface area, i.e., $4\pi r^2$, at the same distance.

Then, the relationship between $g(r)$ and $G$ function is:

\begin{itemize}
    \item Two-dimensional system:
    \begin{equation*}
        \lim_{\delta \rightarrow 0} \frac{\langle G_I (i;r,\delta) \rangle_i}{2 \pi r} = \frac{1}{2 \pi r} \left \langle \sum_{j \in I} \delta_D (R_{ij}-r) \right \rangle_i = g(r)
    \end{equation*}

    \item Three-dimensional system:
    \begin{equation*}
        \lim_{\delta \rightarrow 0} \frac{\langle G_I (i;r,\delta) \rangle_i}{4 \pi r^2} = \frac{1}{4 \pi r^2} \left \langle \sum_{j \in I} \delta_D (R_{ij}-r) \right \rangle_i = g(r).
    \end{equation*}
\end{itemize}

Therefore, the $G$ function is a parametric representation of $g(r)$, which means this particular flavour of feature engineering - using different $G$ functions derived from varying $r$, $\delta$ and $I$ - provides a structural characterization of the system which is equivalent to that provided by the $g(r)$. Although $G$ functions, with their parameterization and high-dimensional vector representation, can describe local structural features in more detail and flexibility, they essentially still provide information about particle density distribution. Note that both the radial distribution function $g(r)$ and the $G$ functions are highly sensitive to the phase of matter (e.g., liquid, crystal) and thermodynamic quantities (e.g., temperature, pressure, volume), which explains the inherent limitations of $G$ in terms of their sensitivity, as metric structures, to the specific phase of matter and/or thermodynamic conditions of the system.

Both $G$ and $\Psi$ symmetry functions are formulated within $\mathbb{R}^2$ or $\mathbb{R}^3$ via Euclidean geometry. By quantifying inter-particle distances and angles, $G$ and $\Psi$  map out the local structural environment of particles. The metric structures on $\mathbb{R}^2$ or $\mathbb{R}^3$ are sensitive to phases and thermodynamic quantities, leading to the $G$ and $\Psi$ functions, as well as machine learning models that use them as input features, being sensitive to changes to them. As a result, classifiers trained under specific phases and thermodynamic conditions may not be applicable when transferred to other states and thermodynamic variables. We explicitly verify that this is the case via a specific set of computational experiments reported in section~\ref{subsec:con_exp}.

Cubuk \textit{et al.} noted that when using $D_{\text{min},0}^2$ as the threshold for determining whether a particle is soft or hard, it is possible to adapt to temperature fluctuations based on temperature $T$ (e.g., through $T \sigma$), thereby preserving the predictive accuracy of machine learning models across different thermal conditions~\cite{cubuk2015identifying}. However, when utilizing the behavior of $p_{\text{hop},i}(t)$ (see Section~\ref{subsec:some_metrics}d) as the criterion to determine if a particle is soft or hard, or regarding its applicability across changes in volume or state of matter, this modification is of limited utility. Moreover, this modification based on $T \sigma$ also alters the criteria for classifying a particle as soft or hard (see Section~\ref{subsec:some_metrics}d).

In contrast, the topological quantities we have used here to construct our graph-based descriptors (which we will introduce in the next section) offers a distinct and possibly more fundamental viewpoint to capture the structural intricacies of condensed matter systems, notwithstanding their state and/or thermodynamic conditions.

\subsection{Graph-based approach}
\label{sec:graph}

Exploring many-body systems from a topological viewpoint involves two key steps. Firstly, we need to map these systems onto graphs, as detailed in section~\ref{subsec:cons_graph}. Secondly, we need to construct some functions - defined on graphs specifically - as descriptors, as discussed in section~\ref{subsec:des_graph}. This process shares some aspects with the work of Chapman \textit{et al.}~\cite{chapman2022efficient}, who previously introduced the so-called ``VGOP'' method, i.e., a graph-based approach to describe ``atomic structures". However, there are several, fundamental differences between our approach and the VGOP method, which we discuss in detail in Appendix~\ref{app:alternative}. We also stress that the graph-based metrics we have constructed and utilised in this work are by no means specific to the Lennard-Jones system we have investigated here. In fact, graph-based metrics are effectively transferable across majorly different length scales. The fact that - as we will show in the section~\ref{subsec:results_ml} - we are leveraging in this work some metrics that were originally designed to characterise complex networks is a testament to the versatility of this approach - and the potential it holds in condensed matter physics and beyond.

\subsubsection{Basic Concepts}
\label{subsec:basic_concepts}

\textbf{Topology and Graphs.} Topology examines how objects retain their properties through continuous deformation~\cite{armstrong2013basic}, offering deep insights into the invariant properties of complex systems, especially disordered or crowded systems. A graph is a typical example of low-dimensional topology.

A topological space $U$ is a CW complex~\cite{whitehead1949combinatorial} if it can be expressed as a union of a series of increasing subspaces $U = \bigcup_{n=0}^\infty U^n$, where each $U^n$ is called the n-skeleton and satisfies the following conditions in recursion:
\begin{itemize}
    \item $U^0$ is a discrete set of points;
    \item For $n \geq 1$, $U^n$ is constructed by attaching some n-dimensional spheres $B^n$ to $U^{n-1}$, i.e., $U^n = U^{n-1} \cup \bigcup_{\alpha \in A_n} e_\alpha^n$, where each $e_\alpha^n$ is an n-dimensional cell, $e_\alpha^n \cong B^n$, and the attaching map is $\phi_\alpha: S^{n-1} \to U^{n-1}$. The set $A_n$ denotes the index set for constructing the n-dimensional cells.
\end{itemize}

A graph $\mathcal{G}$ can be seen as a special 1-dimensional CW complex, where the 0-dimensional cell is the element of node set $\mathcal{V}(\mathcal{G})$, and the 1-dimensional cell is the element of edge set $\mathcal{E}(\mathcal{G})$. The graph is formed by ``gluing" 0-dimensional cells onto a 1-dimensional skeleton~\cite{hatcher2002algebraic}. In other words, the edge set $\mathcal{E}(\mathcal{G})$ is the result of selecting a topology of the given node set $\mathcal{V}(\mathcal{G})$. This topological information is described via the adjacency matrix $\mathcal{M}(\mathcal{G}) = (\mathcal{M}_{i,j})_{|\mathcal{V}(\mathcal{G})| \times |\mathcal{V}(\mathcal{G})|}$, which describes the way 0-dimensional cells are glued together. If $e_i,e_j \in \mathcal{V}(\mathcal{G})$ and $(e_i,e_j) \notin \mathcal{E}(\mathcal{G})$ then $\mathcal{M}_{i,j} = 0$; else if $e_i,e_j \in \mathcal{V}(\mathcal{G})$ and $(e_i,e_j) \in \mathcal{E}(\mathcal{G})$ then $\mathcal{M}_{i,j} = 1$.

From the aspect of topology, the $0$-dimensional cells (which are the basic building blocks of a CW complex) correspond to the vertex set $\mathcal{V}(\mathcal{G})$ of the graph. Each vertex can be considered an independent point, i.e., a $0$-dimensional space. The 1-dimensional cells correspond to the edge set $\mathcal{E}(\mathcal{G})$ of the graph. Each edge can be considered a line segment connecting two vertices, i.e., a 1-dimensional space. Formally, each edge is a continuous map from the interval of $[0,1]$ to the topological space, with the endpoints of this interval mapped to two 0-dimensional cells (vertices). In the graph, if there is an edge between vertices $e_i$ and $e_j$, it is denoted as $(e_i, e_j) \in \mathcal{E}(\mathcal{G})$. The graph $\mathcal{G}$ is a 1-dimensional CW complex because it is constructed by gluing 0-dimensional cells (vertices) and 1-dimensional cells (edges).

In the construction of CW complexes, gluing higher-dimensional (larger than 1) cells to build higher-dimensional complexes is common practice in persistent homology theory~\cite{edelsbrunner2002topological,bubenik2015statistical}. However, graph theory usually only concerns itself with 0-dimensional cells (vertices) and 1-dimensional cells (edges), i.e., the structure of graphs, and does not involve these higher-dimensional structures.

\textbf{The Instantiation of Complexes.} The definition of a CW complex only describes the process of constructing higher-dimensional structures by gluing lower-dimensional cells. It serves as a general framework and does not specify which lower-dimensional cells need to be glued together, or the criteria for determining whether there is an edge between nodes. However, for a graph, the specific criteria for determining which nodes have edges between them is crucial. Therefore, mathematicians typically instantiate the CW complex as specific complexes, such as Vietoris-Rips Complex (VR Complex)~\cite{vietoris1927hoheren}, Čech Complex~\cite{vcech1932theorie,bourbaki1966elements}, or Alpha Complex~\cite{edelsbrunner1983shape,edelsbrunner1994three}, which clearly define the criteria for determining whether there is an edge between two nodes.

\vspace{0.1cm}
1. \textbf{\textit{Vietoris-Rips Complex (VR Complex)}}:
   Given a set of points $\mathcal{V}(\mathcal{G})$ and a distance threshold $\epsilon$, the Vietoris-Rips Complex determines whether there is an edge between two nodes as follows:
     \begin{equation*}
         (e_i, e_j) \in \mathcal{E}(\mathcal{G}) \iff d(e_i, e_j) \leq \epsilon
     \end{equation*}
     where $d(e_i, e_j)$ represents the distance between $e_i$ and $e_j$.
     
\vspace{0.1cm}
2. \textbf{\textit{Čech Complex}}:
   Given a set of points $\mathcal{V}(\mathcal{G})$ and a distance threshold $\epsilon$, the Čech Complex determines whether there is an edge between two nodes as follows:
     \begin{equation*}
         (e_i, e_j) \in \mathcal{E}(\mathcal{G}) \iff B(e_i, \epsilon/2) \cap B(e_j, \epsilon/2) \neq \emptyset
     \end{equation*}
     where $B(e_i, \epsilon/2)$ denotes the ball centered at node $e_i$ with radius $\epsilon/2$.
     
\vspace{0.1cm}
3. \textbf{\textit{Alpha Complex}}:
   Given a set of points $\mathcal{V}(\mathcal{G})$ and a parameter $\alpha$, the Alpha Complex is based on the Delaunay triangulation and Alpha shapes. It determines whether there is an edge between two nodes as follows:
  \begin{align*}
    (e_i, e_j) \in \mathcal{E}(\mathcal{G}) &\iff \exists \ \text{Delaunay edge} \ (e_i, e_j) \\
    &\text{such that} \ r_{e_i e_j} \leq \alpha
  \end{align*}

Here, for any four points $e_i, e_j, e_k, e_l \in \mathcal{V}(\mathcal{G})$, if there exists a circle $\mathcal{C}$ passing through $e_i, e_j, e_k$ such that $e_l$ is not inside $\mathcal{C}$, then $(e_i, e_j)$ is a Delaunay edge:

\begin{align*}
(e_i, e_j) \in \mathcal{E}(\mathcal{G}) \iff \forall e_k \in \mathcal{V}(\mathcal{G}) \setminus \{e_i, e_j\}, \\ \exists \ \mathcal{C} \ \text{such that} \ e_i, e_j, e_k \in \partial \mathcal{C} \text{ and} \\ \forall e_l \in \mathcal{V}(\mathcal{G}) \setminus \{e_i, e_j, e_k\}, \ e_l \notin \mathcal{C}
\end{align*}

The parameter $\alpha$ determines the maximum allowed Alpha radius $r_{e_i e_j}$ when constructing the Alpha Complex, and the Alpha radius $r_{e_i e_j}$ is defined as the radius of the smallest circumcircle passing through $e_i$ and $e_j$:
\begin{equation*}
    r_{e_i e_j} = \min \{ r \mid \exists \ x \in \mathbb{R}^2 \ \text{such that} \ d(x, e_i) = d(x, e_j) = r \}
\end{equation*}

Exceptionally, in the construction of 1-dimensional Alpha complex, if the parameter 
$\alpha$ for truncation is not introduced (or is infinity), then the constructed structure is the Voronoi graph, which can be seen as a non-parametric spatial partitioning on the whole space, with the Delaunay triangulation as its dual structure. Specifically, given a set of points $\mathcal{V}(\mathcal{G})$, the Voronoi cell $\mathcal{O}(e_i)$ for each point $e_i \in \mathcal{V}(\mathcal{G})$ is defined as:

\begin{align*}
    \mathcal{O}(e_i) = \{ x \in \mathbb{R}^2 \mid d(x, e_i) < d(x, e_k) \ \forall e_k \in \mathcal{V}(\mathcal{G}), k \neq i \}
\end{align*}

Two nodes $e_i, e_j \in \mathcal{V}(\mathcal{G})$ have an edge between them if and only if their Voronoi cells share a line segment:

\begin{align*}
    (e_i, e_j) \in \mathcal{E}(\mathcal{G}) \iff \exists \, \epsilon > 0 \ \text{such that } \mathcal{O}(e_i) \cap \mathcal{O}(e_j) \\ = \{ x \in \mathbb{R}^2 \mid \| x - (1 - t)e_i - t e_j \| = 0, \ 0 \leq t \leq 1 \}    
\end{align*}

In conclusion, both the VR Complex and the Čech Complex rely on a simple distance truncation, which is controlled by the parameter $\epsilon$. The Alpha Complex uses distance truncation according to $\alpha$ to select which detected Voronoi neighbors will have edges between them. The Voronoi graph is a special case of the 1-dimensional Alpha Complex when $\alpha \rightarrow +\infty$, which considers all candidate points in the whole space (no truncation) to determine which nodes have edges between them, making it a non-parametric method.

\textbf{Multi-scaling Features.} Without loss of generality, we use VR complexes as an example. Suppose we have a point cloud containing the coordinates of each particle, denoted as $\mathfrak{P} = \{ p_1, p_2, \ldots, p_n \}$, where each point $p_i = (x_i, y_i, z_i)$ is a point in 3-dimensional space (i.e., a subspace of $\mathbb{R}^3$). Based on a simple cutoff distance $\epsilon$, the VR complex corresponding to $\mathfrak{P}$ can be constructed, denoted as $\mathrm{VR}(\mathfrak{P}, \epsilon)$.

In practical applications, we typically select a monotonically increasing parameter sequence: 

\begin{equation}
    0< \epsilon_0 < \epsilon_1 < \epsilon_2 < \ldots < \epsilon_n
    \label{equ:filtrations}
\end{equation}

\noindent thereby obtaining a series of complexes which can characterize multi-scale structural information as follows:

\begin{equation}
    \emptyset = K_0 \subseteq K_1 \subseteq K_2 \subseteq \ldots \subseteq K_n = \mathrm{VR}(\mathfrak{P}, \epsilon_n)
    \label{equ:filtered_complexes}
\end{equation}

\noindent where $K_i = \mathrm{VR}(\mathfrak{P}, \epsilon_i)$, where $i \in [0,n] \cap \mathbb{N}$.

In persistent homology,~(\ref{equ:filtrations}) and~(\ref{equ:filtered_complexes}) are the so-called filtration variables and filtered complexes, respectively. The difference is that, in persistent homology theory, researchers focus on each order of homology groups, i.e., $H_0$, $H_1$ and $H_2$, which correspond to connected components, cycles, and caves, respectively.  

However, in the context of graph theory and network science, researchers are chiefly concerned with graph metrics such as degree, node centralities~\cite{rodrigues2019network,landherr2010critical}, clustering coefficient~\cite{watts1998collective}, etc. By taking a set of descriptors from graphs corresponding to different $\epsilon$ values and concatenating them, one can obtain a multi-scale feature engineering for structural characterization.

\textbf{The VGOP Method.} The VGOP (Chapman \textit{et al.})~\cite{chapman2022efficient} we mentioned earlier leverages the methodology we discussed in Appendix~\ref{app:alternative}. In this case, the filtration variables lead to:

\begin{equation}
    0< r_{c,1} < r_{c,2} < \cdots < r_{c,n}
    \label{equ:fv_example}
\end{equation}

After constructing the 1-dimensional VR complex (graphs) corresponding to these filtration variables in (\ref{equ:fv_example}):

\begin{equation}
    \emptyset = S_0 \subseteq S_1 \subseteq S_2 \subseteq \ldots \subseteq S_n = \mathrm{VR}(\mathfrak{P}, r_{c,n})
    \label{equ:fg_example}
\end{equation}

one can calculate the SGOP value for node $i$ in the $S_w=\mathrm{VR}(\mathfrak{P}, r_{c,w})$, where $w \in [1,n] \cap \mathbb{N}$:

\begin{equation}
    \theta_{\text{SGOP}}^{(w,i)} = \left( \sum_{k \in deg(S_w)} P^{(w)}(k) \mathop{\log} P^{(w)}(k) + k \cdot P^{(w)}(k) \right)^3
    \label{eq:entropy_degree_dist}
\end{equation}

Here, $P^{(w)}(k)$ is the probability of degree $k$ occurring in subgraph $S_w$, and $k \in deg(S_w)$ refers to the summation for any node within $S_w$ that has a degree of $k$.

By taking the different SGOP values corresponding to particle $i$ from the series of graphs in (\ref{equ:fg_example}) and concatenating them, one can obtain the multi-scale descriptors used to characterize the local environment of particle $i$, namely the VGOP:

\begin{equation*}
    \theta_{\text{VGOP}}^{(i)} = [\theta_{\text{SGOP}}^{(w_1,i)},\theta_{\text{SGOP}}^{(w_2,i)},\cdots,\theta_{\text{SGOP}}^{(w_n,i)}]
\end{equation*}

We provide a detailed introduction to this approach in Appendix~\ref{app_sub_vgop}. The criterion for constructing the graph is such that if the distance between two particles is less than $r_c$, then there is an edge between their corresponding nodes in the graph. Thus, in order to calculate VGOP it is necessary to use the degree distribution of the graphs corresponding to different cutoff radii. Note that the degree distribution in (\ref{eq:entropy_degree_dist}) is expressed in the form of entropy instead.

In Appendix~\ref{app:sub_vgop_vs_rdf}, we rigorously prove that by using the filtration variables in Eq.(3) as different cutoff radii $r_c$ to construct graphs, the resulting entropy expression of the degree distribution is, in fact, still a discrete version of the radial distribution function. Most importantly, this method cannot avoid the usage of a metric structure from the geometric information, and thus cannot rely solely on the topological structure. In fact, the VGOP approach is still metric-dependent, so that it remains sensitive to different thermodynamic quantities and phase of matter. In Appendix~\ref{app:sub_res_vgop}, we compared VGOP with the descriptors defined in section~\ref{subsec:des_graph}, and the results demonstrate that our descriptors comprehensively outperform VGOP in machine learning tasks.

\textbf{Our Approach.} Several modifications of the Voronoi tessellation have been proposed, such as the edge-weighted edition of the Voronoi method~\cite{wang2009edge}. Here, we have chosen the modified Voronoi method proposed by Malins~\textit{et al.}, which has been proven to be effective in the context of topological cluster classification ~\cite{malins2013identification}. 

The need for a modified Voronoi method arises from the following considerations~\cite{malins2013identification}:

\begin{enumerate}
    \item The original Voronoi method is highly sensitive to thermal fluctuations, which can lead to significant changes in the structure of the Voronoi graph.

    \item The original Voronoi method may incorrectly identify some particles as neighbors even though they are not in direct contact, leading to errors in identifying multi-member ring structures.

    \item Multi-member rings are very common structures in physical chemistry, but the original Voronoi method performs poorly in detecting these rings with high accuracy.

\end{enumerate}

Malins' modified Voronoi method essentially builds upon the original method by adding the definitions of direct Voronoi neighbors and the quadrilateral asymmetry control. The former means that only if the Voronoi cells of two particles share a face and the line connecting the particle positions intersects this shared face, will they be considered as candidate nodes for bonding. The latter involves introducing a dimensionless parameter $\mathcal{A} \in (0.51,1]$, which determines the maximum distortion of a four-member ring in a plane. When a bond forms between two opposite particles, this parameter is used to check the integrity of the four-member ring. If the distortion exceeds the set value, the four-member ring is considered to have been split into two three-member rings~\cite{malins2013identification}.

In summary, the filtration variables here are no longer the truncation radius, but the dimensionless parameters for the quadrilateral asymmetry control:

\begin{equation*}
    0.5 < \mathcal{A}_1 < \mathcal{A}_2 < \cdots < \mathcal{A}_n \leq 1
\end{equation*}

And the following filtered complexes are obtained:

\begin{equation}
    G_1 \subseteq G_2 \subseteq \ldots \subseteq G_n = \mathrm{MVon}(\mathfrak{P}, \mathcal{A}_n)
    \label{equ:fg_A_von_graphs}
\end{equation}

Here $G_i = \mathrm{MVon}(\mathfrak{P}, \mathcal{A}_i)$, where $i \in [0,n] \cap \mathbb{N}$.

This method is detailed in the next section, i.e., section~\ref{subsec:cons_graph}. We select one or several graphs from (\ref{equ:fg_A_von_graphs}) to calculate the descriptors defined in section~\ref{subsec:des_graph}.

\subsubsection{Graph Generation}
\label{subsec:cons_graph}

Several classes of different objects within a given system can be represented by the vertices of a graph, and the interactions between these objects are represented by the edges of the graph. In this work, we used a modified Voronoi method from computational geometry, which has been proven effective in topological cluster detection~\cite{malins2013identification}, to construct graphs from the particle coordinates of a many-body system.

Consider a set of points $\mathcal{V} = \{ v_1, \cdots, v_{\mathcal{N}}\}$ with coordinates in $\mathbb{R}^d$, where $d = 3$ is the dimensionality of the systems we are interested in. Here, we use $\overrightarrow{v_i}$ to represent the coordinate of the $i$-th point, and $\overrightarrow{r_{ij}} = \overrightarrow{v_j} - \overrightarrow{v_i}$ to represent the vector from the position of point $v_i$ to the position of point $v_j$. From the abstract perspective of algebraic topology and homology theory in section~\ref{subsec:basic_concepts} (Alpha Complex), the definition under the original Voronoi graph~\cite{voronoi1908nouvelles} can be expressed in the terminology of graph theory as follows:

\begin{enumerate}
    \item Voronoi Cell:
    The Voronoi cell of point $v_i$ is
    \begin{equation*}
    \mathcal{O}(v_i) = \begin{aligned}
    & \{ \overrightarrow{v} \in \mathbb{R}^d : ||\overrightarrow{v}-\overrightarrow{v_i}|| \leq ||\overrightarrow{v}-\overrightarrow{v_j}|| \\
    & \text{with } \forall{j} \neq i, v_j \in \mathcal{V} \}
    \end{aligned}
    \end{equation*}
    where $||.||$ represents the Euclidean distance.
    
    \item Voronoi Graph: 
    The set of Voronoi cells, represented by $\mathcal{O}$, generated by all points in a given $\mathcal{V}$ forms a Voronoi graph. The latter is denoted as $\mathcal{G}=(\mathcal{V},\mathcal{E})$, where $\mathcal{E}$ is the set of edges that are constructed based on the shared boundaries of Voronoi cells. These boundaries of $\mathcal{O}$ are segments for $d=2$ and planes for $d=3$.
\end{enumerate}

\begin{figure}[htbp]
\includegraphics[width=0.5\textwidth]{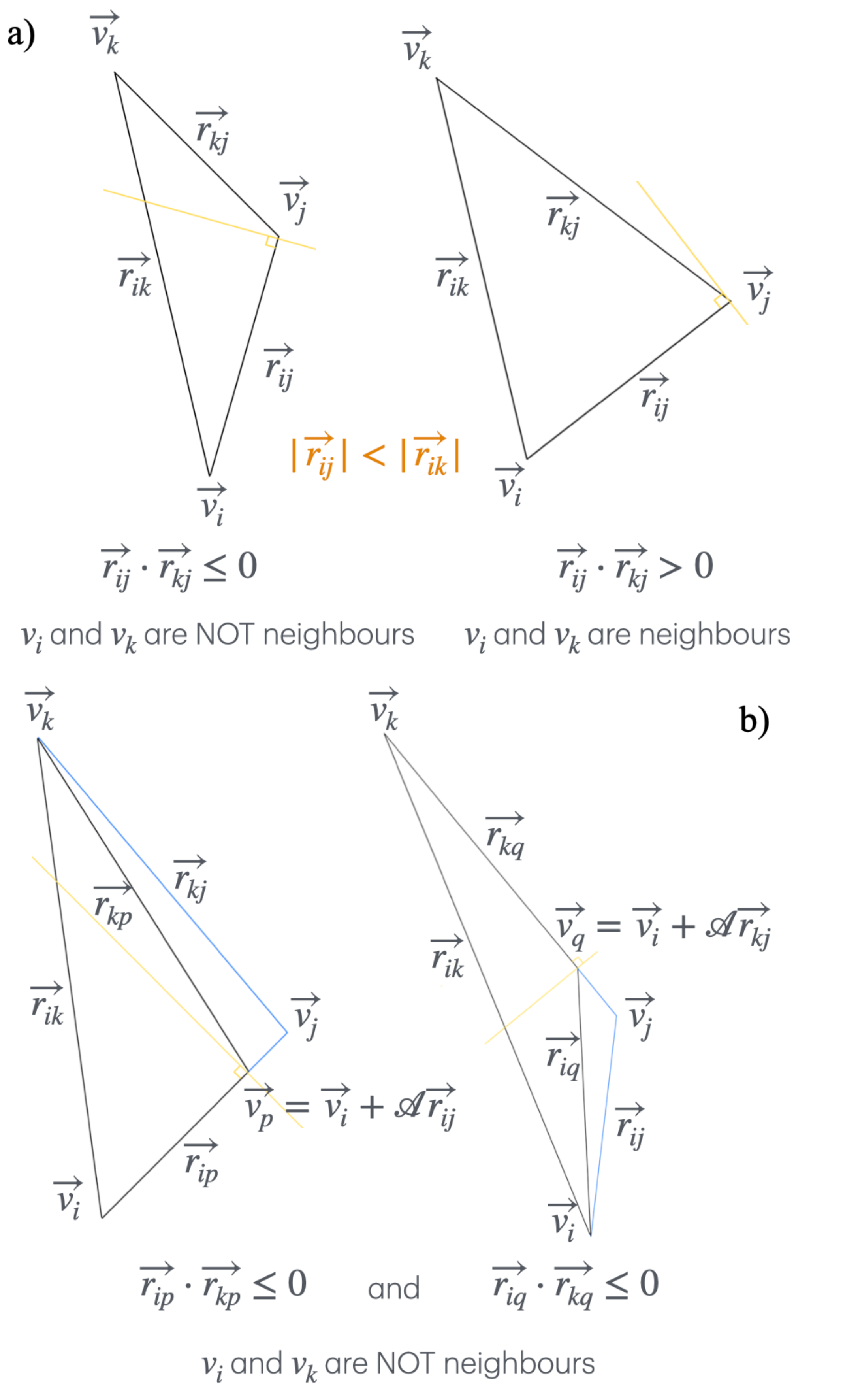}
\caption{The two key steps of the modified Voronoi Method: a) Direct Voronoi Neighbors Determination, b) Quadrilateral Asymmetry Control according to $\mathcal{A}$.}
\label{fig:U}
\end{figure}

The modified Voronoi method~\cite{malins2013identification} we have chosen to use in this work relies on the following definitions instead:

\begin{enumerate}

    \item Direct Voronoi Neighbors Determination:    
    Refer to Fig.~\ref{fig:U}a). For any two points $v_i$ and $v_k$, if (i.) there exists a shared Voronoi face and (ii.) the line segment connecting these two points passes through said shared face, then the two points they constitute ``direct" Voronoi neighbors. The inner product $\overrightarrow{r_{ij}} \cdot \overrightarrow{r_{kj}} > 0$ is used to verify these requirements, where $v_j$ is any other point and $|\overrightarrow{r_{ij}}| < |\overrightarrow{r_{kj}}|$.
    
    \item Quadrilateral Asymmetry Control:
    Refer to Fig.~\ref{fig:U}b). A dimensionless parameter $\mathcal{A}$ is introduced to control the maximum asymmetry of a quadrilateral, which determines the maximum distortion of a four-member ring in a plane. For points $v_i$ and $v_k$ and the intermediary point $v_j$, we can asses whether the asymmetry of the quadrilateral exceeds the threshold set by $\mathcal{A}$ via:
    
    \begin{itemize}
        \item Considering the adjusted positions $\overrightarrow{v_p}=\overrightarrow{v_i} + \mathcal{A} \cdot (\overrightarrow{v_j}-\overrightarrow{v_i})$ and $\overrightarrow{v_q}=\overrightarrow{v_k} + \mathcal{A} \cdot (\overrightarrow{v_j}-\overrightarrow{v_k})$.
        \item Checking whether for all $v_j$, the two conditions: (i.) $\overrightarrow{r_{ip}} \cdot \overrightarrow{r_{kp}} > 0$ and (ii.) $\overrightarrow{r_{kq}} \cdot \overrightarrow{r_{iq}} > 0$ are satisfied.
    \end{itemize}
    
    \item Construction of Edge Set $\mathcal{E}$:
    Based on the above conditions, we can determine whether the pair $(v_i,v_k)$ constitutes a pair of direct Voronoi neighbors and whether that pair obeys the constraints of quadrilateral asymmetry control. At which point, we can construct the edge set $\mathcal{E}$.
\end{enumerate}

Thus, the algorithm to construct the modified Voronoi graph (or network)~\cite{malins2013identification} can be summarised follows:

\begin{enumerate}
    \item Loop over all $\mathcal{N}$ particles with index $i$.
    \item Identify all particles within a distance $r_c$ of $\overrightarrow{r_i}$, ensuring $r_c$ exceeds the maximum bond length in the network, and include these in the set $\mathcal{S}_i$.
    \item Sort the particles in $\mathcal{S}_i$ in ascending order, according to their distance from the $i$-th particle.
    \item Loop over all $v_j \in \mathcal{S}_i$, i.e., over particles with increasing distance from the $i$-th particle.
    \item For each $j$, loop over all $k > j$ in $\mathcal{S}_i$ and eliminate the $k$-th particle from $\mathcal{S}_i$ if the following inequality is not satisfied:
    \begin{equation*}
        \mathcal{A} > \frac{|\overrightarrow{r_{ik}}|^2}{|\overrightarrow{r_{ij}}|^2+|\overrightarrow{r_{jk}}|^2}
    \end{equation*}
\end{enumerate}

The dimensionless parameter $\mathcal{A} \in (0.5,1]$ plays an important role in the modified Voronoi method. To be specific, $\mathcal{A}$ is used to determine the maximum asymmetry that a four-membered ring of particles can exhibit before it is identified as comprising two three-membered rings. Thus, $\mathcal{A}$ enables a precise control of structure recognition: it dictates the extent to which asymmetry is allowed within a four-membered ring before the bond between two of the particles in the ring is considered broken, which involves determining whether a bond (edge) between particles $i$ and $k$ exists when there is a particle $j$ that is bonded to $i$ and is closer to  $i$ than $k$ is. Adjusting the value of $\mathcal{A}$ modifies the position of a plane that affects whether a bond is formed between particles. When $\mathcal{A}<1$, this corresponds to moving a plane that contains $\overrightarrow{v_j}$, perpendicular to $\overrightarrow{r_{ij}}$, towards $\overrightarrow{v_i}$~\cite{malins2013identification}. 

The modified Voronoi graph, which represents the bonding network of a multi-particle system, not only preserves the foundational structure of the original Voronoi graph but also enhances the identification of specific structural features. This is achieved through direct neighbor determination and control of quadrilateral asymmetry~\cite{malins2013identification}. The graph highlights selected interactions among particles and the pathways through which these interactions are transmitted. This is crucial because in reality, one particle can influence others indirectly through intermediary particles.

It is important to note that the overall properties of a given system ultimately originate from the interactions between its constituents particle (or atoms, or molecules). Graphs offer an enticing strategy to capture these interaction. In fact, the criteria for forming edges, i.e., the ``strictness" parameter $\mathcal{A}$, is fundamentally more significant than the cutoff distance $r_c$. The only criterion for $r_c$ is to maintain high graph connectivity, facilitating the transmission of actions between particles. Hence, in section~\ref{subsec:des_graph}, we will fix $r_c=2.5$ and systematically modify $\mathcal{A}$ within the interval of $(0.5,1]$.

\subsubsection{Graph-based descriptors}
\label{subsec:des_graph}

Using functions acting on the edges and/or vertices of a given graph, i.e. $h: (\mathcal{V},\mathcal{E}) \rightarrow \mathbb{R}$, to explore the properties of the latter is common practice in graph theory or network science. For instance, some functions~\cite{biggs1993algebraic,jukna2006graph} 
can be applied to graph, similarly to manifolds, to solve combinatorial problems such as coloring~\cite{jensen2011graph,karger1998approximate,doshi2010functional}, partitioning~\cite{bae2009graph}, and connectivity~\cite{margulis1974probabilistic}. A directed graph can be defined with homology groups like an orientable manifold~\cite{giblin2013graphs}, and it can be equipped with sheaves and cohomology~\cite{friedman2015sheaves}. To control a variable, a common method is to define a function on the structure and perform the gradient estimation, such as Li-Yau inequality (representing Weiguang Li and Shing-Tung Yau, respectively) for control of heat kernel on graphs~\cite{li1986parabolic,bauer2015li}.

In the context of this work, the properties of the system are determined by a minor fraction of the particles within the system - particles characterised by specific topological environments~\cite{cubuk2015identifying}. Metrics such as the node centrality within a graph~\cite{rodrigues2019network,landherr2010critical} can be used to pinpoint such particles. In addition, the hierarchical structure of the neighbors of a node is crucial for understanding the particle influence on the graph as a whole, which in turn can reveal the patterns by which interactions propagate between different nodes and their immediate neighbors - and thus their impact on the wider network~\cite{clauset2008hierarchical,oberg2009hierarchical,clauset2006structural,barabasi2003scale,yu2006genomic}.

Therefore, we designed two sets of descriptors: one based on node centrality and the clustering coefficient (section~\ref{subsec:des_graph}.1), and the other based on the hierarchical structure, which include information about the angles between triplets of nodes (section~\ref{subsec:des_graph}.2). 

\vspace{0.25cm}
\textbf{3.1 - Graph Centrality and Clustering Coefficient.} We have built this set of descriptors by leveraging the concept of network centrality~\cite{rodrigues2019network,landherr2010critical}, inspired by previous research on complex networks~\cite{barabasi2013network,albert2002statistical} - with emphasis on social networks~\cite{borgatti2024analyzing,landherr2010critical}, which, we aregue, are often overlooked by the condensed matter community. In the context of complex networks, vertices are often called nodes, and graphs are often called networks. Node centrality is a critical concept in social network analysis which quantifies the level of influence, control, or significance a particular node holds within the network~\cite{borgatti2024analyzing}. Here, we selected the following node centrality measures, which offer a clear physical meanings that we can use to characterize the ``importance" of any particle in the system. We remind the reader that this is key, as several dynamical properties of a given system as a whole are often determined by a small fraction of particles within it. A good example is provided by the concept of dynamical heterogeneity in supercooled liquids, where the mobility of the whole system is dictated by small pockets of specific particles, or atoms, or molecules~\cite{sosso2014dynamical}.

For a unweighted, undirected graph $\mathcal{G}(\mathcal{V}, \mathcal{E})$, we define:

\vspace{0.25cm}
\paragraph{Degree Centrality} The Degree Centrality~\cite{freeman2002centrality} of node $v_i \in \mathcal{V}$ can be defined as:

\begin{equation*}
    C_D(i)=\frac{\text{deg(i)}}{|\mathcal{V}|-1}
\end{equation*}

where $\text{deg}(i)$ denotes the degree of node $v_i$ and $|\mathcal{V}|$ is the number of nodes in graph $\mathcal{G}$. Degree Centrality is a rather straightforward metric, based on the assumption that the more important a node is, the more edges should be connected to it. The concept of Degree Centrality can be analogous to the coordination number in physical chemistry or materials science.

\vspace{0.25cm}
\paragraph{H-index Centrality}
The H-index Centrality~\cite{hirsch2005index} of a node reflects not just its number of connections but also the significance of these connections, based on the premise that connections to highly connected nodes are more valuable. The H-index Centrality $H_i(i)$ of node $v_i \in \mathcal{V}$ is defined through the following steps:

\begin{enumerate}
    \item Consider the degrees of all neighbors of node $v_i$.
    \item Sort these degrees in descending order.
    \item The H-index $H_i(i)$ for node $v_i$ is $h$ if $h$ is the highest number such that node $v_i$ has at least $h$ neighbors with a degree of $h$ or more.
\end{enumerate}

The H-index provides a nuanced view of the importance of a node in a network, taking into account both the quantity and quality of its connections. In a bond network, the H-index of a particle measures the particle's impact on the structure of the system.

\vspace{0.25cm}
\paragraph{Closeness Centrality} The Closeness Centrality evaluates the significance of a node by measuring its average distance to all other nodes in the network~\cite{bavelas1950communication,freeman1977influence}. Nodes with higher Closeness Centrality are deemed more central, which implies quicker access or connectivity to other nodes - compared to those with lower Closeness Centrality scores. The Closeness Centrality (which definition was improved by Wasserman and Faust~\cite{wasserman1994social}) $C_C(i)$ of node $v_i \in \mathcal{V}$ can be defined as:

\begin{equation*}
    C_C(i) = \frac{(\text{n}(j)-1)^2}{(|\mathcal{V}|-1) \sum_{j \neq i} \text{d}(i,j)}
\end{equation*}

where $\text{d}(i,j)$ is the shortest-path length between nodes $v_i$ and $v_j$, $\text{n}(j)$ is the number of nodes reachable from $v_j$, and $|\mathcal{V}|$ denote the number of nodes in $\mathcal{G}$. The Closeness Centrality suggests that the strategic importance of a position within social networks is quantifiable by the average distance to others. In the context of our Lennard-Jones system a particle's significance is determined by its average path length to all the other particles.

\vspace{0.25cm}
\paragraph{Betweenness Centrality}
The Betweenness Centrality quantifies a node's importance by the frequency with which is encountered across all shortest paths of the network~\cite{bavelas1948mathematical,freeman1977set}. Nodes with high Betweenness act as pivotal bridges, or connectors, and they are crucial for information flow within the network. The Betweenness Centrality $C_B(i)$ of node $v_i \in \mathcal{V}$ is defined as:

\begin{equation*}
    C_B(i) = \sum_{s \in \mathcal{V}, s \neq i \neq t} \frac{\sigma_{\text{st}}(i)}{\sigma_{\text{st}}}
\end{equation*}

where $\sigma_{\text{st}}$ is the total number of shortest paths from node $v_s$ to node $v_t$, and $\sigma_{\text{st}}(i)$ is the number of those paths passing through node $v_i$. In the context of our Lennard-Jones system, the Betweenness Centrality characterizes the ability of a particle to transfer the influences from other particles to another.

\vspace{0.25cm}
\paragraph{Eigenvector Centrality} The Eigenvector Centrality measures the importance of a node by accounting for both the quantity and quality of its connections, emphasizing that links to highly central nodes are more valuable than those to less central ones~\cite{bonacich1972factoring,bonacich2007some}. The Eigenvector Centrality $C_E(i)$ of node $v_i \in \mathcal{V}$ is defined as:

\begin{equation*}
    C_E(i) = \frac{1}{\lambda} \sum_{j \in \mathcal{V}} \mathcal{M}_{i,j} C_E(j)
\end{equation*}

where $\lambda$ is the maximum eigenvalue of the adjacency matrix $\mathcal{M}(\mathcal{G})$ of graph $\mathcal{G}$. In the context of our Lennard-Jones system, Eigenvector Centrality assesses particles by considering both their immediate neighbors and the global influence of these neighbors, offering a metric that combined local and global information.

\vspace{0.25cm}
\paragraph{K-shell Centrality} The K-shell Centrality $C_{KS}(i)$ quantifies node significance, using K-shell decomposition to reveal hierarchical structures and coreness~\cite{carmi2007model,kitsak2010identification}. This method peels the network layer by layer, assigning K-shell values to indicate a node's hierarchical position and importance efficiently. The definition process for K-shell Centrality of all nodes in $\mathcal{V}$ is as follows:

\begin{enumerate}
    \item Initialization: All nodes are unassigned to any K-shell value.
    \item Step $1$ ($K=1$): Remove all nodes with degree $1$, continuing until no nodes of degree $1$ remain. These removed nodes are assigned a K-shell value of $1$.
    \item Step $K$: After removing ($K-1$) shells of nodes, remove all nodes of degree $K$, continuing until no nodes of degree less than or equal to $K$ remain in the network. These nodes are assigned a K-shell value of $K$.
    \item Iteration: Repeat the above process, increasing $K$ at each step, until all nodes are removed and assigned a K-shell value.
\end{enumerate}

K-shell Centrality is a metric leveraging K-shell decomposition to highlight the hierarchical significance of nodes, categorizing them by connectivity density in a layered structure. In the context of our Lennard-Jones system, the K-shell Centrality essentially uses the density information of particles in a layered structure as a discrete version of the radial distribution function.

\vspace{0.25cm}
\paragraph{Clustering Coefficient}
It is important to note that the Clustering Coefficient is not a measure of centrality. Instead, we only use it alongside the other centrality measures for feature engineering. The Clustering Coefficient of a node measures the actual versus potential connections among its neighbors, indicating their inter-connectivity level~\cite{watts1998collective}. The Clustering Coefficient $C_{\text{CC}}(i)$ of a node $v_i\in \mathcal{V}$ can be defined as:

\begin{equation*}
    C_{\text{CC}}(i) = \frac{E_i}{k_i (k_i-1)}
\end{equation*}

where $E_i$ is the number of edges actually existing between the neighbors of node $v_i$ and $k_i = \text{deg(i)}$ is the degree of node $v_i$. In the context of our Lennard-Jones system, a high Clustering Coefficient of a particle $v_i$ typically indicates the presence of tightly connected sub-structures near the particle $v_i$ within a certain structure.

\vspace{0.25cm}
\paragraph{Subgraph Centrality} The Subgraph Centrality measures the number and quality of subgraphs a node participates in, particularly emphasizing how a node connects to itself through cycles~\cite{estrada2005subgraph}. The Subgraph Centrality $C_{\text{SG}}(i)$ of node $v_i \in \mathcal{V}$ can be defined using the adjacency matrix $\mathcal{M}(\mathcal{G})$ of $\mathcal{G}$ and its spectral properties as:

\begin{equation*}
    C_{\text{SG}}(i) = \sum_{k=0}^{\infty} \frac{(\mathcal{M}(\mathcal{G})^k)_{i,i}}{k!}
\end{equation*}

where $\mathcal{M}(\mathcal{G})^k$ represents the $k$-th power of the adjacency matrix $\mathcal{M}(\mathcal{G})$, and $(\mathcal{M}(\mathcal{G})^k)_{i,i}$ is the element of $\mathcal{M}(\mathcal{G})^k$ at row $i$, column $i$, indicating the number of closed walks of length $k$ starting and ending at node $v_i$, and $k!$ serves as a normalization factor. In the context of our Lennard-Jones system, the Subgraph Centrality characterizes the cyclic structures within bonding networks.

\vspace{0.25cm}
\paragraph{Harmonic Centrality}
Harmonic Centrality quantifies the importance of a node through the sum of its inverse distances to all other nodes, effectively addressing unconnected components~\cite{latora2007measure}. The harmonic centrality $C_H(i)$ of node $v_i \in \mathcal{V}$ is defined as:

\begin{equation*}
    C_H(i) = \sum_{j \in \mathcal{V}, j \neq i} \frac{1}{\text{d}(i,j)}
\end{equation*}

where $\text{d}(i,j)$ denotes the shortest path length between nodes $v_i$ and $v_j$, allowing us to study unconnected sub-clusters within our system.

\vspace{0.25cm}
\paragraph{LocalRank Centrality} The LocalRank Centrality gauges the importance of a node based on the connectivity within its immediate neighborhood, highlighting those influential in their vicinity despite their overall network centrality~\cite{chen2012identifying}. The definition of LocalRank Centrality $C_{\text{LR}}(v)$ for node $v$ involves the following steps:

\begin{enumerate}
    \item Neighborhood Determination: For each node $v$ in $\mathcal{V}$, identify its neighborhood $N_{\text{nbh}}(v)$, which typically includes the node itself and its directly connected neighbors.

    \item Local Degree Calculation: Calculate the degree of each node $v$ within this neighborhood $N_{\text{nbh}}(v)$.

    \item Ranking: Rank the nodes based on their local degrees within $N_{\text{nbh}}(v)$. The LocalRank Centrality of node $v$ is determined by its rank within $N_{\text{nbh}}(v)$.
\end{enumerate}

In the context of our Lennard-Jones system, the LocalRank Centrality excels in scenarios where we need to value the local structure of a particle and influence over global metrics, which is an aspects crucial for cluster detection and the spread of localized interactions.

By choosing $p$ different centrality functions and $q$ different values of $\mathcal{A}$ within $(0.5,1]$ (typically $p=10$, which means selecting all above functions), as well as a fixed $r_c$, we can create $pq$ unique predictors, i.e. $X$, to be used as the input for the machine learning model $f$.

\vspace{0.25cm}
\textbf{3.2 - Angular functions on hierarchical graph structures.}
We also developed a new kind of descriptors by merging the hierarchical information of nodes and their neighbors from bonding networks with angular information of nodes and their neighbors in original $\mathbb{R}^3$ spaces, thus creating a hybrid, yet concise representation of the local environments within the network.

The modified Voronoi graph featuring the selected interactions (bonds), i.e. $\mathcal{G}(\mathcal{V}, \mathcal{E})$, is constructed from a set of points $\mathcal{V} = \{ v_1, \cdots, v_{\mathcal{N}}\}$ with their coordinates in $\mathbb{R}^3$ under the parameter $\mathcal{A}$, where we denote the position of the $i$-th point as $\overrightarrow{v_i}$ and $\overrightarrow{r_{ij}} = \overrightarrow{v_j} - \overrightarrow{v_i}$.

Then, the angular function of the $i$-th point about its $\mathcal{L}$-th neighbors $\mathcal{K}_{\mathcal{L}}$ is constructed as:

\begin{equation}
    \mathcal{F}_{\mathcal{A}}(i, \mathcal{L}) = \sum_{j \neq i, v_j \in \mathcal{K}_{\mathcal{L}}} \sum_{k \neq j \neq i, v_k \in \mathcal{K}_{\mathcal{L}}} (1+\text{cos}(\theta_{ijk})) \text{e}^{-\mathcal{L}}
    \label{equ:angular_func_graphs}
\end{equation}

where $\theta_{ijk}$ is the angle between vectors $\overrightarrow{r_{ij}}$ and $\overrightarrow{r_{ik}}$, the cosine of which is calculated via:

\begin{equation*}
    \text{cos}(\theta_{ijk}) = \frac{\overrightarrow{r_{ij}} \cdot \overrightarrow{r_{ik}}}{|\overrightarrow{r_{ij}}| \cdot |\overrightarrow{r_{ik}}|}
\end{equation*}

$\mathcal{K}_{\mathcal{L}}$ can be calculated via the adjacency matrix $\mathcal{M} = \mathcal{M}(\mathcal{G})$ of graph $\mathcal{G}$, which element at row $i$ and column $j$ is denoted as $\mathcal{M}_{ij}$. In this case, $\mathcal{G}$ is unweighted and undirected, and the first-order neighbors ($\mathcal{L} = 1$) can be determined by observing the non-zero elements in $\mathcal{M}$. For $\mathcal{L}$-th order neighbors ($\mathcal{L} \geq 2$), $\mathcal{K}_{\mathcal{L}}$ can be calculated through the power of $\mathcal{M}$. Specifically, the $\mathcal{L}$-th order neighbors of a node $v_i$ can be found by calculating the $\mathcal{L}$-th power of the adjacency matrix $\mathcal{M}$. In $\mathcal{M}^{\mathcal{L}}$, the value of element at row $i$ and column $j$ represents the number of distinct paths of length $\mathcal{L}$ from nodes $v_i$ to $v_j$. Therefore, if $(\mathcal{M^{\mathcal{L}}})_{ij} > 0$, then node $v_j$ is an $\mathcal{L}$-th order neighbor of node $v_i$, and $(\mathcal{M^{\mathcal{L}}})_{ij}$ represents the number of paths.

We typically start assigning values to $\mathcal{L}$ from $1$, and then increment it one by one up to the required number of layers $p$. By selecting $q$ different values of $\mathcal{A}$ within $(0.5,1]$, as well as a fixed $r_c$, we can get $pq$ independent predictor variables for the machine learning model $f$.

We define this descriptor because this is to form a correspondence with the $\Psi$ function. When calculating the $\Psi$ function, it indeed involves capturing bond angle information by searching for all particles within a sector over a certain range of angles, but it does not classify particles based on their different distances relative to the central particle. Not classifying particles might blur the distinction between local structural features and larger-scale global characteristics, leading to less precise capture of local environmental information. In contrast, the descriptor of $\mathcal{F}_{\mathcal{A}}(i, \mathcal{L})$ we defined here (i.e., angular functions on hierarchical graph
structures) classify these particles based on which order of neighbors they are to the central particle on the graph.

\subsubsection{List of descriptors}

In this section, we summarised all the types of descriptors we have discussed so far, so as to help the reader with the discussion of our results, presented in section~\ref{subsec:results_ml}. For convenience, we denote the radial ($G$) symmetry functions defined in section \ref{sec:trad} as ``Type A" descriptors, and the angular ($\Psi$) symmetry functions as ``Type B" descriptors. Then, we label the centrality or clustering functions defined in section~\ref{subsec:des_graph}.1 as ``Type C" descriptors, and the angular functions on the hierarchical structure defined in section section~\ref{subsec:des_graph}.2 as ``Type D" descriptors.

For each Type of descriptor, we can construct different sets of predictors by choosing different parameters (choosing a part or the whole) as follow:

\begin{itemize}
    \item Type A: $r$, $\delta$, and $I$;
    \item Type B: $\xi$, $\lambda$, $\zeta$ and $I$, $J$;
    \item Type C: $\mathcal{A}$ and $r_c$;
    \item Type D: $\mathcal{A}$ and $r_c$.
\end{itemize}

Note that we can combine different types of descriptors as well. As our aim is to compare the performance of the symmetry functions utilised by Cubuk \textit{et al.} with that of our graph-based approach, we will compare: (i.) Type A with Type C; (ii.) Type B with Type D; (iii.) Type A + Type B with Type C + Type D. It should also be noted that the parameter $\mathcal{A}$ for Type C and Type D descriptors can differ between the two types when combining them together (as Type C + Type D).

\subsection{Molecular Dynamics Simulations}
\label{sec:md}

In this work, we focus on the prototypical Lennard-Jones system, specifically its supercooled liquid phase and its crystallisation. The configurations of the system that we will use to construct the ``dataset'' for our ML models have been generated by means of molecular dynamics (MD) simulations. The system contains 864 particles due to the computational efficiency. We work in the customary Lennard-Jones reduced units~\cite{allen2017computer,rapaport2004art} but we define one ``step" as $100 \times (0.2*(m*\sigma^2/\epsilon)^{1/2})$ for convenience. The parameters of the Lennard-Jones potentials are: $\epsilon = 1$, $\sigma = 1$, cutoff radius $r_c = 3.5\sigma$ and mass $m = 1 M$. We have adopted a tail correction as the truncation scheme. 

Previously work shown~\cite{blow2021seven} that these settings allow us to observe crystal nucleation within a timescale accessible via unbiased MD simulations. The computational protocol we have followed begins with a linearly quench of the liquid from $T_{\text{init}} = 1.25$ to a given temperature $T_{\text{final}}$ in 20 steps. Then, we perform a $1000$ steps equilibration at temperature $T_{\text{final}}$. These simulations have been conducted within the NPT ensemble with an isotropic pressure $P=5.68 (\epsilon/{\sigma^3})$, enforced via a chain of five thermostats coupled to a Nosé–Hoover barostat~\cite{hoover1996kinetic} (which accounts for the Martyna-Tobias-Klein correction~\cite{martyna1994constant}). The damping parameter of the barostat is $0.5t^{*}$, where $t^{*} = 0.002 (m * \sigma^2 /\epsilon)^{1/2}$. For convenience, we denote the trajectory corresponding to a given temperature $T$ as $\text{Traj}(T)$. 
All these trajectories were generated using the same initial condition.

We have considered 20 trajectories corresponding to different temperatures: $T_1 = 0.5$, $T_2 = 0.53$, $T_3 = 0.55$, $T_4 = 0.57$, $T_5 = 0.6$, $T_6 = 0.63$, $T_7 = 0.65$, $T_8 = 0.67$, $T_9 = 0.7$, $T_{10} = 0.73$, $T_{11} = 0.75$, $T_{12} = 0.77$, $T_{13} = 0.8$, $T_{14} = 0.83$, $T_{15} = 0.85$, $T_{16} = 0.88$, $T_{17} = 0.9$, $T_{18} = 0.93$, $T_{19} = 0.95$, $T_{20} = 0.97$, i.e, $\text{Traj}(T_i), i \in [1,20] \cap \mathbb{N}$. We divide these trajectories into two groups: Group $1$ includes temperatures with odd indices, i.e, the trajectories corresponding to $T^{(1)}_i=T_{2i-1}$; conversely, Group $2$ includes temperatures with even indices, i.e, the trajectories corresponding to $T^{(2)}_i=T_{2i}$, where $i \in [1,10] \cap \mathbb{N}$. For each set, we select all $896$ particle environments from the frames corresponding to the $t_j = 100 \cdot j$-th ($j \in [1,9] \cap \mathbb{N}$) steps of all $10$ trajectories as the dataset corresponding to that set. The dataset corresponding to Group $1$ is Set $1$, which is used to investigating the physical properties of the systems and evaluating the performance of the machine learning models; similarly, the dataset corresponding to Group $2$ is denotes as Set $2$, which is used to test the transferability of our ML models.

\subsection{Data Analysis}


\subsubsection{Statistical Quantities}
\label{subsec:some_metrics}

\paragraph{Spearman’s coefficient} The Spearman’s rank correlation coefficient~\cite{spearman1987proof} is a non-parametric measure assessing the association between two variables through their ranks. It gauges if their relationship is monotonic, with values closer to $+1$ or $-1$ indicating stronger monotonic increasing or decreasing relationships, respectively. For two sets of observations $\bold{a} = (a_1, \cdots, a_n)^T$ and $\bold{b} = (b_1, \cdots, b_n)^T$, denoting $R(a_i)$ and $R(b_i)$ as the ranks of $a_i$ and $b_i$, respectively, and $d_i = R(a_i)-R(b_i)$, the Spearman's rank correlation coefficient between $\bold{a}$ and $\bold{b}$ can be calculated as:

\begin{equation*}
    \rho(\bold{a}, \bold{b}) = 1 - \frac{6 \sum_{i=1}^n d_i}{n(n^2-1)}
\end{equation*}

\paragraph{ten Wolde's order parameter} The ten Wolde's global bond order parameter $Q_6$~\cite{ten1995numerical,jungblut2016pathways} quantifies the degree of order within a system and can differentiate disordered phases (i.e. liquid or amorphous) from crystalline phases. The calculation of $Q_6$ follows the following procedure:

\begin{enumerate}
    \item For each particle $i$, calculate its Steinhardt's local bond order parameter $q_{6m}(i)$~\cite{steinhardt1983bond}, which is defined through the spherical harmonics $Y_{lm}$ along the bonds between particle $i$ and its neighbors $j$. Specifically:
    
    \begin{equation*}
        q_{6m}(i) = \frac{1}{N_i} \sum_{j=1}^{N_i} Y_{6m} (\theta_{ij}, \phi_{ij})
    \end{equation*}
    
    where $N_i$ is the number of neighbors of particle $i$, and $\theta_{ij}$ and $\phi_{ij}$ are the polar and azimuthal angles of the direction of the bond between particle $i$ and its neighbor $j$, respectively.
    \item Calculate the global order parameter $Q_6$ by evaluating the average of the $q_{6m}(i)$ values for all particles and calculating the square of its magnitude:
    
    \begin{equation*}
        Q_6 = \sqrt{\frac{4 \pi}{13} \sum_{m=-6}^6 \langle q_{6m} \rangle_i^2}
    \end{equation*}
    
    where $\langle . \rangle_i$ indicates the average on all particles.
\end{enumerate}

\paragraph{Mean Square Displacement} The mean square displacement (MSD)~\cite{einstein1905molekularkinetischen} describes the translation diffusion of the particles and is commonly used to distinguish between solids and liquids. In an $\mathcal{N}$-body system, for each particle $i$, its position at time $t$ (initial time is denoted as $t_0$) can be represented by $\overrightarrow{r_i}$. The MSD is defined as:

\begin{equation*}
    \text{MSD}(t) = \langle |\overrightarrow{r_i}(t)-\overrightarrow{r_i}(t_0)|^2 \rangle = \frac{1}{\mathcal{N}} \sum_{i=1}^{\mathcal{N}} |\overrightarrow{r_i}(t)-\overrightarrow{r_i}(t_0)|^2
\end{equation*}

\paragraph{Identifying particles rearrangements} The positional change of a given particles $p_{\text{hop},i}(t)$~\cite{smessaert2013distribution} within a certain time window $t$, i.e., $[t-t_{R/2}, t+t_{R/2}]$, is used to characterise the extent of the particle rearrangement. If the value of $p_{\text{hop},i}(t)$ exceeds the threshold $p_c$ during the time interval $[t-t_{R/2}, t+t_{R/2}]$, the particles is labelled as soft and hard otherwise~\cite{schoenholz2016structural}. In detail, $p_{\text{hop},i}(t)$ is defined as:

\begin{equation*}
    p_{\text{hop},i}(t) = \sqrt{\langle (\bold{r}_i-\langle \bold{r}_i \rangle_B)^2 \rangle_A    \langle (\bold{r}_i-\langle \bold{r}_i \rangle_A)^2 \rangle_B}
\end{equation*}

where $\langle . \rangle_A$ and $\langle . \rangle_B$ indicate averages over the intervals $A=[t-t_{R/2},t]$ and $B=[t,t+t_{R/2}]$, respectively. Determining $t_{R/2}$ and $p_c$ effectively allow to distinguish between actual particle rearrangements and random positional fluctuations due to thermal noise within a single, coherent framework~\cite{schoenholz2016structural}. Adjusting the threshold value $p_c$ causes a logarithmic shift in the energy scale. However this modification is systematic and does have a qualitative impact on the results~\cite{schoenholz2016structural}.

In section~\ref{subsec:particle_labelling}, we elaborate on the choice of this threshold value, as well as on the relationship between the value of $p_{\text{hop},i}(t)$ and the results of the soft-or-hard classification of particle $i$ at time $t$.

The results reported in this section have been obtained via ML models specifically tailored for Group $1$ - we have not used any of the data in Group $2$ at this stage.

\subsubsection{Identifying the state of the system}
\begin{figure}[htbp]
\includegraphics[width=0.5\textwidth]{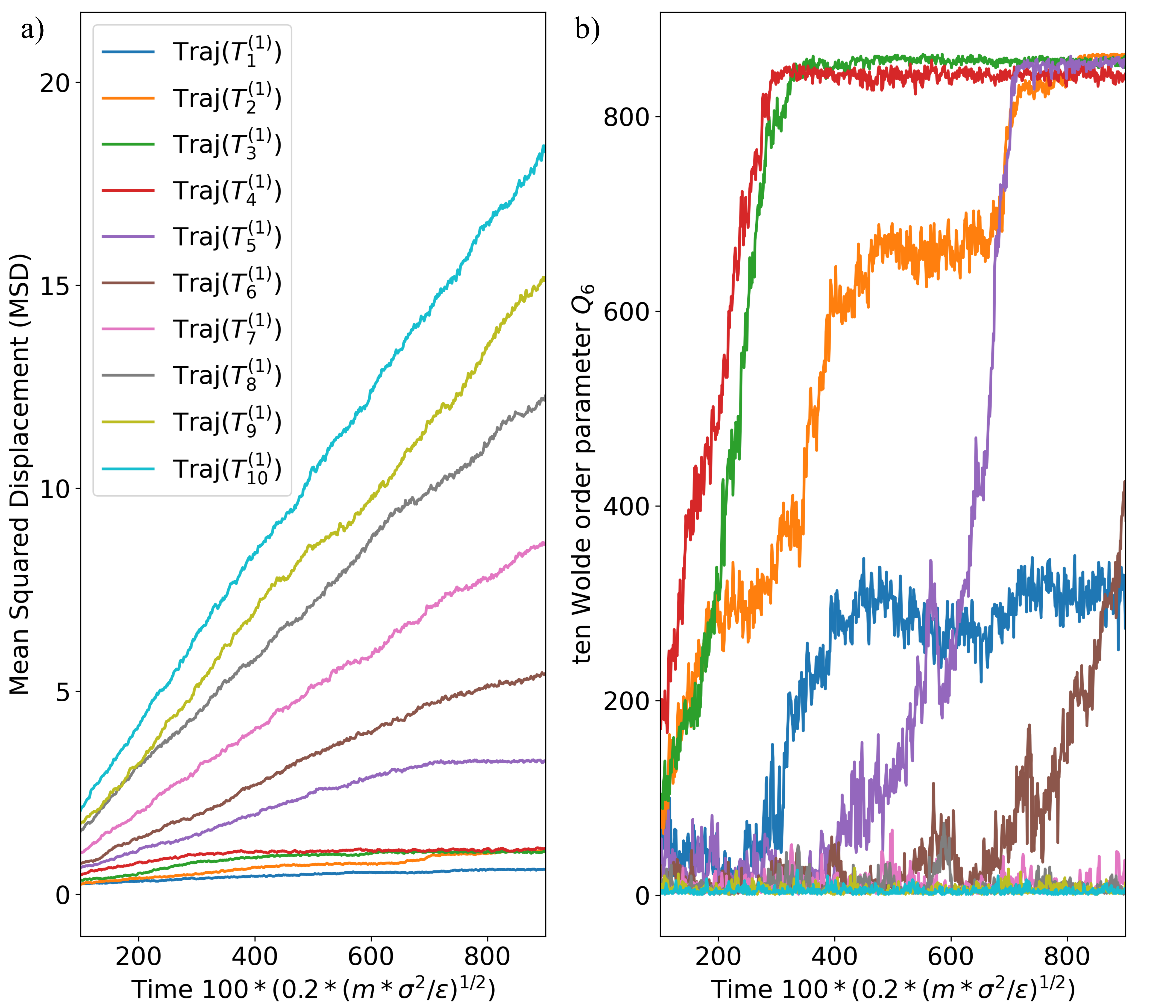}
\caption{The a) mean square displacement (MSD) and b) Global bond order parameter $Q_6$ corresponding to each trajectory in Unit 1, i.e, $\text{Traj}(T^{(1)}_i), i \in [1,10] \cap \mathbb{N}$.}
\label{fig:q6msd_set1}
\end{figure}

We can differentiate between liquid and solid by looking at the MSD as a function of time (Fig.~\ref{fig:q6msd_set1}a), and we can monitor the evolution of the global bond order parameter $Q_6$ over time (Fig.~\ref{fig:q6msd_set1}b) to detect the onset the crystallisation process within the system.

\begin{figure}[htbp]
\includegraphics[width=0.5\textwidth]{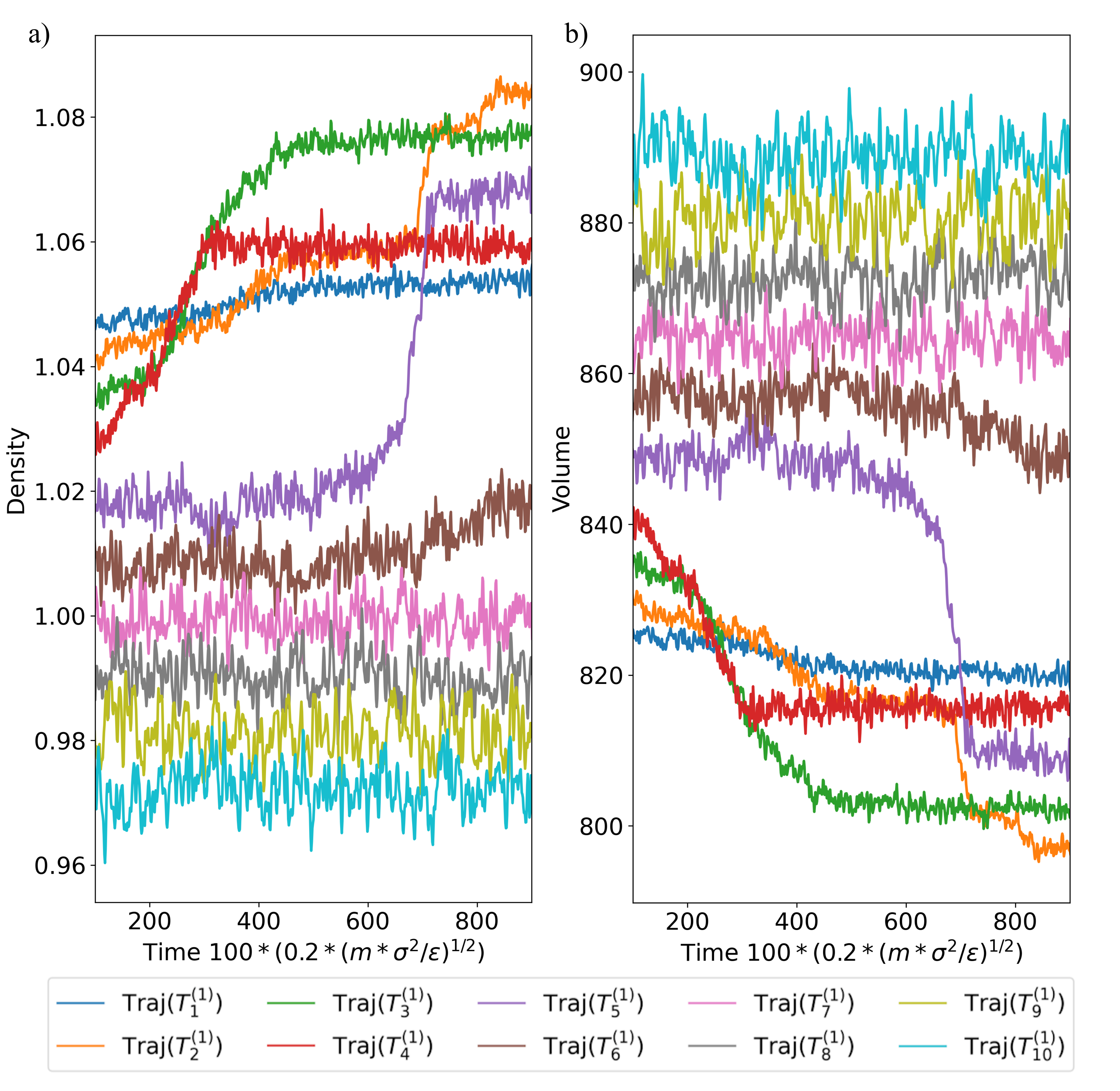}
\caption{The a) density and b) volumn corresponding to each trajectory in Unit 1, i.e, $\text{Traj}(T^{(1)}_i), i \in [1,10] \cap \mathbb{N}$. The Lennard-Jones reduced units are used here.}
\label{fig:thermo_set1}
\end{figure}

According to the results reported in Fig.~\ref{fig:q6msd_set1}, we can classify the different trajectories in Group 1 into the following categories: 

\begin{enumerate}
    \item The system remains liquid at any point in time: $\text{Traj}(T^{(1)}_7)$, $\text{Traj}(T^{(1)}_8)$, $\text{Traj}(T^{(1)}_9)$ and $\text{Traj}(T^{(1)}_{10})$.

    \item The system completely crystallises over the timescale of our simulations: 
    $\text{Traj}(T^{(1)}_3)$, $\text{Traj}(T^{(1)}_4)$, $\text{Traj}(T^{(1)}_5)$.
    
    \item The liquid starts to crystallise, but the system only partially crystallises due to the time bound. However, it can be inferred that as long as the simulation is extended, it will fully crystallize: $\text{Traj}(T^{(1)}_6)$.

    \item Start from a supercooled liquid to complete vitrification, eventually achieving a stable state that is partly vitreous and partly crystalline: $\text{Traj}(T^{(1)}_1)$.

    \item Begin with a supercooled liquid to partial vitrification, but the unstable amorphous state eventually crystallizes: $\text{Traj}(T^{(1)}_2)$.
\end{enumerate}

We can also look at the density (or, equivalently, volume) changes over time to confirm our classification, as reported in Fig.~\ref{fig:thermo_set1}. The phase change from liquid to solid is marked by a significant change in density ($\text{Traj}(T^{(1)}_3)$, $\text{Traj}(T^{(1)}_4)$, $\text{Traj}(T^{(1)}_5)$ and $\text{Traj}(T^{(1)}_{6})$). Transitioning from the amorphous to the crystalline state result in only a slightly density variation ($\text{Traj}(T^{(1)}_1)$ and $\text{Traj}(T^{(1)}_2)$). Conversely, no density changes are observed for those trajectories corresponding to the situation where the liquid survives for the entire duration of our MD simulations ($\text{Traj}(T^{(1)}_7)$, $\text{Traj}(T^{(1)}_8)$, $\text{Traj}(T^{(1)}_9)$ and $\text{Traj}(T^{(1)}_{10})$).

In summary, Set $1$ from Group $1$ comprises particle environments across a diverse portfolio of temperatures, densities, volumes, and phases of matters. This dataset will allows us to explore how effectively a uniform feature engineering strategy can enable a ML model to simultaneously grasp the nuances of these different phases and thermodynamic conditions. In addition, we also aim to examine the transferability of our model by applying the model we will train and test on Set 1 to Set $2$ as well.

\subsubsection{Graph Metrics}
\label{subsec:results_graph_metrics}

The graph metrics, i.e., the Type C and Type D descriptors we defined in section~\ref{subsec:des_graph}, exhibit a significant correlation with the value of $p_{\text{hop},i}(t)$, according to the calculation of them on Set 1.

\begin{figure}[htbp]
\includegraphics[width=0.5\textwidth]{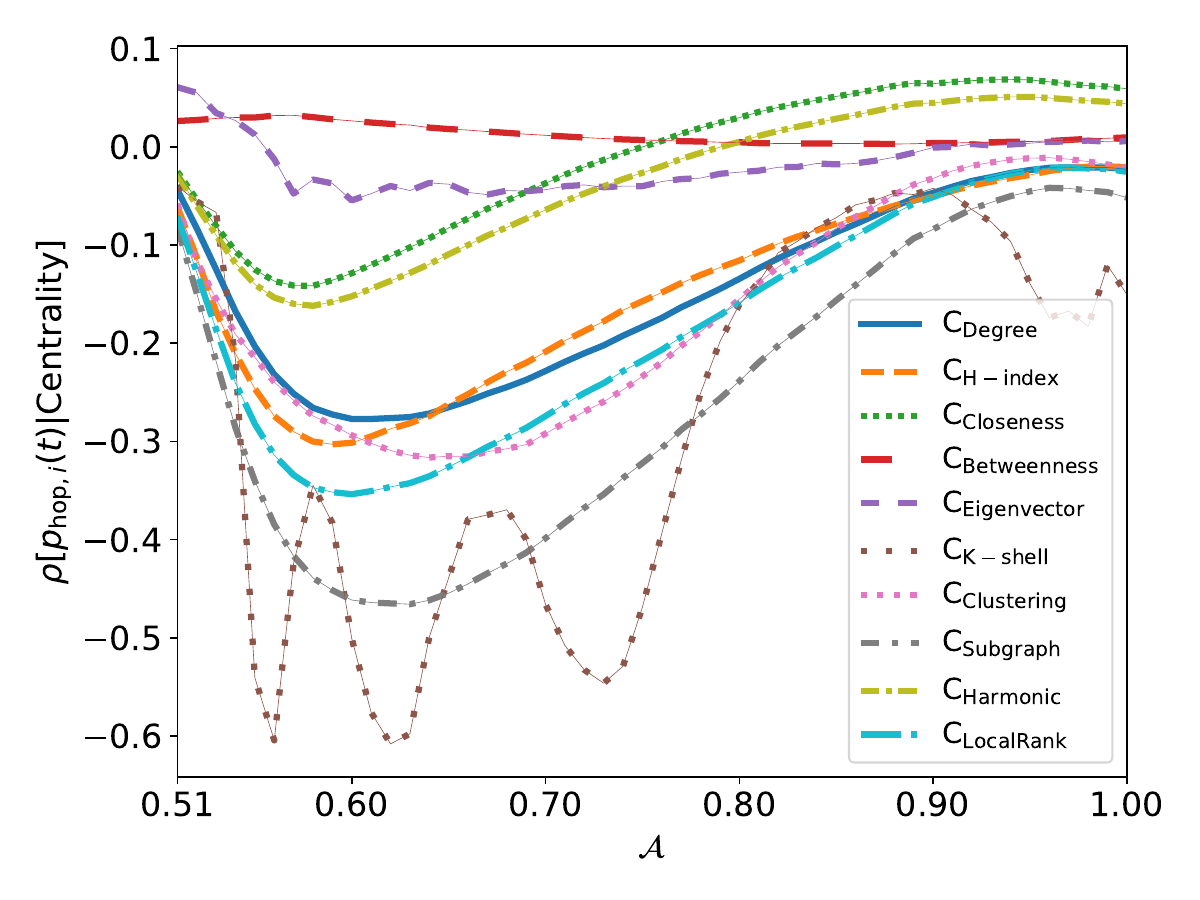}
\caption{The Spearman's rank correlation coefficient between $p_{\text{hop,i}(t)}$ and Type C descriptors, i.e., the centralities and clustering coefficient under different $\mathcal{A}$.}
\label{fig:p}
\end{figure}

\begin{figure}[htbp]
\includegraphics[width=0.5\textwidth]{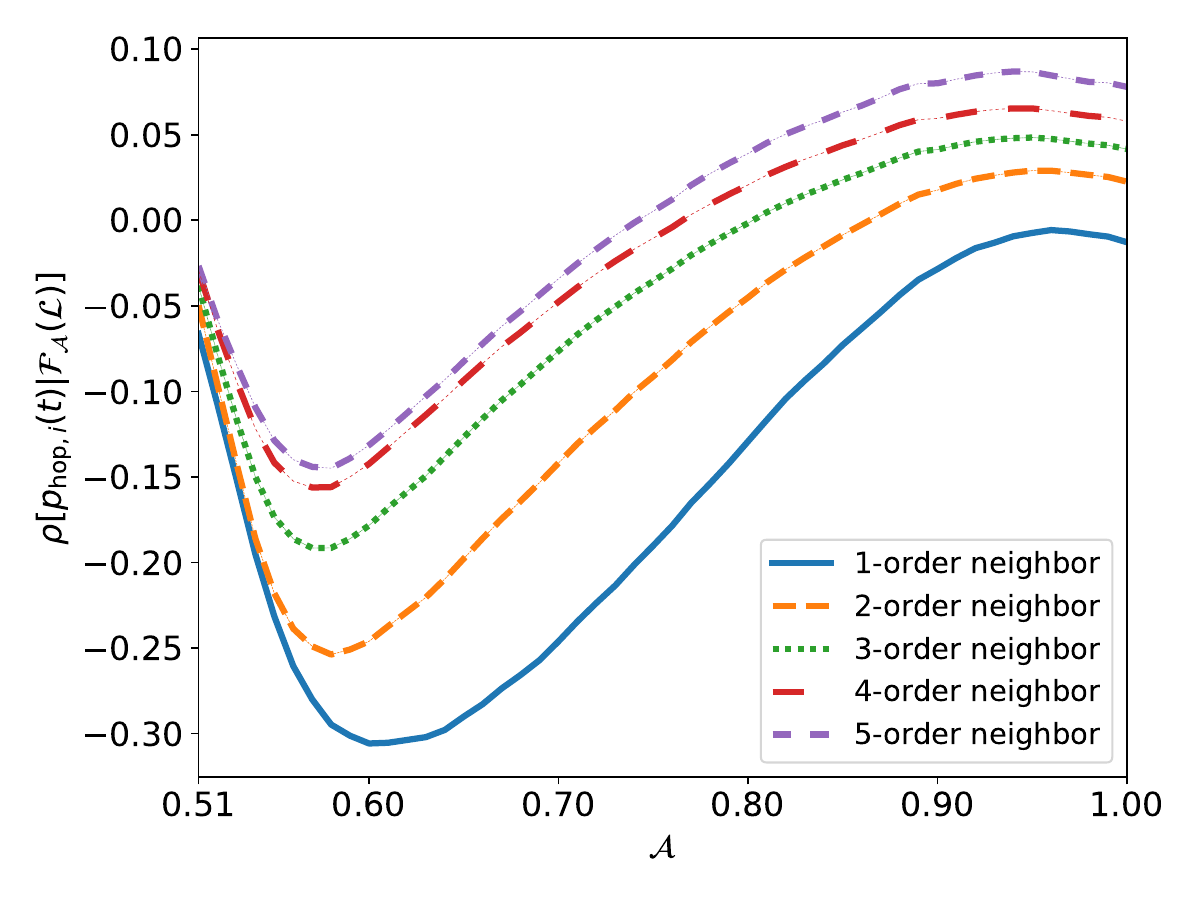}
\caption{The Spearman's rank correlation coefficient between $p_{\text{hop},i}(t)$ and Type D descriptors, i.e., the angular functions on hierarchical structures under different $\mathcal{A}$.}
\label{fig:q}
\end{figure}

In particular, node centrality, clustering coefficient (Type C, seeing Fig.\ref{fig:p}), and angular functions on hierarchical structures (Type D, seeing Fig.\ref{fig:q}) all show a varying degree of negative correlation with $p_{\text{hop},i}(t)$. Essentially, higher values of these descriptors indicate harder particles. Most of these metrics display a minimum for $\mathcal{A} \sim 0.6$, albeit with some differences, namely:

\begin{enumerate}
    \item For any given $\mathcal{A} \in (0.5,1]$, we observe no correlation between either Eigenvector Centrality or Betweenness Centrality with $p_{\text{hop},i}(t)$.

    \item The Closeness Centrality and Harmonic Centrality exhibit a weak correlation with $p_{\text{hop},i}(t)$, which is most pronounced for $\mathcal{A} \in [0.55,0.6]$.

    \item The Degree Centrality, H-index Centrality, LocalRank Centrality, and Clustering Coefficient show moderate correlation with $p_{\text{hop},i}(t)$, most notably within the interval $\mathcal{A} \in [0.55,0.7]$.

    \item The K-shell Centrality and Subgraph Centrality demonstrate strong correlation with $p_{\text{hop},i}(t)$, especially within the interval of $\mathcal{A} \in [0.55,0.75]$. Specifically, we observe the strongest correlation between Subgraph Centrality and $p_{\text{hop},i}(t)$ for $\mathcal{A}=0.6$, while K-shell Centrality exhibits periodic oscillations within the interval of $\mathcal{A} \in [0.55,0.75]$.

    \item Within $\mathcal{A} \in [0.55,0.65]$, angular functions on hierarchical structures and $p_{\text{hop},i}(t)$ exhibit a considerable correlation, which weakens with the inclusion of higher-order neighbors. The correlation difference between successive orders shows a trend towards convergence, indicated by the curves getting closer together. For first-order neighbors, we observe the strongest degree of correlation with $p_{\text{hop},i}(t)$ for $\mathcal{A}=0.6$. The position of this minimum shift toward the right has we increase neighbor order.
\end{enumerate}

As mentioned in section~\ref{subsec:cons_graph}, $\mathcal{A}$ regulates the asymmetry of the topology constructed via the modified Voronoi approach. Based on Fig~\ref{fig:p} and~\ref{fig:q}, we argue that $\mathcal{A} \in [0.55, 0.75]$ corresponds to the most suitable range in terms of feature engineering. Note that we will combine all the Type C functions and the Type D functions for the $1$-st, $2$-nd, $\cdots$ up to the $10$-th order neighbors to construct a group of features corresponding to a given $\mathcal{A}$.

\subsection{Machine Learning}
\label{sec:ml}

\subsubsection{Particle Labelling}
\label{subsec:particle_labelling}
For the purposes of building our ML classification model, we need to label each particle in the system as either hard or soft. Intuitively, most particles in the crystal phase are hard, whilst most particle in the liquid are soft. Based on the results reported in Fig.~\ref{fig:q6msd_set1} and Fig.~\ref{fig:thermo_set1}, we have chosen $t_{R/2} = 20$ and $p_c = 0.1$ to differentiate hard and soft particles. In particular, if $p_{\text{hop},i}(t)$ exceeds the selected threshold $p_c = 0.1$ during the time interval of $[t-20, t+20]$, the $i$-th particle is labelled as soft and hard otherwise. Further details on the choice of $p_c$ can be found in Appendix~\ref{app:p_hop}.
Therefore, in the context of the classification tasks to follow, the dependent variable $\bold{y}$ categorizes particles into two types: soft or hard, i.e., $\bold{y} \in \{\text{soft}, \text{hard} \}^{N}$. The soft category is designated as the positive class, while the hard category is assigned as the negative class.

\subsubsection{Feature Engineering}
\label{subsec:feature_eng}
In the case of symmetry functions, the radial ($G$) functions and the (angular) $\Psi$ functions can be used either separately (``A" or ``B" descriptors, respectively) or in combination with each other (``A+B" descriptor). When working with $G$ functions, we consider two different sets, both initialised for $r=1$ and incrementally choosing $p$ distinct $G$ functions with increments of $\delta$, where $\delta=0.05$ for Group $1$ and $\delta=0.1$ for Group $2$, which are detailed in Appendix~\ref{app:para_g}. Both groups explore scenarios with $p=60$ and $p=80$, respectively. We choose this range of $r$ because in a three-dimensional system of size $L$ obeying periodic boundary conditions, the radial distribution function is defined up to $L/2$. In our MD trajectories, $L$ fluctuates around $15$, and the $r_{\max}$ for each group is between $L/3$ and $L/2$. Concerning the choice of the $\Psi$ functions (see Eq.~\ref{equ:psi_function}), we selected specific $20$ combinations of $\xi$, $\zeta$ and $\lambda$ (as detailed in Appendix~\ref{app:para_psi}) in conjunction with a cutoff distance of $r_{\text{cut}}=4$ directly.

In the case of the graph-based descriptors proposed in this work, Type C (centrality or clustering coefficient) and Type D (angular functions on hierarchical structures) can also be used either separately (``C" or ``D" descriptors, respectively) or in combination with each other (``C+D" descriptor). For the Type C descriptor, all $10$ functions, including $9$ kinds of centralities and clustering coefficient, are included. For the Type D descriptor, we selected the functions corresponding to the $1$-st to $10$-th order neighbors, i.e., $\mathcal{F}_{\mathcal{A}}(i, \mathcal{L})$, where $\mathcal{L} \in [1,10] \cap \mathbb{N}$. As discussed in section~\ref{subsec:cons_graph}, we have chosen $r_c=2.5$. For each of three descriptor types, we select one or more than one values of $\mathcal{A}$ from the set $\{0.55,0.6,0.65,0.7,0.75 \}$ to obtain different combinations of input features.

Further details about the construction of these descriptors, along with a discussion of the performance metrics (accuracy, MCC and AUC) we have utilised to assess the performance of our ML models can be found in Appendix~\ref{app:para_psi}.

\subsubsection{ML models}
\label{subsec:4_models}
In the context of machine learning, the feature matrix is denoted as $\bold{X}$, which consists of $n$ columns for $n$ features (labeled as $\bold{X}^{(l)}$ where $l \in [1,n]$), and $N$ rows for $N$ samples (labeled as $\bold{X}_{(z)}$, where $z \in [1,N]$). The goal is to train a function $f$ to construct a relationship $\bold{y} = f(\bold{X})$, where $\bold{y}=(y_1, \cdots, y_N)^T$ represents the vector of dependent variables, and $f$ is trained to perform a task of binary classification.

In order to compare our results to those obtained by Cubuk \textit{et al.}~\cite{cubuk2015identifying} and to explore the impact of different ML models, we have considered four different classes of ML algorithms (i.e., four different $f$), namely:

\vspace{0.25cm}
\paragraph{Support Vector Machine (Linear Kernel)}
\begin{itemize}
    \item Decision function:
    \begin{equation*}
        f(\bold{X}) = \bold{X} \cdot \bold{w} + b
    \end{equation*}
    where $\bold{w}$ is the weight vector, and $b$ is the bias term.
    \item Objective function:
    \begin{equation*}
        \mathop{\min}_{\bold{w},b} \frac{1}{2} ||\bold{w}||_2^2 + C \sum_{i=1}^N \mathop{\max} (0, 1-y_i (\bold{X}_{(i)} \cdot \bold{w} + b))
    \end{equation*}
    where $C$ is the regularization parameter that controls the trade-off between the regularization term and the loss term. The second term of this expression corresponds to the so-called ``hinge loss'' used to quantify the prediction error.
    \item Hyperparameters: $C \in [0.001,1000]$, log-scaled.
\end{itemize}

In this simplified linear kernel assumption, the directed distance in input space between the sample point $X_{(z)}$ and the decision plane is defined as the softness, i.e., 

\begin{equation}
    \label{equ:softness}
    \mathop{Softness}(z) = \frac{\bold{X}_{(z)} \cdot \bold{w}+b}{||\bold{w}||_2}
\end{equation}

\vspace{0.25cm}
\paragraph{Support Vector Machine (RBF Kernel)}
\begin{itemize}
    \item RBF kernel:
    \begin{equation*}
        K(\bold{X}_{(i)},\bold{X}_{(j)}) = \mathop{\exp}(-\gamma || \bold{X}_{(i)} - \bold{X}_{(j)} ||^2)
    \end{equation*}
    where $\gamma$ is a parameter that controls the extent of the influence of each data point on the similarity measure.

    \item Decision function:
    \begin{equation*}
        f(\bold{X}) = \sum_{i=1}^N \alpha_i y_i K(\bold{X},\bold{X}_{(i)}) + b
    \end{equation*}
    where $\alpha_i$ are Lagrange multipliers quantifying the importance of $X_{(i)}$ within the decision function and $b$ is the bias term.

    \item Objective function:
    \begin{equation*}
        \mathop{\max}_{\alpha} \left[ \sum_{i=1}^N \alpha_i - \frac{1}{2} \sum_{i=1}^N \sum_{j=1}^N \alpha_i \alpha_j y_i y_j K(\bold{X}_{(i)},\bold{X}_{(j)}) \right]
    \end{equation*}
    subject to the constraints:
    \begin{equation*}
        \sum_{i=1}^N \alpha_i y_i = 0 \text{ and } 0 \leq \alpha_i \leq C, \forall i.
    \end{equation*}
    Here, $C$ is a regularization parameter, balancing the trade-off between model complexity and tolerance; $\alpha_i$ are Lagrange multipliers relative to each sample.

    \item Hyperparameters: (1) $C \in [0.001,1000]$, log-scaled; (2) $\gamma \in [0.001,1000]$, log-scaled.
\end{itemize}

In this scenario, $f(\bold{X})$ does not directly depend on the feature vectors themselves but is calculated through interactions between feature vectors, i.e., $\bold{X}^{l}$, via the kernel function $K$, allowing the SVM to handle more complex (and, crucially, non-linear) mappings.

\vspace{0.25cm}
\paragraph{Ensemble Learning}
\begin{itemize}
    \item Decision function:
    \begin{equation*}
        \bold{y} = f(\bold{X}) = \sum_{m=1}^{M} \alpha_m \cdot \eta \cdot T_m\left(SF(\bold{X}, F); S\right)
    \end{equation*}
    \item Objective function:
    \begin{equation*}
        \mathop{\min}_{\{T_m\}} \left(\frac{1}{N} \sum_{i=1}^N L \left(Y_i, \sum_{m=1}^M \alpha_i  \eta  T_m (SF(X_i,F);S)\right) + \lambda R(\{T_m\})\right)
    \end{equation*}
    \item Learner Type: decision tree $T_m, m \in [1,M] \cap \mathbb{N}$.
    \item Hyperparameters: (1) $S \in [2,N-1] \cap \mathbb{N}$, log-scaled; (2) $M \in [10,500] \cap \mathbb{N}$, log-scaled; (3) $\eta \in [0.001,1]$, log-scaled; (4) $F \in [1,\mathop{\max}(2,n)] \cap \mathbb{N}$; (5) Ensemble method $\in \{$AdaBoost, RUSBoost, LogitBoost, GentleBoost, Bagging$\}$.
    
    Further details on the five different ensemble methods we have experimented with can be found in Appendix \ref{subsec:ensemble_methods}.
\end{itemize}

Here: (1) $T_m$ denotes the $m$-th decision tree’s prediction function; (2) $\eta$ is the learning rate, adjusting the influence of each tree on the final model; (3) $\alpha_m$ is the weight of the $m$-th tree; (4) $SF(\bold{X}, F)$ represents the process of feature selection, where $F$ features are selected for training each tree; (5) $S$ indicates the maximum number of splits allowed in training each tree; (6) $L(Y_i, \bold{y}_i)$ is the loss function, capturing the difference between the model predictions and the actual labels; (7) $R({T_m})$ is the regularization term, possibly considering the complexity of the model based on aspects like tree depth or the total number of leaves; (8) $N$ is the total sample count, and is the regularization strength parameter, balancing the impact of the loss and regularization terms; (9) $M$ means the number of base learners.

\vspace{0.25cm}
\paragraph{Neural Network (Multilayer Perceptron)}
\begin{itemize}
    \item Network Structure:
    Consider a multi-layer, fully connected neural network consisting of $P$ layers, where each layer $p$ contains $\mathcal{Q}_p$ neurons. Given a feature matrix $\bold{X}$, which contains multiple samples, each with $n$ features (i.e., the number of columns in $\bold{X}$), the forward propagation of the network can be represented as a series of nested functions. Specifically:

    (1) For the first layer:
    \begin{equation*}
        F^{(1)}(\bold{X}) = \sigma^{(1)} (\bold{W}^{(1)} \bold{X} + \bold{b}^{(1)})
    \end{equation*}
    where $\bold{W}^{(1)}$ is the weight matrix of dimension $\mathcal{Q}_1 \times n$ and $\bold{b}^{(1)}$ is a bias vector of length $\mathcal{Q}_1$.
   
   (2) For the $p$-th layer ($2 \leq p \leq P-1$):
   \begin{equation*}
        F^{(p)}(\bold{A}^{(p-1)}) = \sigma^{(p)} (\bold{W}^{(p)} \bold{A}^{(p-1)} + \bold{b}^{(p)})
    \end{equation*}
    where $\bold{A}^{(p-1)}$ is the output of the $(p-1)$-th layer, $\bold{W}^{(p)}$ is the weight matrix of dimension $\mathcal{Q}_p \times \mathcal{Q}_{p-1}$ and $\bold{b}^{(p)}$ is a bias vector of length $\mathcal{Q}_p$.

    (3) For the output layer:
    \begin{equation*}
        F^{(P)}(\bold{A}^{(P-1)}) = \sigma^{(P)} (\bold{W}^{(P)} \bold{A}^{(P-1)} + \bold{b}^{(P)})
    \end{equation*}
    where $\bold{W}^{(P)}$ is the weight matrix, $\bold{b}^{(P)}$ is a bias vector of length $\mathcal{Q}_P$, and $\sigma^{(P)}$ is a Softmax activation function.
    
    \item Decision Function:
    \begin{equation*}
        \bold{y} = f(\bold{X}) = F^{(P)}( F^{(P-1)}(\cdots F^{(2)} (F^{(1)}(\bold{X})))
    \end{equation*}
    
    \item Objective function:
    \begin{equation*}
        \mathop{\min}_{\bold{W},\bold{b}} L(\bold{Y}, f(\bold{X};\bold{W},\bold{b})) + \lambda ||\bold{W}||_{2}^2
    \end{equation*}
    where $L$ is the loss function, representing the averaged per-sample loss $\ell$ across all samples. In particular:
    \begin{equation*}
        L(\bold{Y}, f(\bold{X};\bold{W},\bold{b})) = \frac{1}{N} \sum_{i=1}^N \ell (\bold{Y}_i, f(\bold{X}_i; \bold{W}, \bold{b}))
    \end{equation*}
    where $\bold{Y}_i$ is the target for the $i$-th sample, and $\ell$ is the cross-entropy loss, which in turn is defined as:
    \begin{equation*}
        \ell(\bold{Y}_i, \hat{y}_i) = -[\bold{Y}_i \mathop{\log}(\hat{y}_i) + (1-\bold{Y}_i) \mathop{\log}(1-\hat{y}_i)]
    \end{equation*}
    where $\hat{y}_i=f(\bold{X}_i; \bold{W}, \bold{b})$ is the predicted probability that the $i$-th sample belongs to the positive class.

    \item Hyperparameters: (1) $P \in \{1,2,3\}$; (2) $\sigma^{(q)} \in \{\text{ReLU}, \text{Tanh}, \text{None}, \text{Sigmoid}\}$, where $q \in [1,P-1]$; (3) $\lambda \in [0.00001/N, 100000/N]$, log-scaled.

    Additional details on the activation functions we have chosen can be found in Appendix \ref{subsec:activation_function}.
\end{itemize}

\subsubsection{Protocol}
All the ML models presented in this work have been built via the Classification Learner app from the Statistics and Machine Learning Toolbox of MATLAB~\cite{app,toolbox,MATLAB}. To ensure the reproducibility of our results, all our training and testing records have been saved as MATLAB session files. These, together with the relevant MATLAB code, are available at: \url{https://github.com/gcsosso/SOFT_GRAPH.git}

For consistency, we have utilised the following setting for every model:

\begin{itemize}
    \item Validation: k-fold cross-validation, $k=10$;
    \item Model Selection: Bayesian optimizer~\cite{snoek2012practical};
    \item Evaluation indicators: Matthews Correlation Coefficient (MCC)~\cite{matthews1975comparison}, accuracy, and AUC~\cite{hanley1982meaning};
    \item Number of steps for the Bayesian optimizer: $200$.
    
    An discussion on the hyper parameters involved with Bayesian optimization is provided in Appendix \ref{subsec:bayes}.
\end{itemize}

We designed four groups of controlled studies, which we discuss in detail in the next section (Section~\ref{subsec:con_exp}). For each group, we employ the four ML algorithm discussed above, i.e., linear SVM, SVM with RBF kernel, ensemble learning, and neural network. Each training is executed as the protocols above.

\subsubsection{The controlled studies}
\label{subsec:con_exp}
To demonstrate the advantages of our graph-based structural descriptors (Type C and D) over symmetry functions (Type A and B), we devised four controlled experiments, which seek to provide a comparison between the two different classes of descriptors form different standpoints. Specifically:

\paragraph{Exp A} The aim of Exp A is to assess how effectively a feature engineering technique can extract structural features from configurations collected across different temperatures, volumes, densities, and phases of matter. Therefore, from Set $1$, we selected $15,000$ balanced samples (the number of positive and negative samples is the same) for the development set and another $15,000$ non-overlapping, balanced samples for the test set. Thus, in this case we are interested in the performance of the models re: Set $1$, in terms of capturing the structural features of the system.

\paragraph{Exp B} The purpose of Exp B is to investigate the impact of feature engineering on the transferability of the model. In particular, we aim to apply the most accurate model obtained from Exp A (trained and tested on Set $1$ only) to Set $2$, so as to assess whether the structural features learned on Set $1$ remain applicable to a different set of structures. To this end, we randomly chose $30,000$ balanced samples from Set $2$ as the test set to evaluate the performance of the most accurate model we have obtained via Exp A on Set $2$.

\paragraph{Exp C} The aim of Exp C is to investigate whether the structural features we have learned in Exp A are robust with respect to the evolution of the system over time. Thus, we randomly selected $10,000$ balanced samples from the set of frames corresponding to $t_1 = 100$-th step in Set $1$ and Set $2$ as the training set, and $30,000$ balanced samples from the set of frames corresponding to $t_j = 100j$-th step (where $j\in[2,9]\cap \mathbb{N}$) as the test set.

\paragraph{Exp D} The aim of Exp D is to assess whether the model can capture the invariant features in a single trajectory. That is to say, we want to investigate whether the features extracted at a single temperature can be generalized to other temperature conditions. To this end, we randomly selected $5,000$ balanced samples from the set of frames at step $t_j=100j$, iterated through all $j\in[1,9]\cap \mathbb{N}$ in $\text{Traj}(T^{(1)}_5)$ from Set $1$ (which contains multiple instances where the system has crystallised) as the training set, and $60,000$ balanced samples were randomly selected from Set $2$ instead as the test set.

\section{Results}
\label{subsec:results_ml}

\subsection{Classifying hard or soft particles: symmetry functions vs graph-based descriptors}

In symmetry function-based methods, the density of the system corresponding to the first peak $r=r_{\text{peak}}$ of the radial distribution function is crucial. According to the notation in section~\ref{sec:trad}, the value of the $G$ function at $r=r_{\text{peak}}$ is denoted as $G_I (i, r_{\text{peak}}, \delta)$. Cubuk~\textit{et al.} pointed out that the value of $G_I (i, r_{\text{peak}}, \delta)$ for hard particles (see section~\ref{subsec:particle_labelling} for the definition of hard/soft particle) at $r = r_{\text{peak}}$ will lead to two different distributions, thus categorizing the hard particles into two classes, i.e., H0 and H1~\cite{cubuk2015identifying}. Specifically: (1) H0-type hard particles have relatively low values of $G_{I}(i, r_{\text{peak}}, \delta)$ at $r = r_{\text{peak}}$, i.e., $G_{I}(i, r_{\text{peak}}, \delta) / r < 1/2$; (2) H1-type hard particles have relatively high values of $G_{I}(i, r_{\text{peak}}, \delta)$ at $r = r_{\text{peak}}$, i.e., $G_{I}(i, r_{\text{peak}}, \delta) / r \geq 1/2$. Cubuk~\textit{et al.} also pointed out that the $G$ function can accurately distinguish between H1-type hard particles and soft particles but cannot accurately distinguish between H0-type hard particles and soft particles. Conversely, the $\Psi$ function can accurately distinguish between H0-type hard particles and soft particles but cannot accurately distinguish between H1-type hard particles and soft particles. The fundamental reason why the $G$ function plays a decisive role in distinguishing particles is that H1-type hard particles are more numerous. However, only by combining the $G$ function and the $\Psi$ function can all particles be accurately distinguished. That is to say, in the symmetry function-based method, the $\Psi$ function clearly provides information complementary to that provided by the $G$ function. 

Conversely, our graph-based method does not depend on the radial distribution function. As such, in the context of our graph-based approached we do not classify hard particles further into H0 or H1 types. Instead, we can directly classify all particles using node centrality and clustering coefficient. Hence, in the context of our method, the angular functions defined on hierarchical graph structures ($\mathcal{F}_{\mathcal{A}}(i, \mathcal{L})$, as defined in Equation~\ref{equ:angular_func_graphs}) simply contribute to an extent to improve the results obtained via Type C descriptors - however, their usage is not essential to ensure the accuracy of our graph-based models.
This is the reason why, in our comparison, we do not consider the results obtained via Type B and Type D descriptors in isolation. In fact, the amount of information provided by the $\Psi$ function is inferior to that provided by the $G$ function: for instance, Ganapathi~\textit{et al.}~\cite{ganapathi2021structure} have used $G$ functions alone to determine the emergence of crystalline nuclei in colloidal glasses, without $\Psi$ functions. Therefore, the results obtained via Type B and Type D descriptors in isolation are presented in Appendix~\ref{app:results} for completeness.

We note that, in comparing traditional descriptors with our graph-based descriptors, we have the opportunity to assess whether univariate feature generation (e.g., node centrality) is superior to parametric feature genera-
tion (e.g., $G$ functions and $\Psi$ functions) in the context of describing the properties of the system. Parametric feature generation relies on a specific principle or formula, thus generating multiple homogeneous features by varying different parameters. In contrast, univariate feature generation derives heterogeneous features from a diverse portfolio of sources of information and a variety of different computational methods as well, each offering a unique perspective on the system's structure. The results presented in this section serve to demonstrate that univariate feature generation is superior to parametric feature generation in the context of this particular system.

All the results relative to the four controlled studies discussed in the previous section \ref{subsec:con_exp} are reported in their entirety in Appendix~\ref{app:results}. Here, we report a selection of those results in Tables~\ref{tab:a_1} to~\ref{tab:d_1}. 
Note that the labels within the ``No." column in Tables\ref{tab:a_1} to~\ref{tab:d_1} refer to the nomenclature discussed in Appendix~\ref{app:results} re: the full set of results. To ensure a fair evaluation, we compare Type A with Type C descriptors, or Type ``A+B" with Type ``C+D". Note the nomenclature utilised in Tables~\ref{tab:a_1} to~\ref{tab:d_1}: specifically, identifiers starting with ``P'' represent groups based on traditional symmetry functions, while identifiers starting with ``Q'' represent groups based on graph descriptors. For convenience, the four ML methods we have used are numbered as follows:

\vspace{0.25cm}
\begin{tabular}{|>
{\centering\arraybackslash}m{1.8cm}|>{\centering\arraybackslash}m{1.8cm}|>{\centering\arraybackslash}m{1.8cm}|>{\centering\arraybackslash}m{1.8cm}|}
  \hline
  1 & 2 & 3 & 4 \\
  \hline
  Linear SVM & SVM with RBF kernel & Ensemble Learning & Neural Networks \\
  \hline
\end{tabular}
\vspace{0.25cm}

Note that in section~\ref{subsec:feature_eng} we introduced two different methods for constructing Type A descriptors, i.e., Group 1 and Group 2. Interestingly, we found that Group 2 descriptors are superior to Group 1 in almost every aspect. As such, the results re: the Type A descriptors (and their combination with B decsriptors,i.e., ``A+B" in tables~\ref{tab:a_1} to~\ref{tab:d_1}) refer to Group 2 Type A descriptors exclusively. These results are in line with the expectations we have set in section~\ref{sec:graph}, as our graph descriptors are not sensitive to the metric structure of the system (and thus to the specific thermodynamic state of the system) but to topological invariance instead.

\begin{table}[htbp]
\footnotesize
\renewcommand{\arraystretch}{1.5}
\centering
\caption{Best Performance Models in Exp A}
\label{tab:a_1}
\begin{tabular}{c|c|c|c|c|c|c}
\hline
\multicolumn{7}{c}{Linear SVM} \\
\hline
Type & No. & Model & Predictors & Accuracy & MCC & Confusion Matrix \\ \hline
\rowcolor{yellow!20}
A & P4 & 1 & $80$ & $76.3\%$ & $0.521$ & $\left(\begin{smallmatrix} 5645 & 1855 \\ 1695 & 5805 \end{smallmatrix}\right)$ \\
C & Q1 & 1 & $10$ & $75.9\%$ & $0.521$ & $\left(\begin{smallmatrix} 5330 & 2170 \\ 1444 & 6056 \end{smallmatrix}\right)$ \\
\rowcolor{gray!20}
C & Q3 & 1 & $50$ & $83.8\%$ & $0.676$ & $\left(\begin{smallmatrix} 6239 & 1261 \\ 1166 & 6334 \end{smallmatrix}\right)$ \\
\hline
\rowcolor{yellow!20}
A+B & P9 & 1 & $100$ & $86.4\%$ & $0.728$ & $\left(\begin{smallmatrix} 6399 & 1101 \\ 938 & 6562 \end{smallmatrix}\right)$ \\
C+D & Q8 & 1 & $20$ & $81.9\%$ & $0.638$ & $\left(\begin{smallmatrix} 6006 & 1494 \\ 1227 & 6273 \end{smallmatrix}\right)$ \\
\rowcolor{gray!20}
C+D & Q10 & 1 & $100$ & $88.3\%$ & $0.767$ & $\left(\begin{smallmatrix} 6566 & 934 \\ 814 & 6686 \end{smallmatrix}\right)$ \\
\hline
\multicolumn{7}{c}{All models} \\
\hline
Type & No. & Model & Predictors & Accuracy & MCC & Confusion Matrix \\ \hline
\rowcolor{yellow!20}
A & P4 & 4 & $80$ & $84.3\%$ & $0.686$ & $\left(\begin{smallmatrix} 6456 & 1044 \\ 1313 & 6187 \end{smallmatrix}\right)$ \\
C & Q1 & 3 & $10$ & $89.0\%$ & $0.780$ & $\left(\begin{smallmatrix} 6541 & 959 \\ 698 & 6802 \end{smallmatrix}\right)$ \\
\rowcolor{gray!20}
C & Q2 & 3 & $40$ & $90.8\%$ & $0.815$ & $\left(\begin{smallmatrix} 6734 & 766 \\ 620 & 6880 \end{smallmatrix}\right)$ \\
\hline
\rowcolor{yellow!20}
A+B & P9 & 4 & $100$ & $88.5\%$ & $0.770$ & $\left(\begin{smallmatrix} 6599 & 901 \\ 821 & 6679 \end{smallmatrix}\right)$ \\
C+D & Q8 & 3 & $20$ & $89.4\%$ & $0.790$ & $\left(\begin{smallmatrix} 6563 & 937 \\ 646 & 6854 \end{smallmatrix}\right)$ \\
\rowcolor{gray!20}
C+D & Q10 & 3 & $100$ &$90.7\%$ & $0.815$ & $\left(\begin{smallmatrix} 6739 & 761 \\ 629 & 6871 \end{smallmatrix}\right)$ \\\hline

\end{tabular}
\end{table}

\begin{table}[htbp]
\footnotesize
\renewcommand{\arraystretch}{1.5}
\centering
\caption{Best Performance Models in Exp B}
\label{tab:b_1}
\begin{tabular}{c|c|c|c|c|c|c}
\hline
\multicolumn{7}{c}{Linear SVM} \\
\hline
Type & No. & Model & Predictors & Accuracy & MCC & Confusion Matrix \\ \hline
\rowcolor{yellow!20}
A & P4 & 1 & $80$ & $71.7\%$ & $0.434$ & $\left(\begin{smallmatrix} 10332 & 4668 \\ 3832 & 11168 \end{smallmatrix}\right)$ \\
C & Q1 & 1 & $10$ & $76.5\%$ & $0.530$ & $\left(\begin{smallmatrix} 11280 & 3720 \\ 3329 & 11671 \end{smallmatrix}\right)$ \\
\rowcolor{gray!20}
C & Q2 & 1 & $40$ & $83.2\%$ & $0.664$ & $\left(\begin{smallmatrix} 12823 & 2177 \\ 2874 & 12126 \end{smallmatrix}\right)$ \\
\hline
\rowcolor{yellow!20}
A+B & P9 & 1 & $100$ & $76.8\%$ & $0.535$ & $\left(\begin{smallmatrix} 11605 & 3395 \\ 3576 & 11424 \end{smallmatrix}\right)$ \\
C+D & Q8 & 1 & $20$ & $82.6\%$ & $0.652$ & $\left(\begin{smallmatrix} 12547 & 2453 \\ 2770 & 12230 \end{smallmatrix}\right)$ \\
\rowcolor{gray!20}
C+D & Q10 & 1 & $100$ & $87.5\%$ & $0.750$ & $\left(\begin{smallmatrix} 13355 & 1645 \\ 2113 & 12887 \end{smallmatrix}\right)$ \\
\hline
\multicolumn{7}{c}{All models} \\
\hline
Type & No. & Model & Predictors & Accuracy & MCC & Confusion Matrix \\ \hline
\rowcolor{yellow!20}
A & P4 & 2 & $80$ & $82.8\%$ & $0.656$ & $\left(\begin{smallmatrix} 12301 & 2699 \\ 2457 & 12543 \end{smallmatrix}\right)$ \\
C & Q1 & 3 & $10$ & $85.7\%$ & $0.715$ & $\left(\begin{smallmatrix} 13407 & 1593 \\ 2708 & 12292 \end{smallmatrix}\right)$ \\
\rowcolor{gray!20}
C & Q2 & 3 & $40$ & $90.8\%$ & $0.817$ & $\left(\begin{smallmatrix} 13887 & 1113 \\ 1634 & 13366 \end{smallmatrix}\right)$ \\
\hline
\rowcolor{yellow!20}
A+B & P9 & 2 & $100$ & $84.8\%$ & $0.696$ & $\left(\begin{smallmatrix} 12586 & 2414 \\ 2152 & 12848 \end{smallmatrix}\right)$ \\
C+D & Q8 & 3 & $20$ & $86.6\%$ & $0.733$ & $\left(\begin{smallmatrix} 13516 & 1484 \\ 2546 & 12454 \end{smallmatrix}\right)$ \\
\rowcolor{gray!20}
C+D & Q10 & 3 & $100$ &$91.2\%$ & $0.824$ & $\left(\begin{smallmatrix} 13925 & 1075 \\ 1565 & 13435 \end{smallmatrix}\right)$ \\
\hline
\end{tabular}
\end{table}

\begin{table}[htbp]
\footnotesize
\renewcommand{\arraystretch}{1.5}
\centering
\caption{Best Performance Models in Exp C}
\label{tab:c_1}
\begin{tabular}{c|c|c|c|c|c|c}
\hline
\multicolumn{7}{c}{Linear SVM} \\
\hline
Type & No. & Model & Predictors & Accuracy & MCC & Confusion Matrix \\ \hline
\rowcolor{yellow!20}
A & P2 & 1 & $60$ & $53.8\%$ & $0.077$ & $\left(\begin{smallmatrix} 10332 & 4668 \\ 3832 & 11168 \end{smallmatrix}\right)$ \\
\rowcolor{gray!20}
C & Q1 & 1 & $10$ & $76.5\%$ & $0.530$ & $\left(\begin{smallmatrix} 11280 & 3720 \\ 3329 & 11671 \end{smallmatrix}\right)$ \\
\hline
\rowcolor{yellow!20}
A+B & P7 & 1 & $80$ & $69.0\%$ & $0.383$ & $\left(\begin{smallmatrix} 11605 & 3395 \\ 3576 & 11424 \end{smallmatrix}\right)$ \\
\rowcolor{gray!20}
C+D & Q8 & 1 & $20$ & $81.2\%$ & $0.626$ & $\left(\begin{smallmatrix} 12547 & 2453 \\ 2770 & 12230 \end{smallmatrix}\right)$ \\
\hline
\multicolumn{7}{c}{All models} \\
\hline
Type & No. & Model & Predictors & Accuracy & MCC & Confusion Matrix \\ \hline
\rowcolor{yellow!20}
A & P4 & 2 & $80$ & $72.8\%$ & $0.458$ & $\left(\begin{smallmatrix} 12301 & 2699 \\ 2457 & 12543 \end{smallmatrix}\right)$ \\
\rowcolor{gray!20}
C & Q1 & 3 & $10$ & $85.7\%$ & $0.715$ & $\left(\begin{smallmatrix} 13407 & 1593 \\ 2708 & 12292 \end{smallmatrix}\right)$ \\
\hline
\rowcolor{yellow!20}
A+B & P9 & 2 & $100$ & $86.1\%$ & $0.723$ & $\left(\begin{smallmatrix} 12586 & 2414 \\ 2152 & 12848 \end{smallmatrix}\right)$ \\
\rowcolor{gray!20}
C+D & Q8 & 3 & $20$ & $87.8\%$ & $0.757$ & $\left(\begin{smallmatrix} 13516 & 1484 \\ 2546 & 12454 \end{smallmatrix}\right)$ \\
\hline
\end{tabular}
\end{table}

\begin{table}[htbp]
\footnotesize
\renewcommand{\arraystretch}{1.5}
\centering
\caption{Best Performance Models in Exp D}
\label{tab:d_1}
\begin{tabular}{c|c|c|c|c|c|c}
\hline
\multicolumn{7}{c}{Linear SVM} \\
\hline
Type & No. & Model & Predictors & Accuracy & MCC & Confusion Matrix \\ \hline
\rowcolor{yellow!20}
A & P1 & 1 & $60$ & $56.6\%$ & $0.134$ & $\left(\begin{smallmatrix} 10332 & 4668 \\ 3832 & 11168 \end{smallmatrix}\right)$ \\
C & Q1 & 1 & $10$ & $73.1\%$ & $0.463$ & $\left(\begin{smallmatrix} 11280 & 3720 \\ 3329 & 11671 \end{smallmatrix}\right)$ \\
\rowcolor{gray!20}
C & Q2 & 1 & $40$ & $75.8\%$ & $0.518$ & $\left(\begin{smallmatrix} 11280 & 3720 \\ 3329 & 11671 \end{smallmatrix}\right)$ \\
\hline
\rowcolor{yellow!20}
A+B & P9 & 1 & $100$ & $66.7\%$ & $0.336$ & $\left(\begin{smallmatrix} 11605 & 3395 \\ 3576 & 11424 \end{smallmatrix}\right)$ \\
C+D & Q8 & 1 & $20$ & $75.4\%$ & $0.517$ & $\left(\begin{smallmatrix} 12547 & 2453 \\ 2770 & 12230 \end{smallmatrix}\right)$ \\
\rowcolor{gray!20}
C+D & Q9 & 1 & $80$ & $82.0\%$ & $0.649$ & $\left(\begin{smallmatrix} 12547 & 2453 \\ 2770 & 12230 \end{smallmatrix}\right)$ \\
\hline
\multicolumn{7}{c}{All models} \\
\hline
Type & No. & Model & Predictors & Accuracy & MCC & Confusion Matrix \\ \hline
\rowcolor{yellow!20}
A & P4 & 2 & $80$ & $71.5\%$ & $0.438$ & $\left(\begin{smallmatrix} 12301 & 2699 \\ 2457 & 12543 \end{smallmatrix}\right)$ \\
C & Q1 & 3 & $10$ & $80.7\%$ & $0.616$ & $\left(\begin{smallmatrix} 13407 & 1593 \\ 2708 & 12292 \end{smallmatrix}\right)$ \\
\rowcolor{gray!20}
C & Q2 & 3 & $40$ & $81.7\%$ & $0.636$ & $\left(\begin{smallmatrix} 13407 & 1593 \\ 2708 & 12292 \end{smallmatrix}\right)$ \\
\hline
\rowcolor{yellow!20}
A+B & P7 & 2 & $80$ & $78.8\%$ & $0.583$ & $\left(\begin{smallmatrix} 12586 & 2414 \\ 2152 & 12848 \end{smallmatrix}\right)$ \\
C+D & Q8 & 3 & $20$ & $81.7\%$ & $0.636$ & $\left(\begin{smallmatrix} 13516 & 1484 \\ 2546 & 12454 \end{smallmatrix}\right)$ \\
\rowcolor{gray!20}
C+D & Q9 & 1 & $80$ & $82.0\%$ & $0.649$ & $\left(\begin{smallmatrix} 13516 & 1484 \\ 2546 & 12454 \end{smallmatrix}\right)$ \\ \hline
\end{tabular}
\end{table}

We begin the discussion of these results by focusing on Table~\ref{tab:a_1}, which summarises the main outcomes of Exp A (see Section~\ref{subsec:con_exp}). When using linear SVM, our graph-based features (Type C, i.e., the graph centrality or clustering descriptors) outperform the symmetry functions (Type A, i.e. radial symmetry functions). Interestingly, we achieve significant improvements (MCC = 0.676) by utilising 50 Type C descriptors, compared to 80 Type A descriptors (MCC = 0.521). In fact, by using only 10 Type C descriptors we can achieve the same accuracy (MCC = 0.521) that we obtain by using 80 Type A descriptors. 

The same arguments hold when considering the performance of the combined Type A+B descriptors compared to that of the combined Type C+D descriptors, albeit the gains in accuracy are less dramatic (MCC or 0.728 and 0.767 for Type A+B and Type C+D descriptors, respectively) and in this case we needed to use the same (large) number of descriptors (100) to obtain our best performance. Similar considerations apply to the case where we utilised non-linear ML algorithms (i.e., ``All models" section in Table~\ref{tab:a_1}), which consistently boost the performance of all our ML models. We note that our best model re: Exp A has an accuracy of 90.7\% and a MCC of 0.815: this is an impressive performance that really give us confidence in the applicability of graph-based metrics to capture the structural features of condensed matter systems.

Moving to Exp B (Table~\ref{tab:b_1}), where we apply the models obtained via  Exp A to a different dataset, the accuracy gains we observe when moving from symmetry functions to graph-based descriptors are stark. For instance, using 80 Type A descriptors yields a MCC of 0.434, whilst using only 10 Type C descriptors we obtain a MCC of 0.530. Substantial improvements can be observed when considering the combined (Type A+B and Type C+D) descriptors as well, notwithstanding the usage of linear SVM or any other ML model. Our best result for this set of studies has been obtained via a combination of Type C and Type D (i.e., angular functions on the hierarchical graph structure) descriptors, and corresponds to an accuracy of 91.2\% with an MCC of 0.824. In contrast, the best results for this set obtained via (the same number of) symmetry functions yields an accuracy of 84.8\% and a MCC of 0.696. These results are especially significant in that they demonstrate the superior transferability of the ML models we have built via our graph-based metrics.

Exp C (Table~\ref{tab:c_1}), in which we probe the capabilities of our models in monitoring the time evolution of the structural features of the system, is where we observe the most striking differences in terms of the predictive power of our graph-based features compared to the symmetry functions. In particular, only a very limited amount of either Type C or Type C + Type D descriptors is needed to outmatch the performance of Type A or Type A + Type B descriptors, respectively. As a representative example, when using linear SVM we obtain an accuracy of 81.2\% and a MCC of 0.626 when using only 20 Type C + Type D descriptors whilst using a much larger number (100) of Type A + Type B descriptors only results in an accuracy of 69\% and a MCC of 0.383. We note that whilst we observe a massive improvement in accuracy for the symmetry functions when switching from linear SVM to the non-liner models (for instance, we see an increase in accuracy from 69\% to 86.1\% for the Type A + Type B descriptors), our graph-based metrics are less sensitive to this change - for instance we move from an accuracy of 81.2\% to 87.8\% for the Type C + Type D descriptors. Overall, these results demonstrate that graph-based features are capable to follow the structural changes of the system in time with an accuracy superior to that of symmetry functions and - crucially - needing to leverage a significantly lower number of descriptors as well.

Finally, in Exp D (Table~\ref{tab:d_1}), we explored the ability of the models to learn the topological invariance of the system from a single thermodynamic state, which is another area where our graph-based approach significantly outperforms traditional methods. When using a linear SVM, we achieved an accuracy of 73.1\% and an MCC of 0.463 using only 10 Type C descriptors, surpassing all combinations of Type A, Type B, and Type A + Type B descriptors when using the same linear SVM. When the non-linear models were employed, the accuracy and MCC using 10 Type C descriptors increased to 81.7\% and 0.636, respectively, similarly surpassing all combinations of Type A, Type B, and Type A + Type B descriptors.

Notably, when switching from a linear SVM to non-linear models, the performance improvement of traditional descriptor methods was greater than that of our graph descriptors, even though the overall performance of our graph descriptors remained higher than that obtained via the symmetry function-based approach. This is to be expected, as non-linear models captured complex patterns that linear SVMs cannot, particularly those influenced by the changes of thermodynamic quantities (such as temperatures) that affect classification results. This indicates these patterns are highly predictable, which is consistent with the results of Cubuk \textit{et al}~\cite{cubuk2015identifying,schoenholz2016structural}.

\subsection{Application of the ML models to study the dynamical properties of the system}
\label{subsec:applications}
We use the models trained via Exp A (on Set 1) to study the behavior of the supercooled Lennard-Jones liquid within the trajectories of Unit $2$, i.e., $\{\text{Traj} (T^{(2)}_p) | p\in[1,10] \cap \mathbb{N}\}$ For the trajectories in Unit 2, the sampling steps are $t_p = 100 + 10p$, where $p$ iterates over the interval of $[0, 80] \cap \mathbb{N}$. In particular, we are interested in identifying the onset of the crystallisation process within the supercooled liquid phase via the graph-based ML models discussed in the previous section. To this end, we will use two quantities, namely the hard particle counter and the average softness.

\vspace{0.25cm}
\paragraph{Hard Particle Counter} We choose model Q10 (ensemble learning, see Table~\ref{tab:b_1}), which yielded the best performance, to determine whether a particle is soft or hard. Then, we define the number of hard particles in the system at time $t$ as the hard particle counter $\mathfrak{N}(t)$:

\begin{equation*}
    \mathfrak{N}(t) = \sum_{i=1}^{\mathcal{N}} \mathbbm{1}_{\text{particle } i \text{ is hard}}(t)
\end{equation*}

where $\mathbbm{1}_{\text{particle is hard}}$ is the indicator function, and $\mathcal{N}$ is the number of particles.

\paragraph{Average Softness} According to formula (\ref{equ:softness}), when using a linear SVM the softness of particle $i$ at time $t$ is defined as its directed distance from the decision plane in the feature space, which is a subset of $\mathbb{R}^n$ (where $n$ is the number of predictors). The dynamical properties of the system as a whole, at time $t$, are characterized by the average softness $\mathfrak{S}(t)$, which is defined as:

\begin{equation*}
    \mathfrak{S}(t) = \frac{1}{\mathcal{N}} \sum_{i=1}^{\mathcal{N}} \frac{\bold{X}_{(i)}(t) \cdot \bold{w} + b}{||\bold{w}||_2}
\end{equation*}

where $\mathcal{N}$ is the number of particles, $\bold{w}$ is the trained weights of linear SVM, and $\bold{X}_{(i)}(t)$ is the feature matrix of particle $i$ at time $t$. Here, we choose the best-performance linear SVM (Q10, see Table~\ref{tab:b_1}).

The hard particle counter $\mathfrak{N}(t)$ effectively counts the number of hard particles in the system. As we expect a strong correlation between hard particles and crystalline particles, the time evolution of $\mathfrak{N}(t)$ along a given MD trajectory should follow the time evolution of $Q_6(t)$, i.e., the order parameter we have used to quantify the crystallinity of the system.
    
On the other hand, the computation of $\mathfrak{S}(t)$ requires a linear SVM. Linear SVMs typically yield less accurate results compared to non-linear ML models (assuming the same set of input features have been used) such as SVMs leveraging non-linear kernel functions or neural networks. However, the usage of our graph-based metrics has resulted in a linear SVM classifier for hard or soft particles characterised by a respectable accuracy of 87.5\% (see Q10 in Table~\ref{tab:b_1}), which should allow us to reliably capture the evolution of the dynamical properties of the system via $\mathfrak{S}(t)$. The main difference between $\mathfrak{N}(t)$ and $\mathfrak{S}(t)$ is that the former is a discrete function, whilst the latter is the arithmetic mean of (continuous) distance functions. We will compare both $\mathfrak{N}(t)$ and $\mathfrak{S}(t)$ with $Q_6(t)$.

Without loss of generality, we report the results relative to trajectories $\text{Traj}(T^{(2)}_6)$ and $\text{Traj}(T^{(2)}_7)$. This is because the onset temperature for the crystallisation of the system is located between $T^{(2)}_6$ and $T^{(2)}_6$. Thus, the system will remain in its liquid phase in $\text{Traj}(T^{(2)}_7)$ and it crystallises instead during $\text{Traj}(T^{(2)}_6)$.
Note that the results discussed below for $\text{Traj}(T^{(2)}_7)$ and $\text{Traj}(T^{(2)}_6)$ hold for any other MD trajectory, as demonstrated by the full set of results (i.e., $\mathfrak{N}(t)$, $\mathfrak{S}(t)$, $Q_6(t)$, MSD, as well as the relevant thermodynamic quantities (density, volume) for all trajectories in Unit 2, i.e., $\{\text{Traj} (T^{(2)}_p) | p\in[1,10] \cap \mathbb{N}\}$), which we report in Appendix~\ref{app:extra_res_apps}.

\begin{figure}[htbp]
\includegraphics[width=0.5\textwidth]{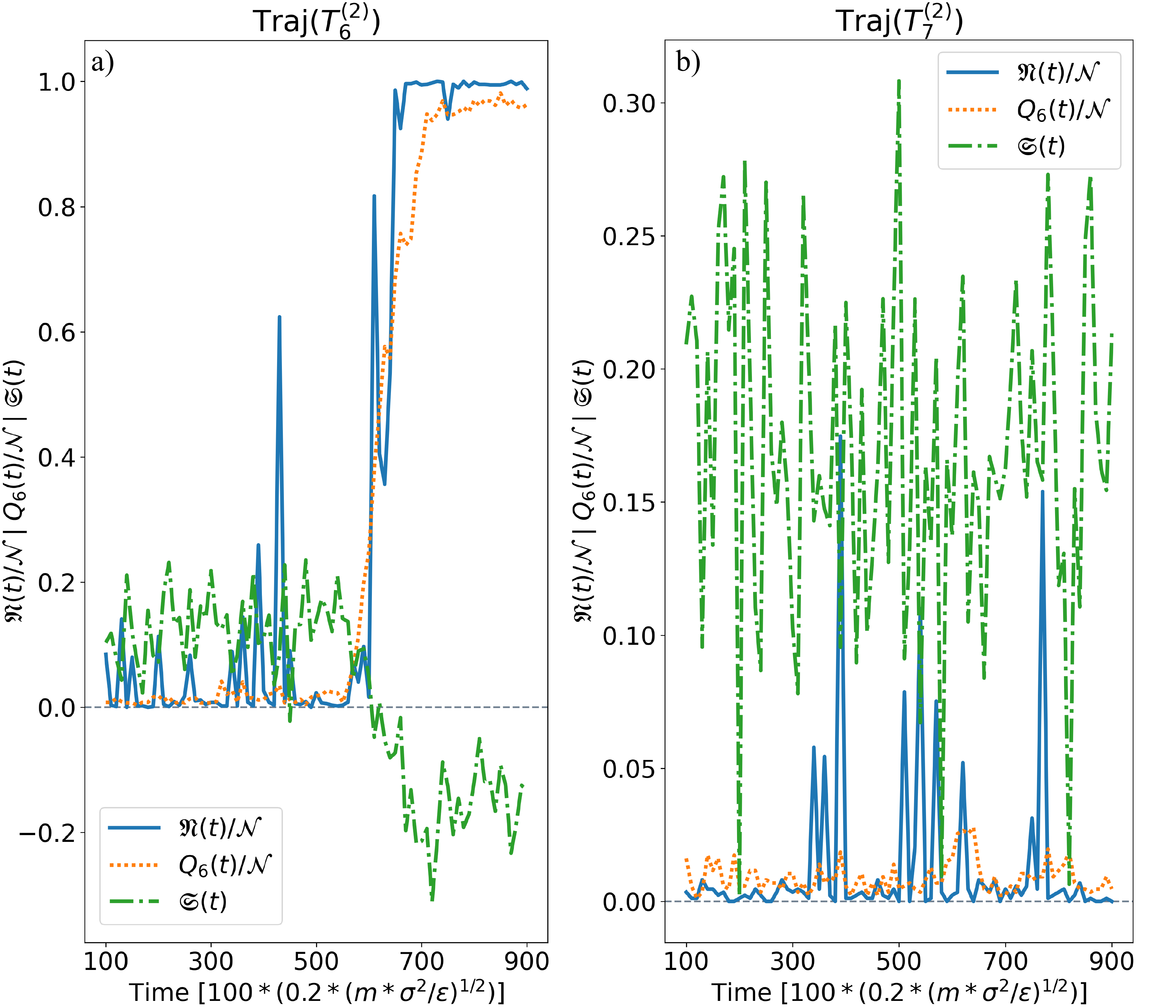}
\caption{Normalized hard particle counter $\mathfrak{N}(t)/\mathcal{N}$, normalized ten Wolde (global bond) order parameter $Q_6(t)/\mathcal{N}$ and average softness $\mathfrak{S}(t)$ for two selected trajectories corresponding to the temperatures near the onset temperature to crystallize: a) $T^{(2)}_6$ and b)$T^{(2)}_7$. Here, $\mathcal{N} = 864$ is the number of particles in the system.}
\label{fig:ntq6}
\end{figure}

We begin by investigating the correlation between $\mathfrak{N}(t)$ and $Q_6(t)$, which we report in Fig.~\ref{fig:ntq6} for $\text{Traj}(T^{(2)}_6)$ and $\text{Traj}(T^{(2)}_7)$. In the case of $\text{Traj}(T^{(2)}_7)$, both $Q_6(t)$ and $\mathfrak{N}(t)$ fluctuate around a constant, very low value. This is expected, as in $\text{Traj}(T^{(2)}_7)$ the system remains in its supercooled liquid state - hence the number of crystalline or hard particles remains low. Note that $\mathfrak{N}(t)$ displays larger fluctuations compared to $Q_6(t)$. This is because the impact of the few crystalline particles (which are always present to an extent within the supercooled liquid phase in the form of pre-critical crystalline nuclei) in the system on the value of $Q_6(t)$ is minimal, given that $Q_6(t)$ involves an average of particle-based quantities over all the particles in the system. Conversely, $\mathfrak{N}(t)$ simply counts the number of hard particles, hence it is more sensitive to the emergence of even a handful of crystalline / hard particles within the system.

In the case of $\text{Traj}(T^{(2)}_6)$, both $Q_6(t)$ and $\mathfrak{N}(t)$ identify in a consistent fashion the onset of crystallisation (around Time = 600 tu). Interestingly, monitoring $\mathcal{N}$ provides some insight into the pre-critical crystalline fluctuations within the system, with a significant fluctuation in terms of the number of hard particles observed between 300 and 500 tu - a fluctuation that is not registered by $Q_6(t)$ - as, again, the latter ``averages out'' the contribution of a small number of crystalline particles.

Moving onto the relationship between $Q_6(t)$ and the softness $\mathfrak{S}(t)$ of the system, we report in Fig.~\ref{fig:ntq6}a) the results we have obtained for the same two MD trajectories, $\text{Traj}(T^{(2)}_6)$ and $\text{Traj}(T^{(2)}_7)$. Consistently with both $Q_6(t)$ and  $\mathfrak{N}(t)$, $\mathfrak{S}(t)$ oscillates around a constant value along the whole trajectory. The fluctuations in terms of $\mathfrak{S}(t)$ are less pronounced than those observed in the case of $\mathfrak{N}(t)$, as $\mathfrak{S}(t)$, similarly to $Q_6(t)$, is a quantity averaged over all the particles within the system. The softness also successfully captures the onset of the crystallisation process within $\text{Traj}(T^{(2)}_7)$, albeit the change is less stark to what we observe for either $\mathfrak{N}(t)$ or $Q_6(t)$. This is reasonable because in liquid systems, highly mobile particles dominate, whereas in solids, the opposite is true. Note that the crystallisation is accompanied by a change in softness from positive to negative values - which is to be expected, given the system intuitively becomes ``less soft'' upon crystallisation.

Overall, these results demonstrate that the concept of hard or soft particles - or, equivalently, the concept of softness, offer robust metrics to characterise the evolution of the dynamical properties of the system, even across different states of matter. Crucially, the graph-based descriptors that we have used to craft the ML models utilised to classify a given particle as either hard or soft allow us to apply $\mathfrak{N}(t)$ and/or $\mathfrak{S}(t)$ to study a given system across a diverse portfolio of external conditions (temperature, pressure, etc.). This is a significant advantage compared to the usage of symmetry functions to build the same classifiers, as these descriptors are far more sensitives to volume changes or equivalently density fluctuations. As a result, the ML models leveraging our graph-based descriptors are consistently far more transferable than those utilising symmetry functions - as illustrated in Table~\ref{tab:b_1}.

\section{Conclusions}

In this work, we present a set of graph-based descriptors that can be used to accurately capture the particle local environment of condensed matter systems. We rigorously follow the tenets of graph theory to construct a robust framework that does not rely on Euclidean geometry - but on the topological features of the system (graph) instead. To demonstrate the applicability of our framework, we focus on the atomic environment in the prototypical Lennard-Jones system. Specifically, we compare the accuracy of our methodology with previous results obtained by means of symmetry functions - a class of descriptors widely used in the context of ML-based interatomic potentials. 

We have built a set of ML models aimed at classifying each particle within the system, within MD trajectories collected across a diverse portfolio of external conditions, as either ``hard'' or ``soft'', which in turns enables the characterisation of the local environment. We have found that our graph-based descriptors consistently outperform symmetry functions in terms of both accuracy and - crucially - transferability. This is because the topology of the system (which we leverage to build our graph-based descriptors) is far less sensitive to, e.g., volume fluctuations than metrics based on the Euclidean distance between particles pairs. In addition, we find that a much smaller number of graph-based descriptor is required to obtain an accuracy equivalent or often even superior to that obtain via symmetry functions. We remark that, whilst we have chosen to compare our results to symmetry functions specifically - as they have been used to study the same problem in the recent literature - it is reasonable to expect that the same arguments in terms of superior accuracy, transferability and low-dimensionality would hold when considering any other descriptor based on Euclidean geometry. 

We have also extended our method to the global properties and applied the ML models we have constructed in this work to study the time evolution of the ``softness'' of the entire system, i.e., the arithmetic mean of the ``softness'' of each particle across different temperatures, including some for which we observe crystal nucleation and growth. We find that this metric can reliably monitor the structural and dynamical changes of the system, thus offering an intriguing - and, possibly, superior - alternative to the metrics conventional used in the context of condensed matter physics (such as ten Wolde order parameters). 

Overall, this work showcases the potential of graph theory-based methods to make important contributions in condensed matter physics and materials science, providing new computational tools and solid theoretical foundations to shed new light onto the structural and dynamical properties of both ordered and disordered phases. We are hoping to expand on this framework by focusing further on local topological metrics, for the purposes of pinpointing the origin of the structural and dynamical changes that precede the onset of phase transitions, with emphasis on crystal nucleation and growth.

\begin{acknowledgments}
We gratefully acknowledge the use of the high-performance computing (HPC) facilities provided by the Scientific Computing Research Technology Platform (SCRTP) at the University of Warwick, particularly the Avon HPC cluster.

\end{acknowledgments}

\clearpage
\appendix
\section{On the choice of $t_{R/2}$ and $p_c$}
\label{app:p_hop}

In Section~\ref{subsec:some_metrics} and~\ref{subsec:particle_labelling} we mentioned the settings for the thresholds $t_{R/2}$ and $p_c$. Here, we intend to discuss in detail the principles for their settings.

\begin{figure}[htbp]
\includegraphics[width=0.5\textwidth]{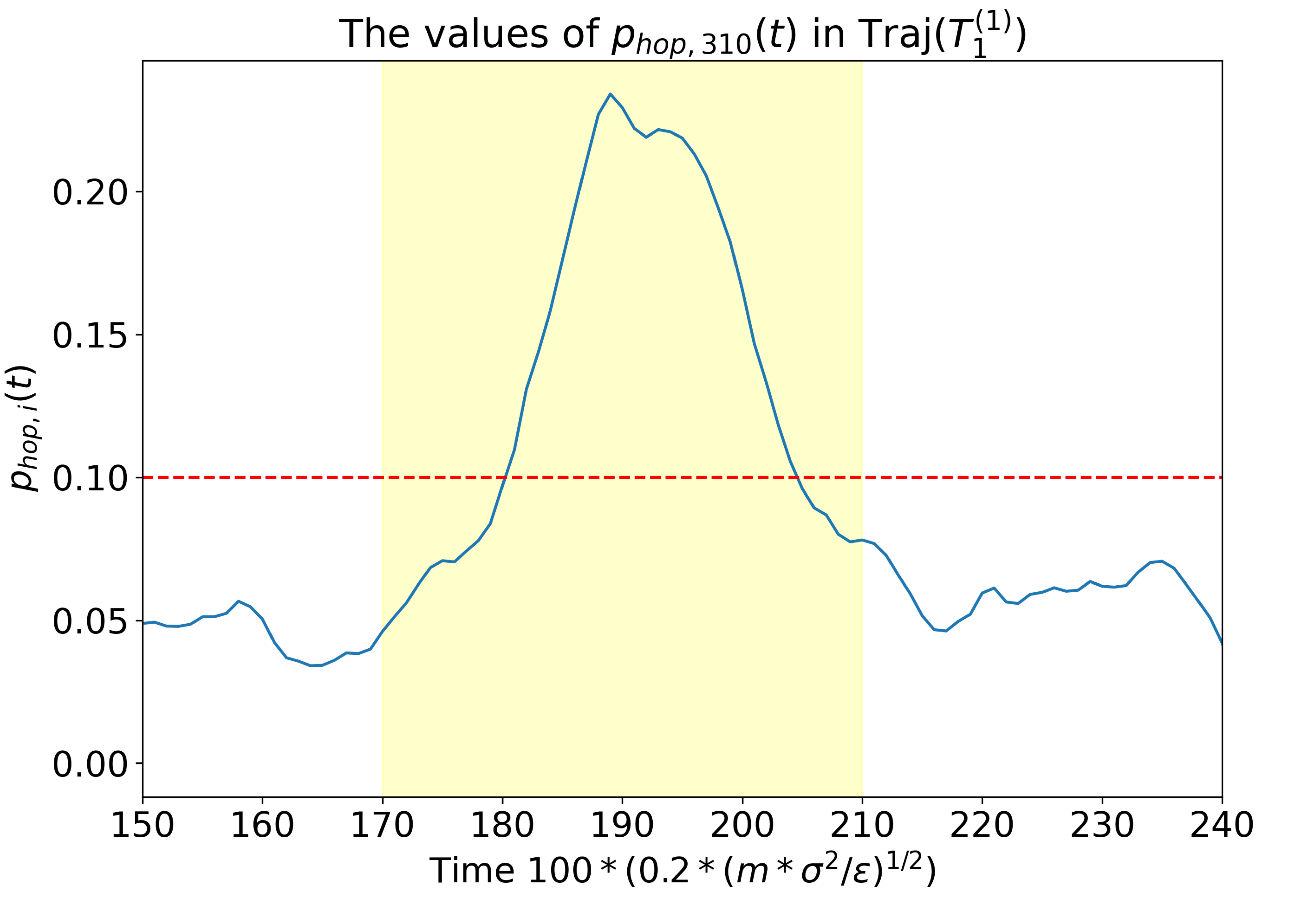}
\caption{The evolution of $p_{\text{hop}}$ for a selected particle (No.310) over a period of time under the lowest temperature (i.e., $T^{(1)}_1$). The yellow region represents the interval with the time length of $2t_{R/2}$, encompassing a complete rearrangement, which means $p_{hop}$ rises above $p_c$ and then falls back below $p_c$. The red-dotted line is the selected $p_c=0.1$.}
\label{fig:p_hop_t}
\end{figure}

\textbf{Underlying principle.} In the beginning, we should note that the difference between a complete rearrangement and the occurrence of a rearrangement. The former means that from the point when $p_{hop}$ rises above $p_c$ until it falls back below $p_c$, the entire process is considered complete. Conversely, as long as $p_{hop}$ exceeds $p_c$, we say that rearrangement has occurred, regardless of whether it falls back below $p_c$. The criterion for identifying the soft particles and hard particles is the latter. The purpose of the former is merely to enable the machine learning model to have better performance and to provide a complete characterization of the system dynamics of rearrangement. In the original literature~\cite{cubuk2015identifying,schoenholz2016structural}, the principle for setting $t_{R/2}$ is to ensure that the interval $[t-t_{R/2}, t+t_{R/2}]$ encompasses a complete rearrangement, which means covering the entire process from when the value of $p_{hop}$ exceeds $p_c$ until it falls back below $p_c$, as shown in Fig.~\ref{fig:p_hop_t}. The primary purpose of covering the complete rearrangement process is to comprehensively analyze the relationship between dynamics and changes in softness during rearrangement. However, if the goal is only to identify particles prone to rearrangement, covering the entire process is not necessary, which is our situation. The setting of $p_c$ is based on experience, observation and comparative experiments under various assignments. It is important that the setting of the threshold will not alter the underlying physical laws uncovered~\cite{cubuk2015identifying,schoenholz2016structural}. It is evident because the underlying physical laws are based on overall statistical behavior and the dynamical characteristics of systems, rather than the specific identification of individual events, and different choices of the threshold only affect the identification of specific events.

Therefore, we combined physical intuition and data-driven considerations to set the two thresholds $t_{R/2}$ and $p_c$ in an interpretable manner. Note that, considering the controlled variable method, the way we set the thresholds does not affect our comparison results of descriptors used for classification tasks under the same settings.

\begin{figure}[htbp]
\includegraphics[width=0.5\textwidth]{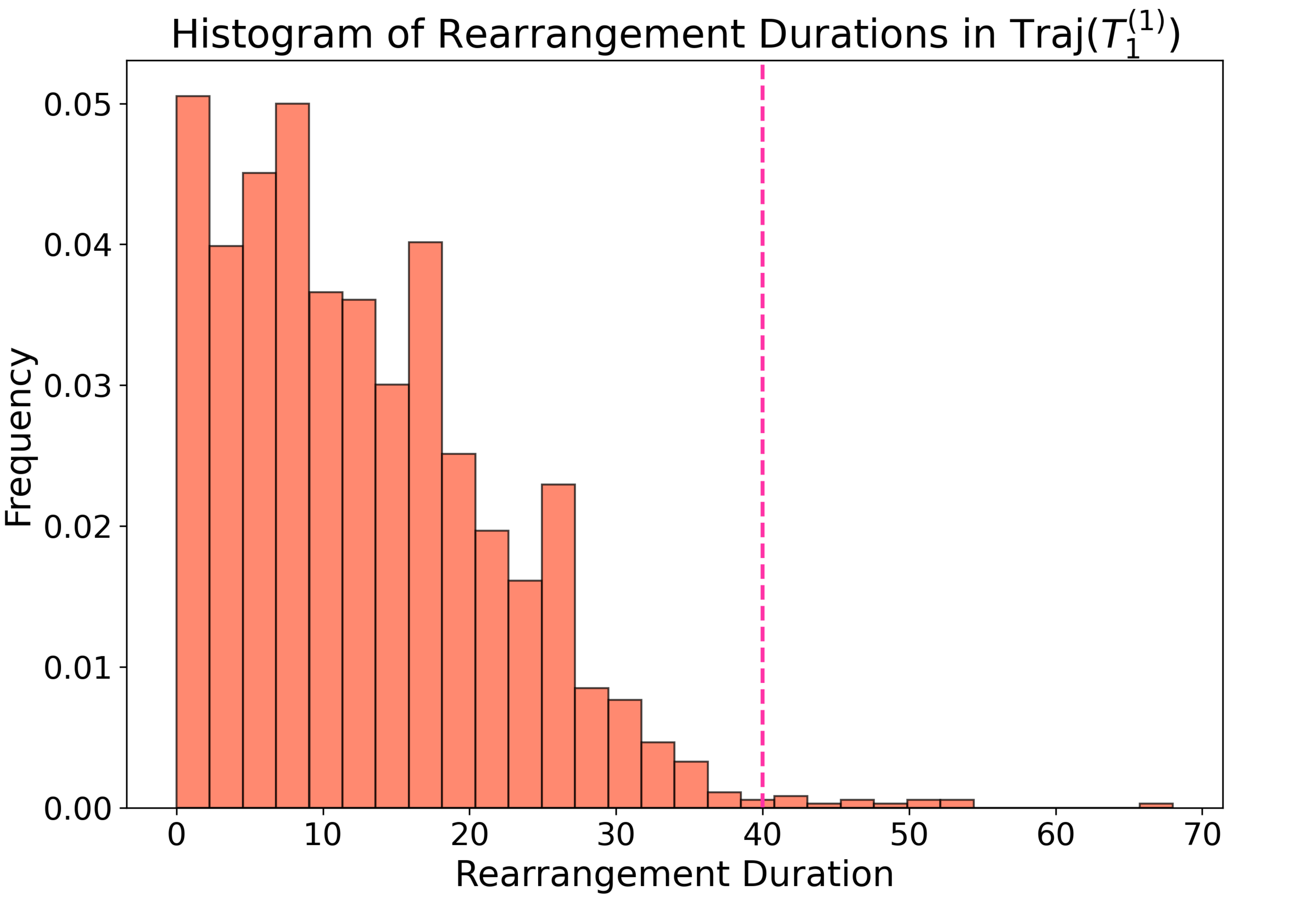}
\caption{The histogram of rearrangement duration in $\text{Traj}(T^{(1)}_1)$, corresponding to the lowest temperature and the system with the lowest mobility. The red-dotted line is the selected $t_{R}=40$, which equals to $2t_{R/2}$.}
\label{fig:durations}
\end{figure}

\vspace{0.1cm}
\textbf{The choice of $t_{R/2}$.} In the selection of $t_{R/2}$, we choose the lowest temperature $T^{(1)}_1$ as the reference temperature, because its corresponding system has the lowest mobility. We need to ensure that the interval $[t-t_{R/2}, t+t_{R/2}]$ covers as many complete rearrangements as possible in the situation of the lowest mobility, as shown in Fig.~\ref{fig:p_hop_t}. We can see that the yellow area covers a complete process where $p_{hop}$ rises above $p_c$ and then falls back below $p_c$, indicating a complete rearrangement. As depicted in Fig.~\ref{fig:durations}, at this reference temperature, the vast majority of rearrangements occur within the range of $t_{R}=40$, so we directly set $t_{R/2}=\frac{1}{2}t_{R}=20$.

\begin{figure}[htbp]
\includegraphics[width=0.5\textwidth]{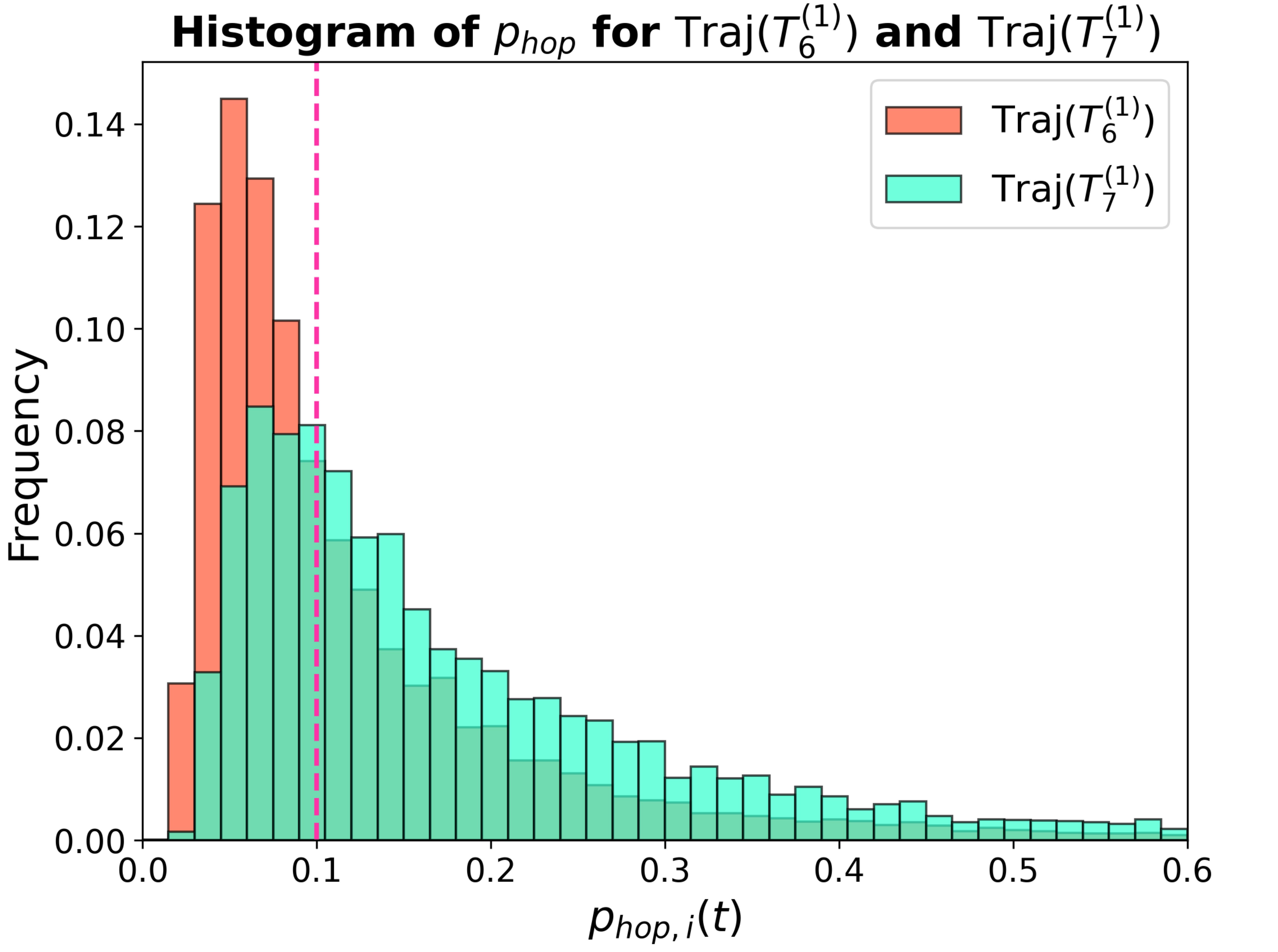}
\caption{The histogram of the value of $p_{hop}$ in $\text{Traj}(T^{(1)}_6)$ and $\text{Traj}(T^{(1)}_7)$. The onset temperature $T^{(1)}_c$ of crystallization separates $T^{(1)}_6$ and $T^{(1)}_7$, i.e., $T^{(1)}_6 < T^{(1)}_c < T^{(1)}_7$. The red-dotted line is the selected $p_c=0.1$. The reason for setting $p_c=0.1$ is when $p_{hop} < p_c$, the frequency of $p_{hop}$ in $\text{Traj}(T^{(1)}_6)$ is larger than that in $\text{Traj}(T^{(1)}_7)$; when $p_{hop} \geq p_c$, the frequency of $p_{hop}$ in $\text{Traj}(T^{(1)}_6)$ is smaller than that in $\text{Traj}(T^{(1)}_7)$. $p_c=0.1$ is the value at which the relative size of two frequency of $\text{Traj}(T^{(1)}_6)$ and $\text{Traj}(T^{(1)}_7)$ reverse.}
\label{fig:p_hop}
\end{figure}

\vspace{0.1cm}
\textbf{The choice of $p_c$.} As shown in Fig.~\ref{fig:p_hop}, the reason for choosing $p_c = 0.1$ is that it neatly separates the majority of particles in two trajectories located above and below the onset temperature of crystallization, i.e., $T^{(1)}_6$ and $T^{(1)}_7$, with their $p_{\text{hop},i}(t)$ values mostly being significantly greater or less than $p_c = 0.1$, respectively. That is to say, $T^{(1)}_6$ is the highest temperature at which crystallization occurs, and $T^{(1)}_7$ is the lowest temperature at which the liquid phase is maintained, which are both exactly what we are concerned with. From a physical intuition perspective, in the study of spontaneous crystallization of supercooled liquids, most particles in the liquid are naturally expected to be soft, while most particles in the solid are expected to be hard. Therefore, by plotting the histograms of $p_{\text{hop}}$ in $\text{Traj}(T^{(1)}_6)$ and $\text{Traj}(T^{(1)}_7)$, we can determine $p_c=0.1$ as the value at which the relative sizes of the two distribution histograms reverse.

\section{On the parameters of $G$ and $\Psi$}

\subsection{For $G$ Functions (Type A)}
\label{app:para_g}

\begin{itemize}
    \item Group 1:

    In this group, $\delta = 0.05$ (fixed).
    
    $r \in \{1+\delta (k-1) | k \in [1,n] \cap \mathbb{N} \}$.
    
    \item Group 2:

    In this group, $\delta = 0.1$ (fixed).
    
    $r \in \{1+\delta (k-1) | k \in [1,n] \cap \mathbb{N} \}$.
\end{itemize}

By setting $n = 60$ and $80$ respectively, we can obtain two descriptor vectors for each group with the number of predictors being 60 and 80, respectively, resulting in a total of four different combinations.

\subsection{For $\Psi$ Functions (Type B)}
\label{app:para_psi}

A selection of $20$ groups of parameters were made, with their corresponding $\Psi$ functions fully utilized to construct Type B input features:

(1) $\xi = 3.5$, $\zeta = 1$, $\lambda = -1$; (2) $\xi = 3.5$, $\zeta = 1$, $\lambda = 1$; (3) $\xi = 3.5$, $\zeta = 2$, $\lambda = -1$; (4) $\xi = 3.5$, $\zeta = 2$, $\lambda = 1$; (5) $\xi = 3.5$, $\zeta = 4$, $\lambda = 1$; (6) $\xi = 3.0$, $\zeta = 1$, $\lambda = 1$; (7) $\xi = 3.0$, $\zeta = 2$, $\lambda = 1$; (8) $\xi = 3.0$, $\zeta = 4$, $\lambda = 1$; (9) $\xi = 2.5$, $\zeta = 1$, $\lambda = 1$; (10) $\xi = 2.5$, $\zeta = 2$, $\lambda = 1$; (11) $\xi = 2.5$, $\zeta = 4$, $\lambda = 1$; (12) $\xi = 2.0$, $\zeta = 1$, $\lambda = 1$; (13) $\xi = 2.0$, $\zeta = 2$, $\lambda = 1$; (14) $\xi = 2.0$, $\zeta = 4$, $\lambda = 1$; (15) $\xi = 1.6$, $\zeta = 1$, $\lambda = 1$; (16) $\xi = 1.6$, $\zeta = 2$, $\lambda = 1$; (17) $\xi = 1.6$, $\zeta = 4$, $\lambda = 1$; (18) $\xi = 1.0$, $\zeta = 1$, $\lambda = 1$; (19) $\xi = 1.0$, $\zeta = 1$, $\lambda = 1$; (20) $\xi = 1.0$, $\zeta = 1$, $\lambda = 1$.

Our study focuses on a single-component homogenized system, thus we did not separate the set of particles into $I$ and $J$ subsets as with mixtures. We only factored in interactions within a distance of $r \leq r_c = 4$ during summations over the particle set.

\section{Additional results re: the crystallisation process}
\label{app:extra_res_apps}

\begin{figure}[htbp]
\includegraphics[width=0.5\textwidth]{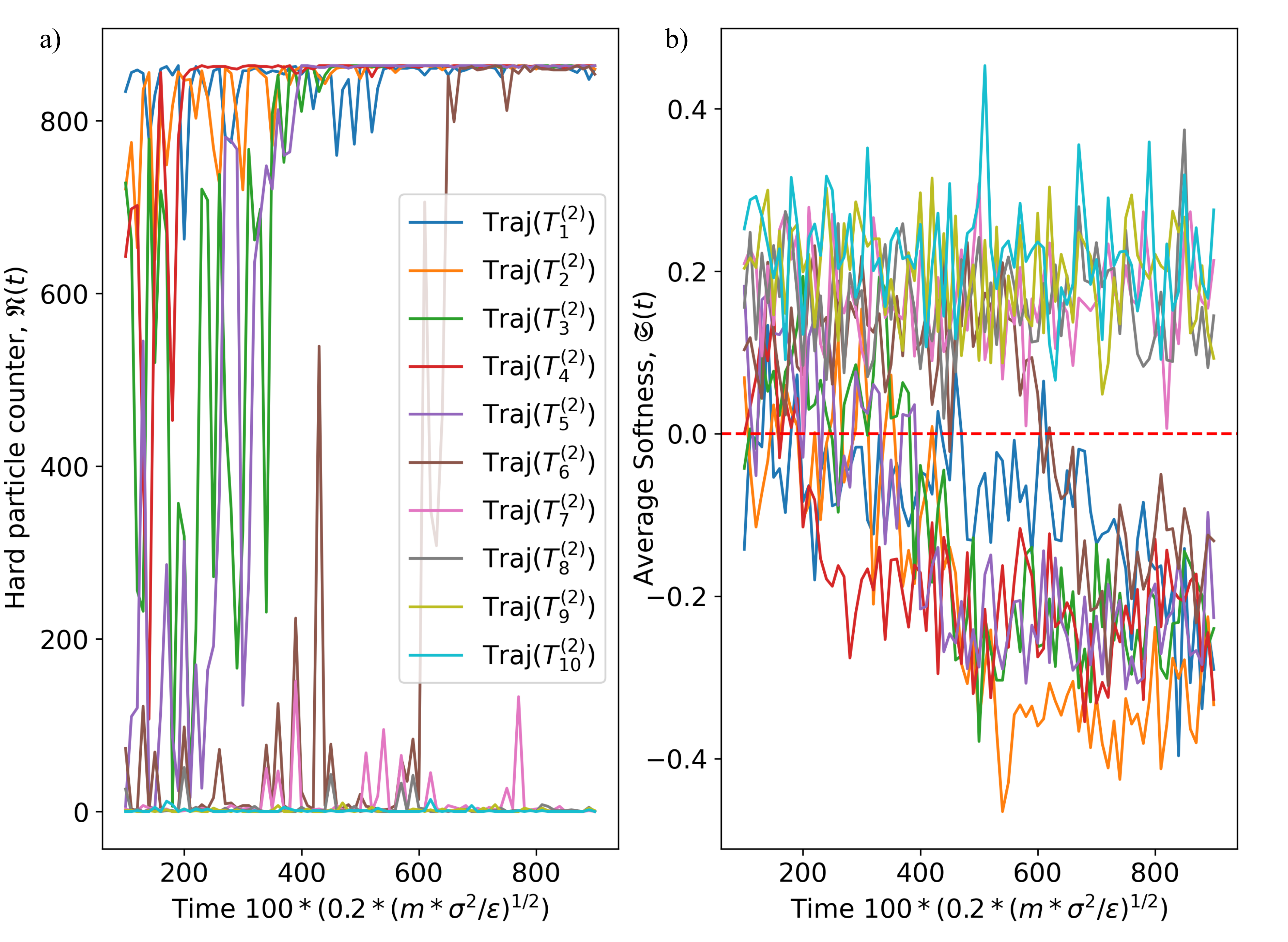}
\caption{a) The hard particle counter $\mathfrak{N}(t)$; b) The average softness $\mathfrak{S}(t)$ on all trajectories from Unit $2$, i.e., $\{\text{Traj} (T^{(2)}_p) | p\in[1,10] \cap \mathbb{N}\}$.}
\label{fig:count_soft}
\end{figure}

\begin{figure}[htbp]
\includegraphics[width=0.5\textwidth]{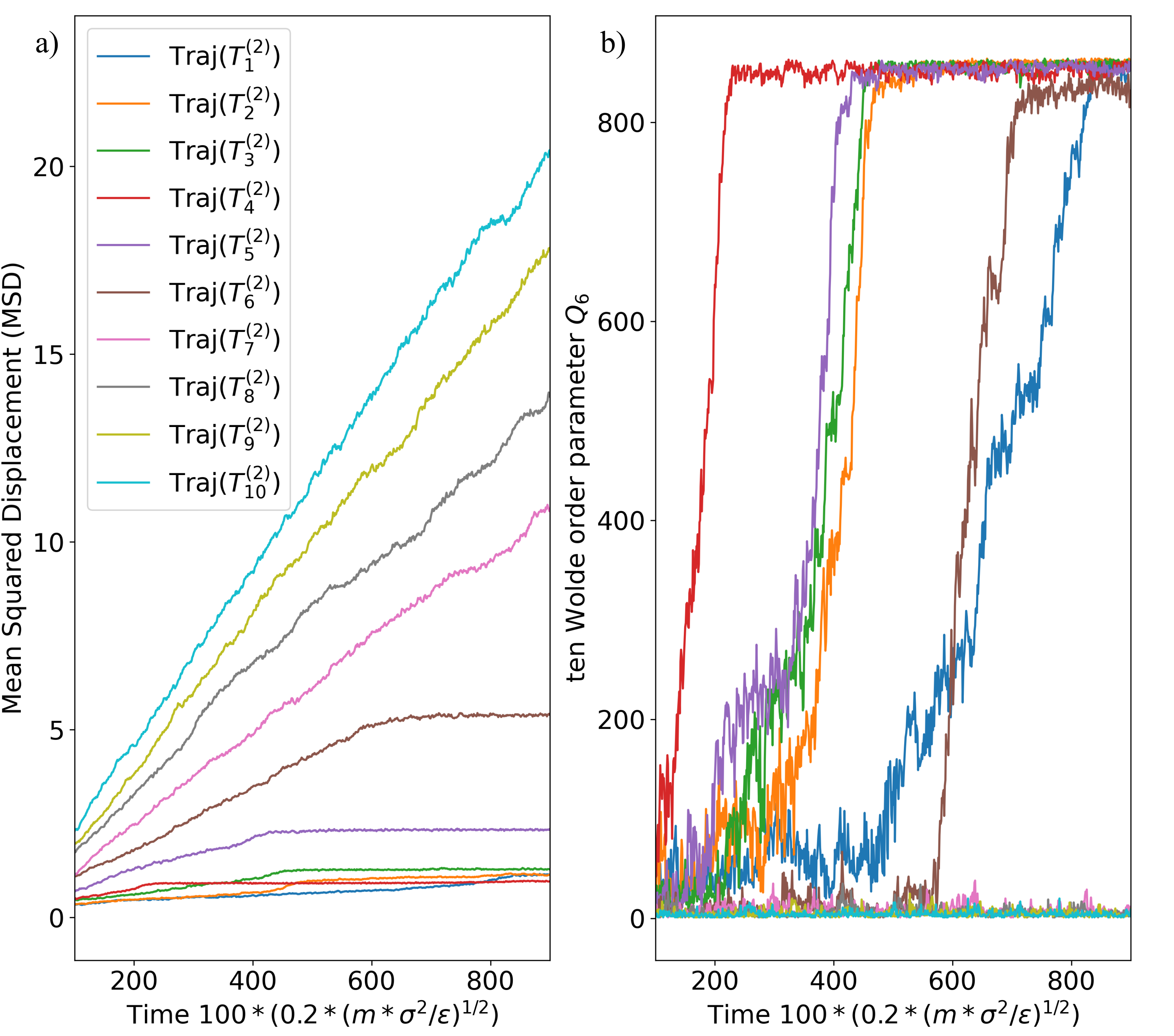}
\caption{The a) mean square displacement (MSD) and b) Global bond order parameter $Q_6$ of all trajectories from Unit $2$, i.e., $\{\text{Traj} (T^{(2)}_p) | p\in[1,10] \cap \mathbb{N}\}$.}
\label{fig:q6msd_set2}
\end{figure}

\begin{figure}[htbp]
\includegraphics[width=0.5\textwidth]{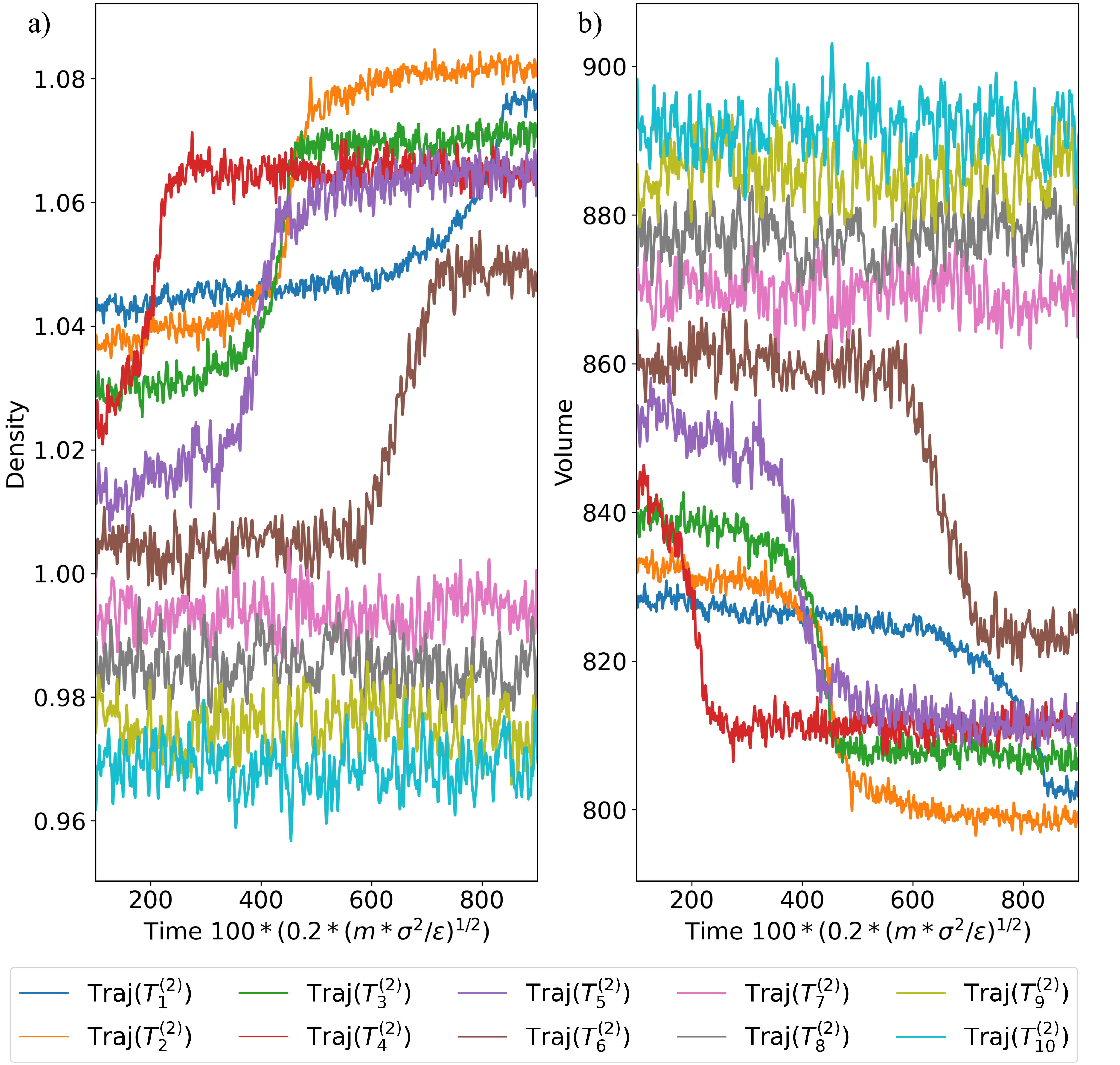}
\caption{The a) density and b) volumn corresponding to all trajectories from Unit $2$, i.e., $\{\text{Traj} (T^{(2)}_p) | p\in[1,10] \cap \mathbb{N}\}$. The Lennard-Jones reduced units are used here.}
\label{fig:thermo_set2}
\end{figure}

This section is the detailed calculation results for all trajectories in Unit 2, i.e., $\{\text{Traj} (T^{(2)}_p) | p\in[1,10] \cap \mathbb{N}\}$. Specifically, as follows:

\begin{enumerate}
    \item $\mathfrak{N}(t)$: Fig.~\ref{fig:count_soft}a);
    \item $\mathfrak{S}(t)$: Fig.~\ref{fig:count_soft}b);
    \item MSD: Fig.~\ref{fig:q6msd_set2}a);
    \item $Q_6(t)$: Fig.~\ref{fig:q6msd_set2}b);
    \item Density: Fig.~\ref{fig:thermo_set2}a);
    \item Volumn: Fig.~\ref{fig:thermo_set2}b).
\end{enumerate}

The results of our defined $\mathfrak{N}(t)$ and $\mathfrak{S}(t)$ across all trajectories, containing the various thermodynamic states and phases of matter are consistent with those obtained by traditional physical methods. Thus, we have bridged local and global properties by $\mathfrak{N}(t)$ and $\mathfrak{S}(t)$, as well as shifted the study of the system's overall temporal dynamics to focus on the evolution of $\mathfrak{N}(t)$ and $\mathfrak{S}(t)$.

\section{Additional Information about the Machine Learning process}

\subsection{The Detail of Ensemble Methods}

\label{subsec:ensemble_methods}

One of the hyperparameters of the ensemble learning is the ensemble methods. Each method uses different strategies to enhance model performance, considering the number of learners $M$, learning rate $\eta$ (if applicable), and feature selection $F$ (if applicable), providing diverse solutions for specific problems as follow:

\paragraph{AdaBoost} AdaBoost~\cite{freund1997decision} integrates multiple weak learners by iteratively adjusting sample weights, emphasizing samples that were misclassified by previous learners. In each round, weights of correctly classified samples decrease, while those of misclassified samples increase, focusing subsequent learners on difficult-to-classify samples. AdaBoost does not directly utilize a learning rate $\eta$ nor feature selection $F$, but adaptively adjusts weights based on the performance of learners in each round:
    \begin{equation*}
        \bold{y} = \sum_{m=1}^M \alpha_m T_m (\bold{X})
    \end{equation*}

    where $M$ is the number of learners, and $\alpha_m$ is the weight calculated based on the performance of the $m$-th learner.

\paragraph{RUSBoost} RUSBoost~\cite{seiffert2009rusboost} combines random under-sampling with AdaBoost, specifically designed for imbalanced datasets. Before each iteration, it achieves class balance by randomly under-sampling the majority class, then applies the AdaBoost. RUSBoost employs a similar learner ensemble strategy to AdaBoost but adds a preprocessing step to handle imbalanced data.

\paragraph{LogitBoost} LogitBoost~\cite{friedman2000additive} is designed for binary classification, iteratively minimizing the binomial logistic loss to train weak learners. In each step, LogitBoost fits the logistic function of residuals, with the optimization process not explicitly setting a learning rate $\eta$ but naturally regulating the learning pace through loss minimization:

    \begin{equation*}
        \bold{y} = \sum_{m=1}^M T_m (\bold{X})
    \end{equation*}

    where $M$ represents the number of learners.

\paragraph{GentleBoost} GentleBoost~\cite{friedman2001greedy} is a variant of AdaBoost, using a more gentle mechanism for updating sample weights, showing better robustness to noise and outliers. It integrates multiple weak learners like AdaBoost but employs a different strategy for gently adjusting weights of misclassified samples.

\paragraph{Bagging} Bagging~\cite{breiman1996bagging} trains multiple base learners by performing bootstrap sampling from the original dataset multiple times, then merges predictions by averaging (for regression problems) or majority voting (for classification problems). Bagging typically doesn't involve a learning rate $\eta$, with each learner contributing equally:

    \begin{equation*}
        \bold{y} = \frac{1}{M} \sum_{m=1}^M T_m (\bold{X})
    \end{equation*}

    where $M$ is the total number of learners, and feature selection $SF(\bold{X},F)$ is not a direct component of Bagging, although similar methods like Random Forests randomly select features at each split.

\subsection{The Activation Functions of Neural Network}
\label{subsec:activation_function}

\paragraph{None (Linear or Identity)}
\begin{equation*}
    \text{Identity}(x)=x
\end{equation*}

\paragraph{Sigmoid}~\cite{rumelhart1986learning}
\begin{equation*}
    \sigma(x) = \frac{1}{1+\text{e}^{-x}}
\end{equation*}

\paragraph{Tanh (Hyperbolic Tangent)}
\begin{equation*}
    \mathop{\tanh}(x) = \frac{\text{e}^{x}-\text{e}^{-x}}{\text{e}^{x}+\text{e}^{-x}}
\end{equation*}

\paragraph{ReLU (Rectified Linear Unit)}~\cite{hahnloser2000digital,nair2010rectified}

\begin{equation*}
    \text{ReLU}(x) = \mathop{\max}(0,x)
\end{equation*}

\paragraph{Softmax} For a vector of raw class scores from the final layer $\bold{z} = (z_1, \cdots, z_s)$, the Softmax function~
\cite{bridle1990probabilistic} for the $i$-th score is defined as:

\begin{equation*}
    \text{Softmax}(z_i) = \frac{\text{e}^{z_i}}{\sum_{j=1}^s \text{e}^{z_j}}
\end{equation*}

\subsection{About the Bayesian Optimization for Hyperparameters}
\label{subsec:bayes}

We have agreed that the notation convention for this section is only for this section. The Bayesian Optimization for Hyperparameters contains the following steps:

\begin{enumerate}
    \item \textbf{Define Problem Space}:
    
    Define the value space for each hyperparameter within an observation, denoted as $p_{ij}$, where $i$ represents the index of the observation and $j$ represents the index of the hyperparameter.
    \item \textbf{Acquisition Function}:

    We choose Expected Improvement per second plus (EIps) as the acquisition function, i.e, $\text{EIps}(\mathbf{p}) = \frac{\text{EI}(\bold{p})}{\text{cost}(\bold{p})}$. Here, $\text{EI}(\bold{p})$ signifies the expected improvement, and $\text{cost}(\bold{p})$ denotes the evaluation cost.
    
    \item \textbf{Initialize Observations}:

    Establish the initial dataset $D = \{(\bold{p}_i, q_i)\}_{i=1}^N$, where $\bold{p}_i$ represents a vector of specific hyperparameter values for the $i$-th observation or evaluation, and $q_i$ is the corresponding performance metric, measured through k-fold cross-validation.

    \item \textbf{Iterative Optimization Process}:
    \begin{itemize}
        \item \textbf{Update Gaussian Process}:

         Use the current dataset $D$ to refine the Gaussian Process model. The objective function $\mathfrak{F}$, representing the performance metric as a function of hyperparameters, is modeled through a Gaussian Process: $\mathfrak{F}(\bold{p}) \sim GP(\mu, k)$, capturing our beliefs about the relationship between hyperparameters and performance metrics.

         \item \textbf{Select Hyperparameter Combination}:

         Determine the next evaluation point $\bold{p}^{*}$ by maximizing the EIps function. The Expected Improvement, $\text{EI}(\bold{p})$, is based on the improvement over the current best performance $q_{\text{best}}$ and is calculated as:
         \begin{equation*}
             \text{EI}(\bold{p}) = (\mu(\bold{p})-q_{\text{best}}-\xi) \Phi(Z) + \sigma (\bold{p}) \phi (Z)
         \end{equation*}
         with $Z = \frac{\mu(\bold{p})-q_{\text{best}}-\xi}{\sigma (\bold{p})}$, and $\xi$ as a slight non-negative exploration parameter.

         \item \textbf{Evaluate by k-fold Cross-validation}:

         To obtain the performance metric $q^{*}$ for $\bold{p}^{*}$, perform k-fold cross-validation, which divides the data into $k$ subsets, training the model on $(k-1)$ of these subsets and validating it on the remaining subset. This process is repeated $k$ times (folds), each time with a different subset serving as the validation data. The average performance across these $k$ folds serves as the performance metric $q^{*}$.

         \item \textbf{Update Dataset $D$}:
         
         Incorporate the new observation $(\bold{p}^{*}, q^{*})$ into the dataset $D$, enhancing $D$ with this latest observation, i.e., $D \leftarrow D \cup \{\bold{p}^{*}, q^{*}\}$.
    \end{itemize}

    \item \textbf{Repeat Step 4}:

    Until reaching a termination condition, such as a specified number of iterations or an acceptable performance metric threshold.

    \item \textbf{Select the Best Hyperparameter}:

    Choose the hyperparameter combination from the dataset $D$ that shows the highest performance metric observed, ensuring that the selection is based on robust evaluation through k-fold cross-validation.
    
\end{enumerate}

\section{On the Vector Graph Order Parameter (VGOP) method}
\label{app:alternative}

As mentioned in section~\ref{sec:graph}, the Vector Graph Order Parameter VGOP~\cite{chapman2022efficient} is not dissimilar to a discrete form of the radial distribution function. We provide a detailed explanation about this similarity, as well as the results obtained by applying the VGOP on four controlled experiments, i.e., Exp A, Exp B, Exp C and Exp D. Our method consistently outperforms the VGOP approach in terms of accuracy. For convenience, we refer to the VGOP as Type E descriptor.

Performance aside, we remark that there exists a fundamental difference between the VGOP and our graph-based method: namely, the former still relies on a metric structure to extract geometric information from the system, whilst our approach is purely topological in nature and thus entirely independent from any metric structure. 

\subsection{Calculation of the VGOP}

\label{app_sub_vgop}

The VGOP~\cite{chapman2022efficient} characterizes atomic structures, focusing on all particles within a configuration of a given system taken from, e.g., molecular dynamics simulations. Thus, the aim of the VGOP is to capture the system's structural configuration at a specific moment in time, in terms of the spatial and topological relations among the constituents of the system (in this case, atoms).

In the context of the VGOP method, calculating the feature matrix from particle coordinates in a frame involves four key steps: (i.) building the Graph Coordination Networks (GCNs); (ii.) computing the Scalar Graph Order Parameter (SGOP); (iii.) determining the Vector Graph Order Parameter (VGOP), and  (iv.) compiling the feature matrix. In particular:

\begin{enumerate}
    \item \textbf{Construct the GCN}
    
    For each pair of particles $i$ and $j$ in a $\mathcal{N}$-body system, construct an adjacency matrix $G$, where the element $G_{ij}$ represents the connection between atoms $i$ and $j$, which are based on a distance threshold $r_c$, meaning connections exist only if the distance $d_{ij}$ between atoms is less than $r_c$:
    \begin{equation*}
        G_{ij} = 
            \begin{cases} 
            1 & \text{ if } d_{ij} \leq r_c \\
            0 & \text{ otherwise}.
        \end{cases}
    \end{equation*}

    \item \textbf{Apply the SGOP of An Atom}
    \begin{enumerate}
        \item \textbf{Calculate degree of each node}

        The degree $k_i$ of node $i$ is the number of nodes connected to node $i$, which can be obtained by summing the rows or columns of the adjacency matrix $G$.
        \item \textbf{Compute SGOP}

        For each subgraph $S$ (in this case, the entire network is considered a large subgraph), SGOP of particle $i$ is defined as:
        \begin{equation*}
            \theta_{\text{SGOP}}^{(i)} = \sum_{s\in S} \left( \sum_{k \in deg(S)} P(k) \mathop{\log} P(k) + k \cdot P(k) \right)^3
        \end{equation*}
    \end{enumerate}
    
    where $P(k)$ is the probability of degree $k$ occurring in subgraph $S$, and $k \in deg(S)$ refers to the summation for any node within $S$ that has a degree of $k$.
    
    \item \textbf{Build the VGOP of An Atom}
    \begin{enumerate}
        \item \textbf{Choose multiple $r_c$ values}

        Select a set of different cutoff radius $r_{c,1}$, $r_{c,2}$, $\cdots$, $r_{c,m}$ to capture the local structures at different scales.
        \item \textbf{Repeat Steps 1 and 2}

        For each cutoff radius $r_{c,i}$, repeat Steps 1 and 2 to build GCNs and compute SGOPs, obtaining a series of SGOP values.
        \item \textbf{Construct VGOP}

        Combine the obtained SGOP values into a vector, which is the VGOP of particle $i$:
        \begin{equation*}
            \overrightarrow{\theta}^{(i)} = [\theta_{\text{SGOP}}^{(i)}(r_{c,1}),\theta_{\text{SGOP}}^{(i)}(r_{c,2}),\cdots,\theta_{\text{SGOP}}^{(i)}(r_{c,m})]
        \end{equation*}
    \end{enumerate}
    \item \textbf{Get the Feature Matrix of A Frame}

    Finally, combine the calculated $\overrightarrow{\theta}^{(i)}$ of each particle $i \in [1,\mathcal{N}] \cap \mathbb{N}$ as row vectors to form the feature matrix $\bold{X}$, where each row of the matrix represents the feature matrix of these samples:
    \begin{equation*}
        \bold{X} = ( \overrightarrow{\theta}^{(1)}, \overrightarrow{\theta}^{(2)}, \cdots, \overrightarrow{\theta}^{(\mathcal{N})} )^T
    \end{equation*}
\end{enumerate}

\subsection{The relationship between the VGOP and the RDF}
\label{app:sub_vgop_vs_rdf}

Both the radial distribution function $g(r)$ (or, equivalently, RDF) and the VGOP capture information about the system density. However, whilst the $g(r)$ considers the density distribution of the particles in physical space, the VGOP looks at how the degree distribution of network behaves across various cutoff radius scales, which can be thought as a degree density function (as explained below).

The RDF represents the probability density of finding another particle at a distance $r$ from a reference particle and is defined as:
\[
g(r) = \frac{1}{\rho N} \left\langle \sum_{i \neq j} \delta(r - r_{ij}) \right\rangle
\]
where $\rho$ is the average density of the system,  $N$ is the total number of particles, $r_{ij}$ is the distance between particles $i$ and $j$, and $\delta$ is the Dirac delta function.

In practice, the RDF is often approximated by discretising the continuous distance \( r \) into a series of small intervals $\Delta r$:
\begin{equation*}
    g(r) \approx \frac{1}{\rho N \Delta r} \sum_{i \neq j} \left[ \Theta(r + \Delta r - r_{ij}) - \Theta(r - r_{ij}) \right],
\end{equation*}
\noindent where $\Theta$ is the Heaviside step function. These intervals $\Delta r$ do not overlap, and each segment is independent.

In graph structures, the degree of a node $i$ within a cutoff radius $r$ is defined as:

\begin{equation*}
    k_i(r) = \sum_{j \neq i} \Theta(r - r_{ij}),
\end{equation*}

\noindent which represents the number of nodes connected to node $i$ within the radius $r$.

We define the degree density function between two adjacent cutoff radii \( r \) and \( r + \Delta r \):

\begin{align*}
\Delta k_i(r, \Delta r) &= \sum_{j \neq i} \mathbbm{1}_{[r \leq r_{ij} < r + \Delta r]}(r_{ij}) \\
&= \sum_{j \neq i} \left[ \Theta(r + \Delta r - r_{ij}) - \Theta(r - r_{ij}) \right]
\end{align*}

This expression represents the number of nodes within the distance range $r$ to $r + \Delta r$ from node $i$. We then convert the degree density to local degree density:

\begin{equation*}
    \Delta \rho_i(r, \Delta r) = \frac{\Delta k_i(r, \Delta r)}{V(r, \Delta r)},
\end{equation*}

\noindent where \( V(r, \Delta r) \approx 4 \pi r^2 \Delta r \) is the volume between \( r \) and \( r + \Delta r \). Thus, we can get:

\begin{equation*}
    \label{equ:dens_local}
    \Delta \rho_i(r, \Delta r) \approx \frac{\Delta k_i(r, \Delta r)}{4 \pi r^2 \Delta r}.
\end{equation*}

Averaging over all nodes, we get the average degree density function:

\begin{align*}
\Delta \rho(r, \Delta r) &= \frac{1}{N} \sum_{i=1}^{N} \Delta \rho_i(r, \Delta r) \\
&\approx \frac{1}{4 \pi r^2 \Delta r} \left( \frac{1}{N} \sum_{i=1}^{N} \Delta k_i(r, \Delta r) \right)
\end{align*}

In VGOP, the degree $k_i(r)$ within the cutoff radius $r$ is cumulative. As $r$ increases, the spatial range under consideration also increases, meaning that a larger $r$ includes all information from smaller $r$. Specifically, different cutoff radius $r$ overlap, as larger $r$ encompasses all nodes from $0$ to $r$. This differs from the discretisation of the RDF, where continuous $r$ is segmented into non-overlapping intervals, with each segment independently calculating the probability density, but the two procedures are essentially equivalent.

In fact, the relationship between the degree density function and the RDF can be written as:

\begin{equation*}
    \Delta \rho(r, \Delta r) \approx \frac{1}{4 \pi r^2 \rho} g(r).
\end{equation*}

These considerations demonstrate that the local degree density variation in VGOP at different cutoff radius $r$ is mathematically equivalent to the particle density variation in RDF at different distances $r$. Despite the fact that the VGOP utilises cumulative cutoff radii (encompassing all the information relative to smaller radii) whilst the RDF discretises continuous $r$ into non-overlapping segments, the VGOP can thus be considered as a discrete version of the RDF based on node degree distribution.

\subsection{Experiments and Results}
\label{app:sub_res_vgop}

We conducted comparative tests of Type C (centralities and clustering coefficient) and Type E (VGOP) descriptors across Exp A, B, C, and D.

For our Type C descriptors, we selected the Q1 with the worst performance in most cases, using only $10$ predictors, as a reference. For the VGOP descroptors, we have two setups, i.e., VGOP1 and VGOP2. Specifically, VGOP1 is designed to benchmark against the reference Q1 so it uses $10$ evenly spaced values in the interval of $[1,6)$ as $r_c$, resulting in $10$ predictors. VGOP2, on the other hand, aims for a finer characterization of the interval, using $50$ evenly spaced values in the interval of $[1,6)$ as $r_c$, resulting in a $50$-dimensional feature. Similar to Section~\ref{subsec:feature_eng}, we also determine the maximum value of $r_c$ based on the common knowledge in solid-state physics that the effective range of the radial distribution function for a system of size $L$ lies within $L/3$ to $L/2$. Here, $L$ fluctuates around $15$ and $\max{r_c} \approx 6 \in [5,7.5]$. We have listed the performance of the VGOP descriptors on Exp A, B, C, and D in Fig.~\ref{fig:e_all} of Appendix~\ref{app:results}.

As expected, VGOP2 performs significantly better than VGOP1, which, in most cases, gives results just marginally more accurate than those one would obtained by random selection. Therefore, we compared the better-performing VGOP2 with the reference Q1, which performed the worst in Type C, as shown in Table~\ref{tab:vgop_vs_type_c}. It is clear that Q1 is significantly superior to VGOP2 in all aspects.

In section~\ref{subsec:des_graph}.1, we specifically emphasized that the advantages of Type C descriptors is that the heterogeneous features are maximally intersected to cover the properties of the system as much as possible. Besides, the Type C descriptor has very high interpretability and profound physical insights. If $\mathcal{A}$ is treated as a hyperparameter, then optimizing $\mathcal{A}$ becomes a process of identifying the most important interaction patterns.

On the contrary, whilst the VGOP leverages some concepts of graph theory, it still relies on defining cutoff radii to construct the relevant graphs. As such - in our opinion - the VGOP suffer to an extent from a lack in terms of interpretability and might fail to fully capture the subtleties of interactions determined not just by the distance between the particles but also by factors such as the relative orientation, the thermodynamic conditions, and the particles types. In addition, as mentioned in Appendix~\ref{app:sub_vgop_vs_rdf}, the VGOP is essentially a discrete version of the radial distribution function (except that in the case of the VGOP the degree density is used to estimate particle density). As demonstrated by the results reported in thus sections, such homogeneous features have limited effects in enhancing the performance of machine learning models in the context of characterising the structural and dynamical properties of condensed matter system. In contrast, our graph-based method extensively utilizes graph theory while integrating the underlying physical principles into feature engineering - an approach that we have proved to be rather efficient in predicting with great accuracy the properties of the system, even across different thermodynamic conditions and/or in the presence of phase transitions.

\begin{table}[htbp]
\footnotesize
\renewcommand{\arraystretch}{1.5}
\centering
\caption{Type C (Centralities) VS Type E (VGOP)}
\label{tab:vgop_vs_type_c}
\begin{tabular}{c|c|c|c|c|c|c}
\hline
\multicolumn{7}{c}{Linear SVM} \\
\hline
Type & No. & Exp & Vars & Accuracy & MCC & Confusion Matrix \\ \hline
\rowcolor{yellow!20}
E & VGOP2 & A & $50$ & $67.3\%$ & $0.349$ & $\left(\begin{smallmatrix} 4574 & 2926 \\ 1978 & 5522 \end{smallmatrix}\right)$ \\
\rowcolor{gray!20}
C & Q1 & A & $10$ & $75.9\%$ & $0.521$ & $\left(\begin{smallmatrix} 5330 & 2170 \\ 1444 & 6056 \end{smallmatrix}\right)$ \\
\hline
\rowcolor{yellow!20}
E & VGOP2 & B & $50$ & $67.6\%$ & $0.355$ & $\left(\begin{smallmatrix} 9118 & 5882 \\ 3839 & 11161 \end{smallmatrix}\right)$ \\
\rowcolor{gray!20}
C & Q1 & B & $10$ & $76.5\%$ & $0.530$ & $\left(\begin{smallmatrix} 11280 & 3720 \\ 3329 & 11671 \end{smallmatrix}\right)$ \\
\hline
\rowcolor{yellow!20}
E & VGOP2 & C & $50$ & $63.9\%$ & $0.279$ & $\left(\begin{smallmatrix} 12547 & 2453 \\ 2770 & 12230 \end{smallmatrix}\right)$ \\
\rowcolor{gray!20}
C & Q1 & C & $10$ & $76.5\%$ & $0.530$ & $\left(\begin{smallmatrix} 22809 & 7191 \\ 6898 & 23102 \end{smallmatrix}\right)$ \\
\hline
\rowcolor{yellow!20}
E & VGOP2 & D & $50$ & $55.5\%$ & $0.114$ & $\left(\begin{smallmatrix} 17887 & 12113 \\ 12370 & 17630 \end{smallmatrix}\right)$ \\
\rowcolor{gray!20}
C & Q1 & D & $10$ & $73.1\%$ & $0.463$ & $\left(\begin{smallmatrix} 23337 & 6663 \\ 9502 & 20498 \end{smallmatrix}\right)$ \\
\hline
\multicolumn{7}{c}{All models with the best performance} \\
\hline
Type & No. & Exp & Model & Accuracy & MCC & Confusion Matrix \\ \hline
\rowcolor{yellow!20}
E & VGOP2 & A & $4$ & $80.0\%$ & $0.600$ & $\left(\begin{smallmatrix} 6103 & 1397 \\ 1608 & 5892 \end{smallmatrix}\right)$ \\
\rowcolor{gray!20}
C & Q1 & A & $4$ & $89.0\%$ & $0.780$ & $\left(\begin{smallmatrix} 6524 & 976 \\ 681 & 6819 \end{smallmatrix}\right)$ \\
\hline
\rowcolor{yellow!20}
E & VGOP2 & B & $4$ & $80.7\%$ & $0.615$ & $\left(\begin{smallmatrix} 11684 & 3316 \\ 2475 & 12525 \end{smallmatrix}\right)$ \\
\rowcolor{gray!20}
C & Q1 & B & $3$ & $85.7\%$ & $0.715$ & $\left(\begin{smallmatrix} 13407 & 1593 \\ 2708 & 12292 \end{smallmatrix}\right)$ \\
\hline
\rowcolor{yellow!20}
E & VGOP2 & C & $4$ & $75.2\%$ & $0.511$ & $\left(\begin{smallmatrix} 25124 & 4876 \\ 10021 & 19979 \end{smallmatrix}\right)$ \\
\rowcolor{gray!20}
C & Q1 & C & $3$ & $86.1\%$ & $0.723$ & $\left(\begin{smallmatrix} 25243 & 4757 \\ 3556 & 26444 \end{smallmatrix}\right)$ \\
\hline
\rowcolor{yellow!20}
E & VGOP2 & D & $3$ & $65.5\%$ & $0.311$ & $\left(\begin{smallmatrix} 18743 & 11257 \\ 9420 & 20580 \end{smallmatrix}\right)$ \\
\rowcolor{gray!20}
C & Q1 & D & $3$ & $80.7\%$ & $0.616$ & $\left(\begin{smallmatrix} 25606 & 4394 \\ 7203 & 22797 \end{smallmatrix}\right)$ \\
\hline
\end{tabular}
\end{table}

\newpage

\section{The Detail Results of Controlled Experiments}
\label{app:results}

The detail results of four controlled experiments described in section~\ref{subsec:con_exp} are in Table~\ref{fig:a_1} to~\ref{fig:d_2}.

\begin{table*}[htbp]
\centering
\renewcommand{\arraystretch}{1.2}
\setlength{\tabcolsep}{4pt}
\begin{adjustbox}{width=1\textwidth}
\begin{tabular}{|c|c|c|c|c|c|c|c|c|c|}
\hline
\textbf{No.} & \textbf{Structural functions} & \textbf{Type} & \textbf{Group} & \textbf{Number of features} & \textbf{Evaluation} & \textbf{SVM (Linear)} & \textbf{SVM (RBF kernel)} & \textbf{Ensemble Learning} & \textbf{Neural Network} \\
\hline
\multirow{3}{*}{P1} & \multirow{3}{*}{Only G} & \multirow{3}{*}{A} & \multirow{3}{*}{1} & \multirow{3}{*}{60} & Accuracy & 69.1\% & 72.6\% & 69.9\% & 82.6\% \\
 &  &  &  &  & AUC & 0.7559 & 0.8005 & 0.7768 & 0.9032 \\
 &  &  &  &  & MCC & 0.382 & 0.452 & 0.399 & 0.656 \\
\hline
\multirow{3}{*}{P2} & \multirow{3}{*}{Only G} & \multirow{3}{*}{A} & \multirow{3}{*}{2} & \multirow{3}{*}{60} & Accuracy & 73.5\% & 80.5\% & 78.2\% & 85.0\% \\
 &  &  &  &  & AUC & 0.8115 & 0.8822 & 0.864 & 0.9224 \\
 &  &  &  &  & MCC & 0.470 & 0.611 & 0.565 & 0.700 \\
\hline
\multirow{3}{*}{P3} & \multirow{3}{*}{Only G} & \multirow{3}{*}{A} & \multirow{3}{*}{1} & \multirow{3}{*}{80} & Accuracy & 71.9\% & 74.9\% & 71.6\% & 84.3\% \\
 &  &  &  &  & AUC & 0.7899 & 0.8188 & 0.7901 & 0.9131 \\
 &  &  &  &  & MCC & 0.438 & 0.498 & 0.433 & 0.686 \\
\hline
\multirow{3}{*}{P4} & \multirow{3}{*}{Only G} & \multirow{3}{*}{A} & \multirow{3}{*}{2} & \multirow{3}{*}{80} & Accuracy & \cellcolor{yellow!20} 76.3\% & 81.7\% & 78.3\% & \cellcolor{yellow!50} 85.3\% \\
 &  &  &  &  & AUC & \cellcolor{yellow!20} 0.8327 & 0.891 & 0.8671 & \cellcolor{yellow!50} 0.9254 \\
 &  &  &  &  & MCC & \cellcolor{yellow!20} 0.527 & 0.634 & 0.566 & \cellcolor{yellow!50} 0.706 \\
\hline
\multirow{3}{*}{P5} & \multirow{3}{*}{Only Psi} & \multirow{3}{*}{B} & \multirow{3}{*}{1 and 2} & \multirow{3}{*}{20} & Accuracy & 68.9\% & 63.7\% & 62.8\% & 72.2\% \\
 &  &  &  &  & AUC & 0.7466 & 0.6858 & 0.6768 & 0.789 \\
 &  &  &  &  & MCC & 0.379 & 0.277 & 0.258 & 0.446 \\
\hline
\multirow{3}{*}{P6} & \multirow{3}{*}{G and Psi} & \multirow{3}{*}{A+B} & \multirow{3}{*}{1} & \multirow{3}{*}{60+20} & Accuracy & 82.0\% & 80.8\% & 76.1\% & 85.4\% \\
 &  &  &  &  & AUC & 0.8884 & 0.8833 & 0.8396 & 0.9192 \\
 &  &  &  &  & MCC & 0.640 & 0.617 & 0.522 & 0.709 \\
\hline
\multirow{3}{*}{P7} & \multirow{3}{*}{G and Psi} & \multirow{3}{*}{A+B} & \multirow{3}{*}{2} & \multirow{3}{*}{60+20} & Accuracy & 85.1\% & 85.6\% & 82.5\% & 88.1\% \\
 &  &  &  &  & AUC & 0.9132 & 0.9128 & 0.9002 & 0.9433 \\
 &  &  &  &  & MCC & 0.701 & 0.712 & 0.652 & 0.770 \\
\hline
\multirow{3}{*}{P8} & \multirow{3}{*}{G and Psi} & \multirow{3}{*}{A+B} & \multirow{3}{*}{1} & \multirow{3}{*}{80+20} & Accuracy & 82.6\% & 81.4\% & 76.3\% & 85.0\% \\
 &  &  &  &  & AUC & 0.896 & 0.8884 & 0.8444 & 0.9181 \\
 &  &  &  &  & MCC & 0.652 & 0.628 & 0.526 & 0.701 \\
\hline
\multirow{3}{*}{P9} & \multirow{3}{*}{G and Psi} & \multirow{3}{*}{A+B} & \multirow{3}{*}{2} & \multirow{3}{*}{80+20} & Accuracy & \cellcolor{yellow!20} 86.4\% & 86.6\% & 83.7\% & \cellcolor{yellow!50} 88.5\% \\
 &  &  &  &  & AUC & \cellcolor{yellow!20} 0.9197 & 0.9277 & 0.9117 & \cellcolor{yellow!50} 0.9449 \\
 &  &  &  &  & MCC & \cellcolor{yellow!20} 0.728 & 0.732 & 0.675 & \cellcolor{yellow!50} 0.770 \\
\hline
\end{tabular}
\end{adjustbox}
\caption{Exp A, Traditional Descriptor (Type A and B)}
\label{fig:a_1}
\end{table*}

\begin{table*}[htbp]
\centering
\renewcommand{\arraystretch}{1.2}
\setlength{\tabcolsep}{4pt}
\begin{adjustbox}{width=1\textwidth}
\begin{tabular}{|c|c|c|c|c|c|c|c|c|c|c|}
\hline
\textbf{No.} & \textbf{Structural functions} & \textbf{Type} & \textbf{Selected \(\mathcal{A}\)} & \textbf{Number of features} & \textbf{Evaluation} & \textbf{SVM (Linear)} & \textbf{SVM (RBF kernel)} & \textbf{Ensemble Learning} & \textbf{Neural Network} \\
\hline
\multirow{3}{*}{Q1} & \multirow{3}{*}{Network Centrality} & \multirow{3}{*}{C} & \multirow{3}{*}{0.55} & \multirow{3}{*}{10} & Accuracy & \cellcolor{yellow!20} 75.9\% & 83.9\% & 89.0\% & \cellcolor{yellow!20} 89.0\% \\
 &  &  &  &  & AUC & \cellcolor{yellow!20} 0.8229 & 0.8951 & 0.9251 & \cellcolor{yellow!20} 0.9253 \\
 &  &  &  &  & MCC & \cellcolor{yellow!20} 0.521 & 0.679 & 0.780 & \cellcolor{yellow!20} 0.780 \\
\hline
\multirow{3}{*}{Q2} & \multirow{3}{*}{Network Centrality} & \multirow{3}{*}{C} & \multirow{3}{*}{0.55,0.6,0.65,0.7} & \multirow{3}{*}{40} & Accuracy & 82.7\% & 84.8\% & \cellcolor{yellow!50} 90.8\% & 86.7\% \\
 &  &  &  &  & AUC & 0.8905 & 0.9109 & \cellcolor{yellow!50} 0.9511 & 0.9269 \\
 &  &  &  &  & MCC & 0.654 & 0.698 & \cellcolor{yellow!50} 0.815 & 0.733 \\
\hline
\multirow{3}{*}{Q3} & \multirow{3}{*}{Network Centrality} & \multirow{3}{*}{C} & \multirow{3}{*}{0.55,0.6,0.65,0.7,0.75} & \multirow{3}{*}{50} & Accuracy & \cellcolor{yellow!20} 83.8\% & 85.7\% & 90.4\% & 87.4\% \\
 &  &  &  &  & AUC & \cellcolor{yellow!20} 0.8984 & 0.9204 & 0.9489 & 0.9393 \\
 &  &  &  &  & MCC & \cellcolor{yellow!20} 0.676 & 0.714 & 0.809 & 0.749 \\
\hline
\multirow{3}{*}{Q4} & \multirow{3}{*}{Angle function on graphs} & \multirow{3}{*}{D} & \multirow{3}{*}{0.6} & \multirow{3}{*}{10} & Accuracy & 72.4\% & 78.6\% & 80.3\% & 80.5\% \\
 &  &  &  &  & AUC & 0.7907 & 0.8497 & 0.8807 & 0.8786 \\
 &  &  &  &  & MCC & 0.448 & 0.572 & 0.606 & 0.610 \\
\hline
\multirow{3}{*}{Q5} & \multirow{3}{*}{Angle function on graphs} & \multirow{3}{*}{D} & \multirow{3}{*}{0.55,0.6} & \multirow{3}{*}{20} & Accuracy & 75.0\% & 81.3\% & 82.5\% & 82.5\% \\
 &  &  &  &  & AUC & 0.8187 & 0.8787 & 0.898 & 0.8961 \\
 &  &  &  &  & MCC & 0.501 & 0.625 & 0.651 & 0.651 \\
\hline
\multirow{3}{*}{Q6} & \multirow{3}{*}{Angle function on graphs} & \multirow{3}{*}{D} & \multirow{3}{*}{0.55,0.6,0.65,0.7} & \multirow{3}{*}{40} & Accuracy & 77.9\% & 83.4\% & 84.2\% & 84.0\% \\
 &  &  &  &  & AUC & 0.8528 & 0.9108 & 0.9129 & 0.9108 \\
 &  &  &  &  & MCC & 0.558 & 0.668 & 0.683 & 0.679 \\
\hline
\multirow{3}{*}{Q7} & \multirow{3}{*}{Angle function on graphs} & \multirow{3}{*}{D} & \multirow{3}{*}{0.55,0.6,0.65,0.7,0.75} & \multirow{3}{*}{50} & Accuracy & 79.1\% & 84.0\% & 84.3\% & 84.9\% \\
 &  &  &  &  & AUC & 0.8637 & 0.9069 & 0.9158 & 0.9175 \\
 &  &  &  &  & MCC & 0.583 & 0.680 & 0.686 & 0.699 \\
\hline
\multirow{3}{*}{Q8} & \multirow{3}{*}{Centrality and angle function} & \multirow{3}{*}{C+D} & \multirow{3}{*}{0.55 for centrality and 0.6 for angle function} & \multirow{3}{*}{10+10} & Accuracy & \cellcolor{yellow!20} 81.9\% & 85.2\% & \cellcolor{yellow!20} 89.4\% & 86.2\% \\
 &  &  &  &  & AUC & \cellcolor{yellow!20} 0.8753 & 0.9067 & \cellcolor{yellow!20} 0.9309 & 0.9185 \\
 &  &  &  &  & MCC & \cellcolor{yellow!20} 0.638 & 0.704 & \cellcolor{yellow!20} 0.790 & 0.724 \\
\hline
\multirow{3}{*}{Q9} & \multirow{3}{*}{Centrality and angle function} & \multirow{3}{*}{C+D} & \multirow{3}{*}{0.55,0.6,0.65,0.7} & \multirow{3}{*}{40+40} & Accuracy & 87.6\% & 88.5\% & 90.4\% & 87.3\% \\
 &  &  &  &  & AUC & 0.9348 & 0.9392 & 0.9523 & 0.9307 \\
 &  &  &  &  & MCC & 0.753 & 0.770 & 0.808 & 0.746 \\
\hline
\multirow{3}{*}{Q10} & \multirow{3}{*}{Centrality and angle function} & \multirow{3}{*}{C+D} & \multirow{3}{*}{0.55,0.6,0.65,0.7,0.75} & \multirow{3}{*}{50+50} & Accuracy & \cellcolor{yellow!20} 88.3\% & 88.8\% & \cellcolor{yellow!50} 90.7\% & 89.3\% \\
 &  &  &  &  & AUC & \cellcolor{yellow!20} 0.9378 & 0.9392 & \cellcolor{yellow!50} 0.9536 & 0.9515 \\
 &  &  &  &  & MCC & \cellcolor{yellow!20} 0.767 & 0.776 & \cellcolor{yellow!50} 0.815 & 0.787 \\
\hline
\end{tabular}
\end{adjustbox}
\caption{Exp A, Graph Descriptor (Type C and D)}
\label{fig:a_2}
\end{table*}

\begin{table*}[htbp]
\centering
\renewcommand{\arraystretch}{1.2}
\setlength{\tabcolsep}{4pt}
\begin{adjustbox}{width=1\textwidth}
\begin{tabular}{|c|c|c|c|c|c|c|c|c|c|c|}
\hline
\textbf{No.} & \textbf{Structural functions} & \textbf{Type} & \textbf{Group} & \textbf{Number of features} & \textbf{Evaluation} & \textbf{SVM (Linear)} & \textbf{SVM (RBF kernel)} & \textbf{Ensemble Learning} & \textbf{Neural Network} \\
\hline
\multirow{3}{*}{P1} & \multirow{3}{*}{Only G} & \multirow{3}{*}{A} & \multirow{3}{*}{1} & \multirow{3}{*}{60} & Accuracy & 57.9\% & 71.2\% & 70.6\% & 60.7\% \\
 &  &  &  &  & AUC & 0.6157 & 0.7801 & 0.7792 & 0.6508 \\
 &  &  &  &  & MCC & 0.159 & 0.424 & 0.412 & 0.215 \\
\hline
\multirow{3}{*}{P2} & \multirow{3}{*}{Only G} & \multirow{3}{*}{A} & \multirow{3}{*}{2} & \multirow{3}{*}{60} & Accuracy & 68.4\% & 81.0\% & 79.6\% & 71.9\% \\
 &  &  &  &  & AUC & 0.7485 & 0.8913 & 0.8788 & 0.7928 \\
 &  &  &  &  & MCC & 0.367 & 0.621 & 0.591 & 0.439 \\
\hline
\multirow{3}{*}{P3} & \multirow{3}{*}{Only G} & \multirow{3}{*}{A} & \multirow{3}{*}{1} & \multirow{3}{*}{80} & Accuracy & 56.3\% & 71.3\% & 71.2\% & 57.5\% \\
 &  &  &  &  & AUC & 0.5918 & 0.7819 & 0.7854 & 0.5959 \\
 &  &  &  &  & MCC & 0.127 & 0.427 & 0.425 & 0.151 \\
\hline
\multirow{3}{*}{P4} & \multirow{3}{*}{Only G} & \multirow{3}{*}{A} & \multirow{3}{*}{2} & \multirow{3}{*}{80} & Accuracy & \cellcolor{yellow!20} 71.7\% & \cellcolor{yellow!50} 82.8\% & 80.5\% & 72.4\% \\
 &  &  &  &  & AUC & \cellcolor{yellow!20} 0.7891 & \cellcolor{yellow!50} 0.9043 & 0.886 & 0.803 \\
 &  &  &  &  & MCC & \cellcolor{yellow!20} 0.434 & \cellcolor{yellow!50} 0.656 & 0.610 & 0.448 \\
\hline
\multirow{3}{*}{P5} & \multirow{3}{*}{Only Psi} & \multirow{3}{*}{B} & \multirow{3}{*}{1 and 2} & \multirow{3}{*}{20} & Accuracy & 62.9\% & 62.7\% & 66.2\% & 66.1\% \\
 &  &  &  &  & AUC & 0.6761 & 0.6674 & 0.6772 & 0.7138 \\
 &  &  &  &  & MCC & 0.258 & 0.254 & 0.256 & 0.323 \\
\hline
\multirow{3}{*}{P6} & \multirow{3}{*}{G and Psi} & \multirow{3}{*}{A+B} & \multirow{3}{*}{1} & \multirow{3}{*}{60+20} & Accuracy & 73.3\% & 78.4\% & 78.5\% & 77.9\% \\
 &  &  &  &  & AUC & 0.8121 & 0.8626 & 0.8404 & 0.8618 \\
 &  &  &  &  & MCC & 0.468 & 0.569 & 0.516 & 0.558 \\
\hline
\multirow{3}{*}{P7} & \multirow{3}{*}{G and Psi} & \multirow{3}{*}{A+B} & \multirow{3}{*}{2} & \multirow{3}{*}{60+20} & Accuracy & 76.3\% & 83.6\% & 82.5\% & 81.7\% \\
 &  &  &  &  & AUC & 0.8562 & 0.914 & 0.9046 & 0.9022 \\
 &  &  &  &  & MCC & 0.526 & 0.672 & 0.651 & 0.661 \\
\hline
\multirow{3}{*}{P8} & \multirow{3}{*}{G and Psi} & \multirow{3}{*}{A+B} & \multirow{3}{*}{1} & \multirow{3}{*}{80+20} & Accuracy & 73.8\% & 78.5\% & 75.8\% & 78.1\% \\
 &  &  &  &  & AUC & 0.8137 & 0.8656 & 0.8416 & 0.8662 \\
 &  &  &  &  & MCC & 0.476 & 0.570 & 0.516 & 0.562 \\
\hline
\multirow{3}{*}{P9} & \multirow{3}{*}{G and Psi} & \multirow{3}{*}{A+B} & \multirow{3}{*}{2} & \multirow{3}{*}{80+20} & Accuracy & \cellcolor{yellow!20} 76.8\% & \cellcolor{yellow!50} 84.8\% & 81.4\% & 82.5\% \\
 &  &  &  &  & AUC & \cellcolor{yellow!20} 0.8603 & \cellcolor{yellow!50} 0.9258 & 0.9188 & 0.9128 \\
 &  &  &  &  & MCC & \cellcolor{yellow!20} 0.535 & \cellcolor{yellow!50} 0.696 & 0.682 & 0.651 \\
\hline
\end{tabular}
\end{adjustbox}
\caption{Exp B, Traditional Descriptor (Type A and B)}
\label{fig:b_1}
\end{table*}

\begin{table*}[htbp]
\centering
\renewcommand{\arraystretch}{1.2}
\setlength{\tabcolsep}{4pt}
\begin{adjustbox}{width=1\textwidth}
\begin{tabular}{|c|c|c|c|c|c|c|c|c|c|c|}
\hline
\textbf{No.} & \textbf{Structural functions} & \textbf{Type} & \textbf{Selected \(\mathcal{A}\)} & \textbf{Number of features} & \textbf{Evaluation} & \textbf{SVM (Linear)} & \textbf{SVM (RBF kernel)} & \textbf{Ensemble Learning} & \textbf{Neural Network} \\
\hline
\multirow{3}{*}{Q1} & \multirow{3}{*}{Network Centrality} & \multirow{3}{*}{C} & \multirow{3}{*}{0.55} & \multirow{3}{*}{10} & Accuracy & \cellcolor{yellow!20} 76.5\% & 81.5\% & \cellcolor{yellow!20} 85.7\% & 85.6\% \\
 &  &  &  &  & AUC & \cellcolor{yellow!20} 0.8313 & 0.8713 & \cellcolor{yellow!20} 0.9171 & 0.9175 \\
 &  &  &  &  & MCC & \cellcolor{yellow!20} 0.530 & 0.631 & \cellcolor{yellow!20} 0.715 & 0.715 \\
\hline
\multirow{3}{*}{Q2} & \multirow{3}{*}{Network Centrality} & \multirow{3}{*}{C} & \multirow{3}{*}{0.55,0.6,0.65,0.7} & \multirow{3}{*}{40} & Accuracy & \cellcolor{yellow!20} 83.2\% & 84.9\% & \cellcolor{yellow!50} 90.8\% & 86.2\% \\
 &  &  &  &  & AUC & \cellcolor{yellow!20} 0.8988 & 0.9174 & \cellcolor{yellow!50} 0.9554 & 0.9293 \\
 &  &  &  &  & MCC & \cellcolor{yellow!20} 0.664 & 0.679 & \cellcolor{yellow!50} 0.817 & 0.726 \\
\hline
\multirow{3}{*}{Q3} & \multirow{3}{*}{Network Centrality} & \multirow{3}{*}{C} & \multirow{3}{*}{0.55,0.6,0.65,0.7,0.75} & \multirow{3}{*}{50} & Accuracy & 82.8\% & 85.4\% & 90.6\% & 86.6\% \\
 &  &  &  &  & AUC & 0.8966 & 0.9192 & 0.9541 & 0.9358 \\
 &  &  &  &  & MCC & 0.656 & 0.709 & 0.813 & 0.733 \\
\hline
\multirow{3}{*}{Q4} & \multirow{3}{*}{Angle function on graphs} & \multirow{3}{*}{D} & \multirow{3}{*}{0.6} & \multirow{3}{*}{10} & Accuracy & 75.9\% & 81.4\% & 82.8\% & 83.5\% \\
 &  &  &  &  & AUC & 0.8341 & 0.8766 & 0.9056 & 0.912 \\
 &  &  &  &  & MCC & 0.518 & 0.628 & 0.657 & 0.671 \\
\hline
\multirow{3}{*}{Q5} & \multirow{3}{*}{Angle function on graphs} & \multirow{3}{*}{D} & \multirow{3}{*}{0.55,0.6} & \multirow{3}{*}{20} & Accuracy & 76.7\% & 82.0\% & 83.6\% & 83.4\% \\
 &  &  &  &  & AUC & 0.8411 & 0.8897 & 0.9087 & 0.9101 \\
 &  &  &  &  & MCC & 0.535 & 0.640 & 0.671 & 0.669 \\
\hline
\multirow{3}{*}{Q6} & \multirow{3}{*}{Angle function on graphs} & \multirow{3}{*}{D} & \multirow{3}{*}{0.55,0.6,0.65,0.7} & \multirow{3}{*}{40} & Accuracy & 80.7\% & 85.3\% & 85.9\% & 86.3\% \\
 &  &  &  &  & AUC & 0.8829 & 0.9221 & 0.9309 & 0.9329 \\
 &  &  &  &  & MCC & 0.614 & 0.707 & 0.718 & 0.727 \\
\hline
\multirow{3}{*}{Q7} & \multirow{3}{*}{Angle function on graphs} & \multirow{3}{*}{D} & \multirow{3}{*}{0.55,0.6,0.65,0.7,0.75} & \multirow{3}{*}{50} & Accuracy & 81.4\% & 86.0\% & 86.2\% & 86.5\% \\
 &  &  &  &  & AUC & 0.8936 & 0.9279 & 0.9353 & 0.9371 \\
 &  &  &  &  & MCC & 0.629 & 0.720 & 0.723 & 0.731 \\
\hline
\multirow{3}{*}{Q8} & \multirow{3}{*}{Centrality and angle function} & \multirow{3}{*}{C+D} & \multirow{3}{*}{0.55 for centrality and 0.6 for angle function} & \multirow{3}{*}{10+10} & Accuracy & 
\cellcolor{yellow!20} 82.6\% & 85.2\% & \cellcolor{yellow!20} 86.6\% & 86.2\% \\
 &  &  &  &  & AUC & \cellcolor{yellow!20} 0.893 & 0.9175 & \cellcolor{yellow!20} 0.9278 & 0.9331 \\
 &  &  &  &  & MCC & \cellcolor{yellow!20} 0.652 & 0.704 & \cellcolor{yellow!20} 0.733 & 0.724 \\
\hline
\multirow{3}{*}{Q9} & \multirow{3}{*}{Centrality and angle function} & \multirow{3}{*}{C+D} & \multirow{3}{*}{0.55,0.6,0.65,0.7} & \multirow{3}{*}{40+40} & Accuracy & 87.4\% & 88.6\% & 90.6\% & 87.7\% \\
 &  &  &  &  & AUC & 0.9369 & 0.9444 & 0.9555 & 0.9394 \\
 &  &  &  &  & MCC & 0.748 & 0.772 & 0.813 & 0.755 \\
\hline
\multirow{3}{*}{Q10} & \multirow{3}{*}{Centrality and angle function} & \multirow{3}{*}{C+D} & \multirow{3}{*}{0.55,0.6,0.65,0.7,0.75} & \multirow{3}{*}{50+50} & Accuracy & \cellcolor{yellow!20} 87.5\% & 88.9\% & \cellcolor{yellow!50} 91.2\% & 89.3\% \\
 &  &  &  &  & AUC & \cellcolor{yellow!20} 0.9365 & 0.9454 & \cellcolor{yellow!50} 0.9577 & 0.9535 \\
 &  &  &  &  & MCC & \cellcolor{yellow!20} 0.750 & 0.779 &\cellcolor{yellow!50} 0.824 & 0.787 \\
\hline
\end{tabular}
\end{adjustbox}
\caption{Exp B, Graph Descriptor (Type C and D)}
\label{fig:b_2}
\end{table*}

\begin{table*}[htbp]
\centering
\renewcommand{\arraystretch}{1.2}
\setlength{\tabcolsep}{4pt}
\begin{adjustbox}{width=1\textwidth}
\begin{tabular}{|c|c|c|c|c|c|c|c|c|c|c|}
\hline
\textbf{No.} & \textbf{Structural functions} & \textbf{Type} & \textbf{Group} & \textbf{Number of features} & \textbf{Evaluation} & \textbf{SVM (Linear)} & \textbf{SVM (RBF kernel)} & \textbf{Ensemble Learning} & \textbf{Neural Network} \\
\hline
\multirow{3}{*}{P1} & \multirow{3}{*}{Only G} & \multirow{3}{*}{A} & \multirow{3}{*}{1} & \multirow{3}{*}{60} & Accuracy & 50.4\% & 55.8\% & 56.4\% & 54.1\% \\
 &  &  &  &  & AUC & 0.5056 & 0.5812 & 0.5927 & 0.5605 \\
 &  &  &  &  & MCC & 0.008 & 0.116 & 0.127 & 0.083 \\
\hline
\multirow{3}{*}{P2} & \multirow{3}{*}{Only G} & \multirow{3}{*}{A} & \multirow{3}{*}{2} & \multirow{3}{*}{60} & Accuracy & \cellcolor{yellow!20} 53.8\% & 70.7\% & 68.5\% & 52.7\% \\
 &  &  &  &  & AUC & \cellcolor{yellow!20} 0.5452 & 0.7757 & 0.7524 & 0.5394 \\
 &  &  &  &  & MCC & \cellcolor{yellow!20} 0.077 & 0.414 & 0.370 & 0.055 \\
\hline
\multirow{3}{*}{P3} & \multirow{3}{*}{Only G} & \multirow{3}{*}{A} & \multirow{3}{*}{1} & \multirow{3}{*}{80} & Accuracy & 52.6\% & 58.8\% & 58.0\% & 53.4\% \\
 &  &  &  &  & AUC & 0.5412 & 0.622 & 0.6156 & 0.5526 \\
 &  &  &  &  & MCC & 0.054 & 0.175 & 0.160 & 0.071 \\
\hline
\multirow{3}{*}{P4} & \multirow{3}{*}{Only G} & \multirow{3}{*}{A} & \multirow{3}{*}{2} & \multirow{3}{*}{80} & Accuracy & 52.0\% & \cellcolor{yellow!20} 72.8\% & 69.5\% & 52.9\% \\
 &  &  &  &  & AUC & 0.526 & \cellcolor{yellow!20} 0.8 & 0.7648 & 0.5454 \\
 &  &  &  &  & MCC & 0.041 & \cellcolor{yellow!20} 0.458 & 0.392 & 0.062 \\
\hline
\multirow{3}{*}{P5} & \multirow{3}{*}{Only Psi} & \multirow{3}{*}{B} & \multirow{3}{*}{1 and 2} & \multirow{3}{*}{20} & Accuracy & 60.1\% & 61.0\% & 60.7\% & 61.0\% \\
 &  &  &  &  & AUC & 0.636 & 0.6488 & 0.6514 & 0.6319 \\
 &  &  &  &  & MCC & 0.201 & 0.224 & 0.218 & 0.222 \\
\hline
\multirow{3}{*}{P6} & \multirow{3}{*}{G and Psi} & \multirow{3}{*}{A+B} & \multirow{3}{*}{1} & \multirow{3}{*}{60+20} & Accuracy & 68.1\% & 69.6\% & 62.5\% & 69.6\% \\
 &  &  &  &  & AUC & 0.7677 & 0.7833 & 0.6712 & 0.773 \\
 &  &  &  &  & MCC & 0.378 & 0.408 & 0.253 & 0.402 \\
\hline
\multirow{3}{*}{P7} & \multirow{3}{*}{G and Psi} & \multirow{3}{*}{A+B} & \multirow{3}{*}{2} & \multirow{3}{*}{60+20} & Accuracy & \cellcolor{yellow!20} 69.0\% & 78.4\% & 74.1\% & 76.8\% \\
 &  &  &  &  & AUC & \cellcolor{yellow!20} 0.7576 & 0.8783 & 0.8256 & 0.8477 \\
 &  &  &  &  & MCC & \cellcolor{yellow!20} 0.383 & 0.575 & 0.486 & 0.539 \\
\hline
\multirow{3}{*}{P8} & \multirow{3}{*}{G and Psi} & \multirow{3}{*}{A+B} & \multirow{3}{*}{1} & \multirow{3}{*}{80+20} & Accuracy & 68.4\% & 70.4\% & 62.3\% & 66.1\% \\
 &  &  &  &  & AUC & 0.7655 & 0.7932 & 0.6724 & 0.72 \\
 &  &  &  &  & MCC & 0.380 & 0.424 & 0.248 & 0.323 \\
\hline
\multirow{3}{*}{P9} & \multirow{3}{*}{G and Psi} & \multirow{3}{*}{A+B} & \multirow{3}{*}{2} & \multirow{3}{*}{80+20} & Accuracy & 60.8\% & \cellcolor{yellow!50} 80.6\% & 74.7\% & 62.4\% \\
 &  &  &  &  & AUC & 0.6618 & \cellcolor{yellow!50} 0.8995 & 0.8317 & 0.6865 \\
 &  &  &  &  & MCC & 0.219 & \cellcolor{yellow!50} 0.623 & 0.498 & 0.257 \\
\hline
\end{tabular}
\end{adjustbox}
\caption{Exp C, Traditional Descriptor (Type A and B)}
\label{fig:c_1}
\end{table*}

\begin{table*}[htbp]
\centering
\renewcommand{\arraystretch}{1.2}
\setlength{\tabcolsep}{4pt}
\begin{adjustbox}{width=1\textwidth}
\begin{tabular}{|c|c|c|c|c|c|c|c|c|c|c|}
\hline
\textbf{No.} & \textbf{Structural functions} & \textbf{Type} & \textbf{Selected \(\mathcal{A}\)} & \textbf{Number of features} & \textbf{Evaluation} & \textbf{SVM (Linear)} & \textbf{SVM (RBF kernel)} & \textbf{Ensemble Learning} & \textbf{Neural Network} \\
\hline
\multirow{3}{*}{Q1} & \multirow{3}{*}{Network Centrality} & \multirow{3}{*}{C} & \multirow{3}{*}{0.55} & \multirow{3}{*}{10} & Accuracy & \cellcolor{yellow!20} 76.5\% & 79.5\% & \cellcolor{yellow!50} 86.1\% & 79.5\% \\
 &  &  &  &  & AUC & \cellcolor{yellow!20} 0.8253 & 0.862 & \cellcolor{yellow!50} 0.913 & 0.8609 \\
 &  &  &  &  & MCC & \cellcolor{yellow!20} 0.530 & 0.589 & \cellcolor{yellow!50} 0.723 & 0.590 \\
\hline
\multirow{3}{*}{Q2} & \multirow{3}{*}{Network Centrality} & \multirow{3}{*}{C} & \multirow{3}{*}{0.55,0.6,0.65,0.7} & \multirow{3}{*}{40} & Accuracy & 72.4\% & 78.9\% & 80.6\% & 82.1\% \\
 &  &  &  &  & AUC & 0.7953 & 0.891 & 0.9201 & 0.8691 \\
 &  &  &  &  & MCC & 0.449 & 0.578 & 0.670 & 0.643 \\
\hline
\multirow{3}{*}{Q3} & \multirow{3}{*}{Network Centrality} & \multirow{3}{*}{C} & \multirow{3}{*}{0.55,0.6,0.65,0.7,0.75} & \multirow{3}{*}{50} & Accuracy & 71.7\% & 78.5\% & 82.7\% & 82.9\% \\
 &  &  &  &  & AUC & 0.7872 & 0.8642 & 0.9017 & 0.9043 \\
 &  &  &  &  & MCC & 0.433 & 0.570 & 0.658 & 0.658 \\
\hline
\multirow{3}{*}{Q4} & \multirow{3}{*}{Angle function on graphs} & \multirow{3}{*}{D} & \multirow{3}{*}{0.6} & \multirow{3}{*}{10} & Accuracy & 73.1\% & 77.5\% & 79.6\% & 80.2\% \\
 &  &  &  &  & AUC & 0.8049 & 0.8344 & 0.8703 & 0.8961 \\
 &  &  &  &  & MCC & 0.463 & 0.553 & 0.596 & 0.608 \\
\hline
\multirow{3}{*}{Q5} & \multirow{3}{*}{Angle function on graphs} & \multirow{3}{*}{D} & \multirow{3}{*}{0.55,0.6} & \multirow{3}{*}{20} & Accuracy & 74.3\% & 80.2\% & 81.1\% & 81.2\% \\
 &  &  &  &  & AUC & 0.8135 & 0.869 & 0.888 & 0.888 \\
 &  &  &  &  & MCC & 0.485 & 0.605 & 0.624 & 0.626 \\
\hline
\multirow{3}{*}{Q6} & \multirow{3}{*}{Angle function on graphs} & \multirow{3}{*}{D} & \multirow{3}{*}{0.55,0.6,0.65,0.7} & \multirow{3}{*}{40} & Accuracy & 76.2\% & 82.2\% & 82.6\% & 81.1\% \\
 &  &  &  &  & AUC & 0.8364 & 0.8886 & 0.9007 & 0.8841 \\
 &  &  &  &  & MCC & 0.524 & 0.644 & 0.654 & 0.622 \\
\hline
\multirow{3}{*}{Q7} & \multirow{3}{*}{Angle function on graphs} & \multirow{3}{*}{D} & \multirow{3}{*}{0.55,0.6,0.65,0.7,0.75} & \multirow{3}{*}{50} & Accuracy & 76.4\% & 81.6\% & 82.6\% & 81.5\% \\
 &  &  &  &  & AUC & 0.842 & 0.8845 & 0.897 & 0.8895 \\
 &  &  &  &  & MCC & 0.527 & 0.632 & 0.652 & 0.630 \\
\hline
\multirow{3}{*}{Q8} & \multirow{3}{*}{Centrality and angle function} & \multirow{3}{*}{C+D} & \multirow{3}{*}{0.55 for centrality and 0.6 for angle function} & \multirow{3}{*}{10+10} & Accuracy & \cellcolor{yellow!20} 81.2\% & 82.7\% & \cellcolor{yellow!50} 87.8\% & 85.0\% \\
 &  &  &  &  & AUC & \cellcolor{yellow!20} 0.8826 & 0.8947 & \cellcolor{yellow!50} 0.9228 & 0.9145 \\
 &  &  &  &  & MCC & \cellcolor{yellow!20} 0.626 & 0.654 & \cellcolor{yellow!50} 0.757 & 0.701 \\
\hline
\multirow{3}{*}{Q9} & \multirow{3}{*}{Centrality and angle function} & \multirow{3}{*}{C+D} & \multirow{3}{*}{0.55,0.6,0.65,0.7} & \multirow{3}{*}{40+40} & Accuracy & 80.9\% & 86.4\% & 80.5\% & 81.7\% \\
 &  &  &  &  & AUC & 0.888 & 0.9254 & 0.8978 & 0.8953 \\
 &  &  &  &  & MCC & 0.619 & 0.727 & 0.619 & 0.634 \\
\hline
\multirow{3}{*}{Q10} & \multirow{3}{*}{Centrality and angle function} & \multirow{3}{*}{C+D} & \multirow{3}{*}{0.55,0.6,0.65,0.7,0.75} & \multirow{3}{*}{50+50} & Accuracy & 80.1\% & 85.6\% & 84.7\% & 79.3\% \\
 &  &  &  &  & AUC & 0.8772 & 0.9175 & 0.9177 & 0.8747 \\
 &  &  &  &  & MCC & 0.601 & 0.712 & 0.696 & 0.589 \\
\hline
\end{tabular}
\end{adjustbox}
\caption{Exp C, Graph Descriptor (Type C and D)}
\label{fig:c_2}
\end{table*}

\begin{table*}[htbp]
\centering
\renewcommand{\arraystretch}{1.2}
\setlength{\tabcolsep}{4pt}
\begin{adjustbox}{width=1\textwidth}
\begin{tabular}{|c|c|c|c|c|c|c|c|c|c|c|}
\hline
\textbf{No.} & \textbf{Structural functions} & \textbf{Type} & \textbf{Group} & \textbf{Number of features} & \textbf{Evaluation} & \textbf{SVM (Linear)} & \textbf{SVM (RBF kernel)} & \textbf{Ensemble Learning} & \textbf{Neural Network} \\
\hline
\multirow{3}{*}{P1} & \multirow{3}{*}{Only G} & \multirow{3}{*}{A} & \multirow{3}{*}{1} & \multirow{3}{*}{60} & Accuracy & \cellcolor{yellow!20} 56.6\% & 66.7\% & 65.6\% & 53.2\% \\
 &  &  &  &  & AUC & \cellcolor{yellow!20} 0.5955 & 0.7243 & 0.7156 & 0.5438 \\
 &  &  &  &  & MCC & \cellcolor{yellow!20} 0.134 & 0.337 & 0.316 & 0.063 \\
\hline
\multirow{3}{*}{P2} & \multirow{3}{*}{Only G} & \multirow{3}{*}{A} & \multirow{3}{*}{2} & \multirow{3}{*}{60} & Accuracy & 46.0\% & \cellcolor{yellow!50} 70.3\% & 69.3\% & 48.8\% \\
 &  &  &  &  & AUC & 0.4514 & \cellcolor{yellow!50} 0.7753 & 0.7631 & 0.4919 \\
 &  &  &  &  & MCC & -0.084 & \cellcolor{yellow!50} 0.417 & 0.388 & -0.026 \\
\hline
\multirow{3}{*}{P3} & \multirow{3}{*}{Only G} & \multirow{3}{*}{A} & \multirow{3}{*}{1} & \multirow{3}{*}{80} & Accuracy & 54.3\% & 65.5\% & 65.9\% & 52.7\% \\
 &  &  &  &  & AUC & 0.5642 & 0.7097 & 0.7195 & 0.5357 \\
 &  &  &  &  & MCC & 0.090 & 0.311 & 0.322 & 0.054 \\
\hline
\multirow{3}{*}{P4} & \multirow{3}{*}{Only G} & \multirow{3}{*}{A} & \multirow{3}{*}{2} & \multirow{3}{*}{80} & Accuracy & 52.7\% & 71.5\% & 68.7\% & 50.2\% \\
 &  &  &  &  & AUC & 0.5485 & 0.7895 & 0.7563 & 0.5057 \\
 &  &  &  &  & MCC & 0.057 & 0.438 & 0.376 & 0.003 \\
\hline
\multirow{3}{*}{P5} & \multirow{3}{*}{Only Psi} & \multirow{3}{*}{B} & \multirow{3}{*}{1 and 2} & \multirow{3}{*}{20} & Accuracy & 56.6\% & 54.2\% & 53.8\% & 57.6\% \\
 &  &  &  &  & AUC & 0.5504 & 0.5531 & 0.5507 & 0.5647 \\
 &  &  &  &  & MCC & 0.146 & 0.083 & 0.075 & 0.161 \\
\hline
\multirow{3}{*}{P6} & \multirow{3}{*}{G and Psi} & \multirow{3}{*}{A+B} & \multirow{3}{*}{1} & \multirow{3}{*}{60+20} & Accuracy & 64.1\% & 70.1\% & 66.7\% & 58.2\% \\
 &  &  &  &  & AUC & 0.6811 & 0.7701 & 0.732 & 0.606 \\
 &  &  &  &  & MCC & 0.288 & 0.405 & 0.335 & 0.165 \\
\hline
\multirow{3}{*}{P7} & \multirow{3}{*}{G and Psi} & \multirow{3}{*}{A+B} & \multirow{3}{*}{2} & \multirow{3}{*}{60+20} & Accuracy & 54.7\% & \cellcolor{yellow!50} 78.8\% & 71.6\% & 51.4\% \\
 &  &  &  &  & AUC & 0.5768 & \cellcolor{yellow!50} 0.8464 & 0.7889 & 0.5634 \\
 &  &  &  &  & MCC & 0.097 & \cellcolor{yellow!50} 0.583 & 0.435 & 0.035 \\
\hline
\multirow{3}{*}{P8} & \multirow{3}{*}{G and Psi} & \multirow{3}{*}{A+B} & \multirow{3}{*}{1} & \multirow{3}{*}{80+20} & Accuracy & 62.3\% & 70.5\% & 64.5\% & 59.4\% \\
 &  &  &  &  & AUC & 0.6617 & 0.7745 & 0.7021 & 0.6169 \\
 &  &  &  &  & MCC & 0.255 & 0.411 & 0.291 & 0.189 \\
\hline
\multirow{3}{*}{P9} & \multirow{3}{*}{G and Psi} & \multirow{3}{*}{A+B} & \multirow{3}{*}{2} & \multirow{3}{*}{80+20} & Accuracy & \cellcolor{yellow!20} 66.7\% & 75.7\% & 70.7\% & 51.4\% \\
 &  &  &  &  & AUC & \cellcolor{yellow!20} 0.7276 & 0.8406 & 0.7789 & 0.5235 \\
 &  &  &  &  & MCC & \cellcolor{yellow!20} 0.336 & 0.524 & 0.416 & 0.033 \\
\hline
\end{tabular}
\end{adjustbox}
\caption{Exp D, Traditional Descriptor (Type A and B)}
\label{fig:d_1}
\end{table*}

\begin{table*}[htbp]
\centering
\renewcommand{\arraystretch}{1.2}
\setlength{\tabcolsep}{4pt}
\begin{adjustbox}{width=1\textwidth}
\begin{tabular}{|c|c|c|c|c|c|c|c|c|c|c|}
\hline
\textbf{No.} & \textbf{Structural functions} & \textbf{Type} & \textbf{Selected \(\mathcal{A}\)} & \textbf{Number of features} & \textbf{Evaluation} & \textbf{SVM (Linear)} & \textbf{SVM (RBF kernel)} & \textbf{Ensemble Learning} & \textbf{Neural Network} \\
\hline
\multirow{3}{*}{Q1} & \multirow{3}{*}{Network Centrality} & \multirow{3}{*}{C} & \multirow{3}{*}{0.55} & \multirow{3}{*}{10} & Accuracy & \cellcolor{yellow!20} 73.1\% & 74.2\% & \cellcolor{yellow!20} 80.7\% & 76.6\% \\
 &  &  &  &  & AUC & \cellcolor{yellow!20} 0.798 & 0.8148 & \cellcolor{yellow!20} 0.8745 & 0.8493 \\
 &  &  &  &  & MCC & \cellcolor{yellow!20} 0.463 & 0.491 & \cellcolor{yellow!20} 0.616 & 0.545 \\
\hline
\multirow{3}{*}{Q2} & \multirow{3}{*}{Network Centrality} & \multirow{3}{*}{C} & \multirow{3}{*}{0.55,0.6,0.65,0.7} & \multirow{3}{*}{40} & Accuracy & \cellcolor{yellow!20} 75.8\% & 78.1\% & \cellcolor{yellow!50} 81.7\% & 76.1\% \\
 &  &  &  &  & AUC & \cellcolor{yellow!20} 0.8358 & 0.851 & \cellcolor{yellow!50} 0.8866 & 0.8371 \\
 &  &  &  &  & MCC & \cellcolor{yellow!20} 0.518 & 0.566 & \cellcolor{yellow!50} 0.636 & 0.522 \\
\hline
\multirow{3}{*}{Q3} & \multirow{3}{*}{Network Centrality} & \multirow{3}{*}{C} & \multirow{3}{*}{0.55,0.6,0.65,0.7,0.75} & \multirow{3}{*}{50} & Accuracy & 74.7\% & 77.5\% & 79.4\% & 70.3\% \\
 &  &  &  &  & AUC & 0.8208 & 0.8468 & 0.8681 & 0.7806 \\
 &  &  &  &  & MCC & 0.495 & 0.557 & 0.589 & 0.411 \\
\hline
\multirow{3}{*}{Q4} & \multirow{3}{*}{Angle function on graphs} & \multirow{3}{*}{D} & \multirow{3}{*}{0.6} & \multirow{3}{*}{10} & Accuracy & 67.4\% & 67.3\% & 69.4\% & 69.8\% \\
 &  &  &  &  & AUC & 0.7372 & 0.7405 & 0.7571 & 0.7672 \\
 &  &  &  &  & MCC & 0.348 & 0.349 & 0.388 & 0.395 \\
\hline
\multirow{3}{*}{Q5} & \multirow{3}{*}{Angle function on graphs} & \multirow{3}{*}{D} & \multirow{3}{*}{0.55,0.6} & \multirow{3}{*}{20} & Accuracy & 61.5\% & 67.8\% & 69.5\% & 66.5\% \\
 &  &  &  &  & AUC & 0.6605 & 0.7375 & 0.7595 & 0.7255 \\
 &  &  &  &  & MCC & 0.232 & 0.355 & 0.389 & 0.332 \\
\hline
\multirow{3}{*}{Q6} & \multirow{3}{*}{Angle function on graphs} & \multirow{3}{*}{D} & \multirow{3}{*}{0.55,0.6,0.65,0.7} & \multirow{3}{*}{40} & Accuracy & 68.0\% & 74.8\% & 73.0\% & 64.2\% \\
 &  &  &  &  & AUC & 0.7455 & 0.8116 & 0.799 & 0.7159 \\
 &  &  &  &  & MCC & 0.360 & 0.499 & 0.460 & 0.287 \\
\hline
\multirow{3}{*}{Q7} & \multirow{3}{*}{Angle function on graphs} & \multirow{3}{*}{D} & \multirow{3}{*}{0.55,0.6,0.65,0.7,0.75} & \multirow{3}{*}{50} & Accuracy & 72.3\% & 75.6\% & 73.1\% & 68.7\% \\
 &  &  &  &  & AUC & 0.7952 & 0.8192 & 0.8019 & 0.7367 \\
 &  &  &  &  & MCC & 0.446 & 0.519 & 0.462 & 0.375 \\
\hline
\multirow{3}{*}{Q8} & \multirow{3}{*}{Centrality and angle function} & \multirow{3}{*}{C+D} & \multirow{3}{*}{0.55 for centrality and 0.6 for angle function} & \multirow{3}{*}{10+10} & Accuracy & \cellcolor{yellow!20} 75.4\% & 75.4\% & \cellcolor{yellow!20} 81.7\% & 77.3\% \\
 &  &  &  &  & AUC & \cellcolor{yellow!20} 0.8343 & 0.8294 & \cellcolor{yellow!20} 0.8859 & 0.8707 \\
 &  &  &  &  & MCC & \cellcolor{yellow!20} 0.517 & 0.510 & \cellcolor{yellow!20} 0.636 & 0.566 \\
\hline
\multirow{3}{*}{Q9} & \multirow{3}{*}{Centrality and angle function} & \multirow{3}{*}{C+D} & \multirow{3}{*}{0.55,0.6,0.65,0.7} & \multirow{3}{*}{40+40} & Accuracy & \cellcolor{yellow!50} 82.0\% & 78.6\% & 81.4\% & 79.3\% \\
 &  &  &  &  & AUC & \cellcolor{yellow!50} 0.9118 & 0.8602 & 0.8837 & 0.8874 \\
 &  &  &  &  & MCC & \cellcolor{yellow!50} 0.649 & 0.578 & 0.630 & 0.599 \\
\hline
\multirow{3}{*}{Q10} & \multirow{3}{*}{Centrality and angle function} & \multirow{3}{*}{C+D} & \multirow{3}{*}{0.55,0.6,0.65,0.7,0.75} & \multirow{3}{*}{50+50} & Accuracy & 81.0\% & 77.2\% & 80.6\% & 78.5\% \\
 &  &  &  &  & AUC & 0.8966 & 0.8375 & 0.8723 & 0.8803 \\
 &  &  &  &  & MCC & 0.628 & 0.550 & 0.616 & 0.588 \\
\hline
\end{tabular}
\end{adjustbox}
\caption{Exp D, Graph Descriptor (Type C and D)}
\label{fig:d_2}
\end{table*}

\begin{table*}[htbp]
\centering
\renewcommand{\arraystretch}{1.2}
\setlength{\tabcolsep}{4pt}
\begin{adjustbox}{width=1\textwidth}
\begin{tabular}{|c|c|c|c|c|c|c|c|c|}
\hline
\textbf{Exp} & \textbf{No.} & \textbf{Type} & \textbf{Number of features} & \textbf{Evaluation} & \textbf{SVM (Linear)} & \textbf{SVM (RBF kernel)} & \textbf{Ensemble Learning} & \textbf{Neural Network} \\
\hline
\multirow{6}{*}{A} & \multirow{3}{*}{VGOP1} & \multirow{6}{*}{E} & \multirow{3}{*}{10} & Accuracy & 57.4\% & 61.9\% & 70.3\% & 68.8\% \\
 &  &  &  & AUC      & 0.6028 & 0.6685 & 0.7765 & 0.7488 \\
 &  &  &  & MCC      & 0.149  & 0.238  & 0.407  & 0.376  \\
\cline{2-2} \cline{4-9}
 & \multirow{3}{*}{VGOP2} & & \multirow{3}{*}{50} & Accuracy & \cellcolor{yellow!20} 67.3\% & 73.3\% & 77.8\% & \cellcolor{yellow!50} 80.0\% \\
 &  &  &  & AUC      & \cellcolor{yellow!20} 0.7432 & 0.8047 & 0.8579 & \cellcolor{yellow!50} 0.8803 \\
 &  &  &  & MCC      & \cellcolor{yellow!20} 0.349  & 0.467  & 0.556  & \cellcolor{yellow!50} 0.600  \\
\hline
\multirow{6}{*}{B} & \multirow{3}{*}{VGOP1} & \multirow{6}{*}{E} & \multirow{3}{*}{10} & Accuracy & 59.8\% & 62.1\% & 67.0\% & 69.7\% \\
 &  &  &  & AUC      & 0.6323 & 0.6737 & 0.7334 & 0.766  \\
 &  &  &  & MCC      & 0.198  & 0.243  & 0.340  & 0.394  \\
\cline{2-2} \cline{4-9}
 & \multirow{3}{*}{VGOP2} & & \multirow{3}{*}{50} & Accuracy & \cellcolor{yellow!20} 67.6\% & 74.2\% & 77.0\% & \cellcolor{yellow!50} 80.7\% \\
 &  &  &  & AUC      & \cellcolor{yellow!20} 0.7425 & 0.8164 & 0.8523 & \cellcolor{yellow!50} 0.8919 \\
 &  &  &  & MCC      & \cellcolor{yellow!20} 0.355  & 0.485  & 0.539  & \cellcolor{yellow!50} 0.615  \\
\hline
\multirow{6}{*}{C} & \multirow{3}{*}{VGOP1} & \multirow{6}{*}{E} & \multirow{3}{*}{10} & Accuracy & 56.8\% & 57.3\% & 60.6\% & 64.7\% \\
 &  &  &  & AUC      & 0.5933 & 0.6031 & 0.6416 & 0.7008 \\
 &  &  &  & MCC      & 0.141  & 0.146  & 0.214  & 0.297  \\
\cline{2-2} \cline{4-9}
 & \multirow{3}{*}{VGOP2} & & \multirow{3}{*}{50} & Accuracy & \cellcolor{yellow!20} 63.9\% & 66.4\% & 71.2\% & \cellcolor{yellow!50} 75.2\% \\
 &  &  &  & AUC      & \cellcolor{yellow!20} 0.6907 & 0.7193 & 0.7859 & \cellcolor{yellow!50} 0.8371 \\
 &  &  &  & MCC      & \cellcolor{yellow!20} 0.279  & 0.329  & 0.428  & \cellcolor{yellow!50} 0.511  \\
\hline
\multirow{6}{*}{D} & \multirow{3}{*}{VGOP1} & \multirow{6}{*}{E} & \multirow{3}{*}{10} & Accuracy & 55.5\% & 56.5\% & 58.7\% & 48.6\% \\
 &  &  &  & AUC      & 0.5804 & 0.5882 & 0.6064 & 0.4816 \\
 &  &  &  & MCC      & 0.114  & 0.131  & 0.174  & -0.028 \\
\cline{2-2} \cline{4-9}
 & \multirow{3}{*}{VGOP2} & & \multirow{3}{*}{50} & Accuracy & \cellcolor{yellow!20} 59.2\% & 65.3\% & \cellcolor{yellow!50} 65.5\% & 63.1\% \\
 &  &  &  & AUC      & \cellcolor{yellow!20} 0.6242 & 0.7011 & \cellcolor{yellow!50} 0.7104 & 0.6992 \\
 &  &  &  & MCC      & \cellcolor{yellow!20} 0.184  & 0.309  & \cellcolor{yellow!50} 0.311  & 0.280  \\
\hline
\end{tabular}
\end{adjustbox}
\caption{Both Exp A, B, C, and D, VGOP descriptor (Type E)}
\label{fig:e_all}
\end{table*}

\clearpage
\newpage

%

\end{document}